\pdfoutput=1
\documentclass[11pt,a4paper]{article}
\usepackage{jheppub}
\usepackage{amsmath}
\usepackage{amssymb}
\usepackage{slashed}
\usepackage{multirow}
\usepackage{changepage}
\usepackage{placeins}
\usepackage{bm}

\newcommand{\bctaunu}{b \to c \tau^- {\bar\nu}}

\def\bsll{b \to s \ell^+ \ell^-}
\def\bsnunubar{b \to s \nu {\bar\nu}}

\def \cL{{\cal L}}

\def \oQ{\overline{Q}}
\def \oL{\overline{L}}

\def \NP{{\rm NP}}

\def \eff{{\rm eff}}

\def \ga{\gamma}

\def \De{\Delta}

\def \si{\sigma}

\def \lb{\Lambda_b}
\def \lc{\Lambda_c}
\def \lbt{\Lambda_b \to \Lambda_c \tau \bar{\nu}_{\tau}}
\def \lbl{\Lambda_b \to \Lambda_c \ell \bar{\nu}_{\ell}}

\def\beq{\begin{equation}}
\def\eeq{\end{equation}}
\def\bea{\begin{eqnarray}}
\def\eea{\end{eqnarray}}
\def\ber{\begin{eqnarray*}}
\def\eer{\end{eqnarray*}}
\def\bwt{\begin{widetext}}
\def\ewt{\end{widetext}}
\def\nn{\nonumber}

\def\RD{R({D^{(*)}})}

\def\RDr{R_D^{Ratio}}

\def\RDrstar{R_{D^\ast}^{Ratio}}

\def\Rlc{R(\Lambda_c)}

\def\Rlcr{R_{\Lambda_c}^{Ratio}}

\newcommand{\bctaunutau}{b \to c \tau^- {\bar\nu}_\tau}

\def\bra#1{\left\langle #1\right|}
\def\ket#1{\left| #1\right\rangle}

\def \({\left(}
\def \){\right)}
\def \[{\left[}
\def \]{\right]}
\def \l|{\left|}
\def \r|{\right|}

\def \nn{\nonumber}
\def \nl{\nn \\}

\newcommand{\wm}{\phantom{-}}

\title{\boldmath Phenomenology of $\Lambda_b \to \Lambda_c \tau \bar{\nu}_{\tau}$ using lattice QCD calculations}

\author[a,b]{Alakabha Datta,}
\author[a]{Saeed Kamali,}
\author[c,d]{Stefan Meinel,}
\author[a,e]{and Ahmed Rashed}

\affiliation[a]{Department of Physics and Astronomy, University of Mississippi, Oxford, MS 38677, USA}
\affiliation[b]{Department of Physics and Astronomy, University of Hawaii, Honolulu, HI 96826, USA}
\affiliation[c]{Department of Physics,  University of Arizona, Tucson, AZ 85721, USA}
\affiliation[d]{RIKEN BNL Research Center, Brookhaven National Laboratory, Upton, NY 11973, USA}
\affiliation[e]{Department of Physics, Faculty of Science, Ain Shams University, Cairo, 11566, Egypt}

\emailAdd{datta@phy.olemiss.edu}
\emailAdd{skamali@go.olemiss.edu}
\emailAdd{smeinel@email.arizona.edu}
\emailAdd{amrashed@go.olemiss.edu}

\abstract{In a recent paper we studied the effect of new-physics operators with different Lorentz structures on the semileptonic $\Lambda_b \to \Lambda_c \tau \bar{\nu}_{\tau}$ decay. This decay is of interest in light of the $R({D^{(*)}})$ puzzle in the semileptonic $\bar{B} \to D^{(*)} \tau {\bar\nu}_\tau$ decays. In this work we add tensor operators to extend our previous results and consider both model-independent new physics (NP) and specific classes of models proposed to address the $R({D^{(*)}})$ puzzle. We show that a measurement of $R(\Lambda_c) = {\cal B}[\Lambda_b \to \Lambda_c \tau \bar{\nu}_{\tau}] / {\cal B}[\Lambda_b \to \Lambda_c \ell \bar{\nu}_{\ell}]$ can strongly constrain the NP parameters of models discussed for the $R({D^{(*)}})$ puzzle. We use form factors from lattice QCD to calculate all $\Lambda_b \to \Lambda_c \tau \bar{\nu}_{\tau}$ observables. The $\Lambda_b \to \Lambda_c$ tensor form factors had not previously been determined in lattice QCD, and we present new lattice results for these form factors here.}

\begin{document}

{\flushright
UMISS-HEP-2017-01 \\
RBRC-1228 \\

\vspace{-6ex}
}

\maketitle
\flushbottom

\section{Introduction}
\label{sec:introduction}

A major part of particle physics research is focused on searching for physics beyond the standard model (SM). In the flavor sector a key property of the SM gauge interactions is that they are lepton flavor universal. Evidence for violation of this property would be a clear sign of new physics (NP) beyond the SM. 
In the search for NP, the second and third generation quarks and leptons are 
quite special because they are comparatively heavier and are 
expected to be relatively more sensitive to NP. 
As an example, in certain versions of the two Higgs doublet models (2HDM) the couplings of the new Higgs bosons are proportional to the masses and so NP effects are more pronounced for the heavier generations. Moreover, the constraints on new physics, especially involving the third generation leptons and quarks, are somewhat weaker allowing for larger new physics effects. 

The charged-current decays $\bar{B} \to D^{(*)} \ell^{-} {\bar\nu}_\ell$ have been measured by the BaBar
\cite{Lees:2013uzd}, Belle \cite{Huschle:2015rga,Abdesselam:2016cgx} and LHCb \cite{Aaij:2015yra}
Collaborations. It is found that the values of the ratios
$\RD \equiv {\cal B}(\bar{B} \to D^{(*)} \tau^{-}
{\bar\nu}_\tau)/{\cal B}(\bar{B} \to D^{(*)} \ell^{-}
{\bar\nu}_\ell)$, where $\ell = e,\mu$, considerably exceed their SM
predictions. 
  
This ratio of branching fractions has certain advantages over the absolute branching fraction measurement
of $ B \to D^{(\ast)} \tau  \nu_\tau$ decays, as this is  relatively less sensitive to form factor variations and several systematic uncertainties,
such as those on the experimental efficiency, as well as the dependence on the value of $|V_{cb}|$, cancel in the ratio.

There are lattice QCD predictions for the ratio $R(D)_{SM}$ in the Standard Model \cite{Bailey:2012jg,Bailey:2015rga,Na:2015kha} that are in good agreement with one another,
\bea
R(D)_{SM} &=& 0.299 \pm 0.011 \quad \quad [\mathrm{FNAL/MILC}], \\
R(D)_{SM} &=& 0.300\pm 0.008 \quad \quad [\mathrm{HPQCD}].
\eea
These values are also in good agreement with the phenomenological prediction 
\cite{Sakaki:2013bfa}
\bea
R(D)_{SM} &=& 0.305\pm 0.012, \label{eq:RDpheno}
\eea
which is based on form factors extracted from experimental data for the $B\to D\ell\bar{\nu}$ differential decay
rates using heavy-quark effective theory. See also Ref.~\cite{Wang:2017jow} for a recent analysis of $B\to D$ form factors using light-cone sum rules.

A calculation of ${R}(D^\ast)_{SM}$ is not yet available from lattice QCD. The phenomenological
prediction using form factors extracted from $B\to D^*\ell\bar{\nu}$ experimental data is 
\cite{Fajfer:2012vx}
\bea
R(D^\ast)_{SM}&=& 0.252\pm 0.003.  \label{eq:RDstarpheno}
\eea

The averages of $R(D)$ and $R(D^{\ast})$ measurements, evaluated by the Heavy-Flavor Averaging Group, are \cite{HFAGWinter2016}
\bea
R(D)_{exp}  &=& 0.397 \pm 0.040 \pm 0.028, \label{eq:RDexp} \\
R(D^{\ast})_{exp} &=& 0.316 \pm 0.016 \pm 0.010, \label{eq:RDstexp}
\eea
where the first uncertainty is statistical and the second is
systematic. $R(D^{\ast})$ and $R(D)$ exceed the SM
predictions  by 3.3$\sigma$ and 1.9$\sigma$, respectively.
The combined analysis of $R(D^{\ast})$ and $R(D)$, taking
into account measurement correlations, finds that the deviation
is 4$\sigma$ from the SM prediction \cite{HFAGWinter2016,Ricciardi:2016pmh}.

Since lattice QCD results are not yet available for the $B\to D^*$ form factors at nonzero recoil and for the $B\to D$ tensor form factor,
we use the phenomenological form factors from Ref.~\cite{Sakaki:2013bfa} for both channels in our analysis. For $B \to D$,
we have compared the phenomenological results for $f_0$ and $f_+$ to the results obtained from a joint BGL $z$-expansion fit \cite{DeTar} to the
FNAL/MILC lattice QCD results \cite{Bailey:2015rga} and Babar \cite{Aubert:2009ac} and Belle experimental data \cite{Glattauer:2015teq}, and we
found that the differences between both sets of form factors are below 5\% across the entire kinematic range. The constraints
on the new-physics couplings from the experimental measurement of $R(D)$ obtained with both sets of form factors are practically identical.

We also construct the ratios of the experimental results (\ref{eq:RDexp}) and (\ref{eq:RDstexp}) to the phenomenological SM predictions (\ref{eq:RDpheno}) and (\ref{eq:RDstarpheno}):
\bea
\RDr & = & \frac{R(D)_{exp}}{R(D)_{SM}}= 1.30 \pm 0.17, \label{eq:RDrexp} \\
\RDrstar & =  & \frac{R(D^{\ast})_{exp}} {R(D^{\ast})_{SM}}= 1.25 \pm 0.08. \label{eq:RDstrexp}
\eea
There have been numerous analyses examining NP explanations of the
$R(D^{(*)})$ measurements \cite{Fajfer:2012jt,Crivellin:2012ye,Datta:2012qk,Becirevic:2012jf,Deshpande:2012rr,Celis:2012dk,Choudhury:2012hn,Tanaka:2012nw,Ko:2012sv,Fan:2013qz,Biancofiore:2013ki,Celis:2013jha,Duraisamy:2013kcw,Dorsner:2013tla,Sakaki:2013bfa,Sakaki:2014sea,Bhattacharya:2014wla}. The new physics involves new charged-current interactions.
In the neutral-current sector, data from $ b \to s \ell^+ \ell^-$ decays  also hint at lepton flavor non-universality -- the so called $R_K$ puzzle:
the LHCb Collaboration has found a 2.6$\sigma$ deviation from the SM prediction for the  ratio $R_K \equiv {\cal B}(B^+
\to K^+ \mu^+ \mu^-)/{\cal B}(B^+ \to K^+ e^+ e^-)$ in the dilepton
invariant mass-squared range 1 GeV$^2$ $\le q^2 \le 6$ GeV$^2$
\cite{Aaij:2014ora}.
There are also other, not necessarily lepton-flavor non-universal anomalies in $b\to s\ell^+\ell^-$ decays, most significantly in the $B^0 \to K^{*0} \mu^+ \mu^-$ angular observable $P_5^\prime$ \cite{Aaij:2013qta, Aaij:2015oid}. Global fits of the experimental data prefer a negative shift
in one of the $b\to s\mu^+\mu^-$ Wilson coefficients, $C_9$ \cite{Blake:2016olu}. Common explanations of the $\bctaunutau$ and $b\to s\mu^+\mu^-$
anomalies have been proposed in Refs.~\cite{Bhattacharya:2014wla, Calibbi:2015kma, Greljo:2015mma, Boucenna:2016wpr, Bhattacharya:2016mcc,Barbieri:2016las}.

The underlying quark level transition $\bctaunutau$ in the $\RD$ puzzle can be probed in both
$B$ and $\Lambda_b$ decays. 
Recently, the decay $\lbt$ was discussed in the standard model and with new physics in Ref.~\cite{Gutsche:2015mxa, Woloshyn:2014hka, Shivashankara:2015cta, Dutta:2015ueb, Faustov:2016pal, Li:2016pdv, Celis:2016azn}.
 $\lbt$ decays could be useful to confirm
possible new physics in the $\RD$ puzzle and to point to the correct model of new physics.

In Ref.~\cite{Shivashankara:2015cta} the following quantities were calculated within the SM and with various new physics operators:
\bea
\label{ratio1}
R(\lc) & = & \frac{{\cal B}[ \lbt]}{{\cal B}[\lbl]}
\eea
and
\bea
\label{ratio22}
B_{\lc}(q^2) & = & \frac{\frac{d\Gamma[ \lbt]}{d q^2}}{\frac{ d \Gamma[\lbl]}{d q^2}}, \label{eq:Bqsqr}
\eea
where $\ell$ represents $\mu$ or $e$.
In this paper we work with the ratio $\Rlcr$, defined as
\bea
\label{ratio3}
\Rlcr & = & \frac{R(\lc)^{SM+NP}}{R(\lc)^{SM}}. \
\eea
We also consider the forward-backward asymmetry
\begin{align} \label{eq:AFB}
A_{FB}(q^2) &= \frac{ \int_{0}^{1} ( d^2 \Gamma/dq^2 d \cos\theta_\tau )\:d \cos\theta_\tau-\int_{-1}^{0} ( d^2 \Gamma/dq^2 d \cos\theta_\tau )\:d \cos\theta_\tau  }{ d \Gamma/dq^2 },
\end{align}
where $\theta_\tau$ is the angle between the momenta of the $\tau$ lepton and $\Lambda_c$ baryon in the dilepton rest frame.

This paper improves upon the earlier work \cite{Shivashankara:2015cta} in several ways:
\begin{itemize}
\item{ We add tensor interactions in the effective Lagrangian.}

\item{ Instead of a quark model, we use form factors from lattice QCD to calculate all $\lbt$ observable. The vector and axial vector form factors
are taken from Ref.~\cite{Detmold:2015aaa}, and we extend the analysis of Ref.~\cite{Detmold:2015aaa} to obtain lattice QCD results for the tensor form factors as well.}

\item{ In addition to $R(\lc)$ and $B_{\lc}(q^2)$, we also calculate the forward-backward asymmetry (\ref{eq:AFB}) in the SM and with new physics.}

\item{ We include new constraints from the $B_c$ lifetime \cite{Li:2016vvp,Alonso:2016oyd,Celis:2016azn} in our analysis.}

\item{ In addition to analyzing the effects of individual new physics-couplings, we study specific models that introduce multiple new-physics couplings
simultaneously. We consider a 2-Higgs doublet model, models with new vector bosons, and several leptoquark models.}

\end{itemize}
The paper is organized in the following manner: In Sec.~\ref{sec:formalism} we introduce the effective Lagrangian to parametrize the NP operators and give
the expressions for the decay distribution in terms of helicity amplitudes. In Sec.~\ref{sec:latticeqcd}, we present the new lattice QCD results for
the tensor form factors. The model-independent phenomenological analysis of individual new-physics couplings is discussed in Sec.~\ref{sec:modelindependent},
while explicit models are considered in Sec.~\ref{sec:models}. We conclude in Sec.~\ref{sec:conclusion}.

\section{Formalism}
\label{sec:formalism}

\subsection{Effective Hamiltonian}

In the presence of NP, the effective Hamiltonian for the quark-level transition $b\to c\tau^-\bar{\nu}_\tau$  can be written in the form \cite{Chen:2005gr,Bhattacharya:2011qm}
\bea
\label{eq1:Lag}
 {\cal{H}}_{eff} &=&  \frac{G_F V_{cb}}{\sqrt{2}}\Big\{
\Big[\bar{c} \gamma_\mu (1-\gamma_5) b  + g_L \bar{c} \gamma_\mu (1-\gamma_5)  b + g_R \bar{c} \gamma_\mu (1+\gamma_5) b\Big] \bar{\tau} \gamma^\mu(1-\gamma_5) \nu_{\tau} \nl && +  \Big[g_S\bar{c}  b   + g_P \bar{c} \gamma_5 b\Big] \bar{\tau} (1-\gamma_5)\nu_{\tau} + \Big[g_T\bar{c}\sigma^{\mu \nu}(1-\gamma_5)b\Big]\bar{\tau}\sigma_{\mu \nu}(1-\gamma_5)\nu_{\tau} + h.c \Big\}, \nonumber \\ \label{eq:Heff}
\eea 
where  $G_F$ is the Fermi constant, $V_{cb}$ is the Cabibbo-Kobayashi-Maskawa (CKM) matrix element, and we use $\sigma_{\mu \nu} = i[\gamma_\mu, \gamma_\nu]/2$. We consider that the above Hamiltonian is written at the $m_b$ energy scale. 

If the effective interaction is written at the cut-ff scale $\Lambda$ then running down to the $m_b$ scale will generate new operators and new contributions,
which have been discussed in Refs.~\cite{Feruglio:2016gvd, Feruglio:2017rjo}. These new contributions can strongly constrain models but to really calculate their true impacts we have to consider specific models
 where there might be cancellations between various terms.

The SM effective Hamiltonian corresponds to $g_L = g_R = g_S = g_P = g_T = 0$.
In Eq.~(\ref{eq:Heff}), we have assumed the neutrinos to be always left chiral. In general, with NP the neutrino associated with the $\tau$ lepton does not have to carry the same flavor. In the model-independent analysis of individual couplings (Sec.~\ref{sec:modelindependent}) we will not consider this possibility. Specific models will be discussed in Sec.~\ref{sec:models}.

\subsection{Decay process}

The process under consideration is
$$\Lambda_{b}(p_{\lb})\rightarrow\tau^{-}(p_{\tau})+\bar{\nu}_{\tau}(p_{\bar{\nu}_\tau})+\Lambda_{c}(p_{\lc}).$$
The differential decay rate for this process can be represented as \cite{Tanaka:2012nw}
\bea
\frac{d\Gamma}{dq^2 d \cos\theta_\tau}&=&\frac{G_{F}^2 |V_{cb}|^2}{2048\pi^3}(1-\frac{m_\tau ^2}{q^2})\frac{\sqrt{Q_+ Q_-}}{m_{\lb}^{3}}\sum_{\lambda_{\Lambda_c}}\sum_{\lambda_\tau}|\mathcal{M}_{\lambda_{\Lambda_c}}^{\lambda_\tau}|^2, \label{eq:rate}
\eea
where
\bea
q     &=& p_{\Lambda_b}-p_{\Lambda_c}, \\
Q_\pm &=& (m_{\lb} \pm m_{\lc})^2 - q^2\,,
\eea
and the helicity amplitude $\mathcal{M}_{\lambda_{\lc}}^{\lambda_\tau} $ is written as
\bea
\mathcal{M}^{\lambda_\tau}_{\lambda_{\Lambda_c}}&=&H^{SP}_{\lambda_{\Lambda_c},\lambda_\tau=0}+\sum_{\lambda}\eta_{\lambda}H^{VA}_{\lambda_{\Lambda_c},\lambda}L^{\lambda_\tau}_{\lambda}+\sum_{\lambda,\lambda^{\prime}} \eta_\lambda \eta_{\lambda^{\prime}} H^{(T){\lambda_{\lb}}}_{\lambda_{\Lambda_c},\lambda ,\lambda^{\prime}}L^{\lambda_\tau}_{\lambda,\lambda^\prime}.
\eea
Here, ($\lambda$, $\lambda^\prime$) indicate the helicity of the virtual vector boson (see Appendix \ref{sec:spinorsandvectors}),  $\lambda_{\Lambda_c}$ and $\lambda_\tau$ are the
helicities of the  $\Lambda_c$ baryon and $\tau$ lepton, respectively, and $\eta_\lambda=1$ for $\lambda=t$ and $\eta_\lambda=-1$ for $\lambda=0,\pm 1$.

The scalar-type, vector/axial-vector-type, and tensor-type hadronic helicity amplitudes are defined as
\bea
H^{SP}_{\lambda_{\Lambda_c},\lambda=0}&=&H^S_{\lambda_{\Lambda_c},\lambda=0}+H^P_{\lambda_{\Lambda_c},\lambda=0}, \nn\\
H^S_{\lambda_{\Lambda_c},\lambda=0}&=&g_S \bra{\lc}\bar{c} b\ket{\lb},\nn\\
H^P_{\lambda_{\Lambda_c},\lambda=0}&=&g_P \bra{\lc}\bar{c}\gamma_5 b\ket{\lb},
\eea
\bea
H^{VA}_{\lambda_{\Lambda_c},\lambda}&=&H^V_{\lambda_{\Lambda_c},\lambda}-H^A_{\lambda_{\Lambda_c},\lambda}, \nn\\
H^V_{\lambda_{\Lambda_c},\lambda}&=&(1+g_L+g_R)\,\epsilon^{*\mu}(\lambda)\bra{\lc}\bar{c}\gamma_{\mu} b\ket{\lb}, \nn\\
H^A_{\lambda_{\Lambda_c},\lambda}&=&(1+g_L-g_R)\,\epsilon^{*\mu}(\lambda)\bra{\lc}\bar{c}\gamma_{\mu}\gamma_5 b\ket{\lb},
\eea
and
\bea
H^{(T){\lambda_{\lb}}}_{\lambda_{\lc},\lambda ,\lambda^{\prime}}&=&H^{(T1){\lambda_{\lb}}}_{\lambda_{\lc},\lambda ,\lambda^{\prime}}-H^{(T2){\lambda_{\lb}}}_{\lambda_{\lc},\lambda ,\lambda^{\prime}}, \nn\\
H^{(T1){\lambda_{\lb}}}_{\lambda_{\lc},\lambda ,\lambda^{\prime}}&=&g_T\: \epsilon^{*\mu}(\lambda)\epsilon^{*\nu}(\lambda^{\prime})\bra{\lc}\bar{c}i\sigma_{\mu \nu} b\ket{\lb},\nn\\
H^{(T2){\lambda_{\lb}}}_{\lambda_{\lc},\lambda ,\lambda^{\prime}}&=&g_T\:\epsilon^{*\mu}(\lambda)\epsilon^{*\nu}(\lambda^{\prime})\bra{\lc}\bar{c}i\sigma_{\mu \nu}\gamma_5 b\ket{\lb}.
\eea
The leptonic amplitudes are defined as
\bea
L^{\lambda_\tau}&=&\bra{\tau\bar{\nu}_\tau}\bar{\tau} (1-\gamma_5)\nu_\tau\ket{0}, \nn\\
L^{\lambda_\tau}_{\lambda}&=&\epsilon^\mu (\lambda)\bra{\tau\bar{\nu}_\tau}\bar{\tau}\gamma_\mu (1-\gamma_5)\nu_\tau\ket{0}, \nn\\
L^{\lambda_\tau}_{\lambda ,\lambda^{\prime}}&=&-i\epsilon^\mu (\lambda)\epsilon^\nu (\lambda^\prime)\bra{\tau\bar{\nu}_\tau}\bar{\tau}\sigma_{\mu \nu} (1-\gamma_5)\nu_\tau\ket{0}.
\eea
Above, $\epsilon^{\mu}$ are the polarization vectors of the virtual vector boson (see Appendix \ref{sec:spinorsandvectors}). The explicit expressions for the hadronic and leptonic helicity amplitudes are presented in the following.

\subsubsection{Hadronic helicity amplitudes}

In this paper, we use the helicity-based definition of the $\Lambda_b\to\Lambda_c$ form factors, which was introduced in \cite{Feldmann:2011xf}. 
The matrix elements of the vector and axial vector currents can be written in terms of six helicity form factors $F_+$, $F_\perp$, $F_0$, $G_+$, $G_\perp$, and $G_0$
as follows:
\bea
\bra{\lc}\bar{c}\gamma^\mu b\ket{\lb}&=&\bar{u}_{\lc}\Big[ F_0 (q^2)(m_{\lb} - m_{\lc})\frac{q^\mu}{q^2} \nonumber\\
&&+F_+ (q^2)\frac{m_{\lb} + m_{\lc}}{Q_+}(p_{\lb}^{\mu} +p_{\lc}^{\mu}-(m_{\lb} ^2 - m_{\lc} ^2)\frac{q^\mu}{q^2}) \nonumber\\
&&+F_\perp (q^2)(\gamma^\mu - \frac{2m_{\lc}}{Q_+}p_{\lb}^{\mu} - \frac{2m_{\lb}}{Q_+}p_{\lc}^{\mu})\Big]u_{\lb}, \label{eq:VFF} \\
\bra{\lc}\bar{c}\gamma^\mu \gamma_5 b\ket{\lb}&=&-\bar{u}_{\lc}\gamma_5\Big[ G_0 (q^2)(m_{\lb} + m_{\lc})\frac{q^\mu}{q^2} \nonumber\\
&&+G_+ (q^2)\frac{m_{\lb} - m_{\lc}}{Q_-}(p_{\lb}^{\mu} +p_{\lc}^{\mu}-(m_{\lb} ^2 - m_{\lc} ^2)\frac{q^\mu}{q^2})\nonumber\\
&&+G_\perp (q^2)(\gamma^\mu + \frac{2m_{\lc}}{Q_-}p_{\lb}^{\mu} - \frac{2m_{\lb}}{Q_-}p_{\lc}^{\mu})\Big]u_{\lb} \label{eq:AFF}.
\eea
The matrix elements of the scalar and pseudoscalar currents can be obtained from the vector and axial vector matrix elements using the equations of motion:
\bea
\nonumber \bra{\lc}\bar{c} b\ket{\lb} &=& \frac{q_\mu}{m_b-m_c}\bra{\lc}\bar{c}\gamma^\mu b\ket{\lb} \\
&=& F_0(q^2)  \frac{m_{\lb} - m_{\lc}}{m_b-m_c} \bar{u}_{\lc}u_{\lb}, \\
\nonumber \bra{\lc}\bar{c}\gamma_5 b\ket{\lb} &=& \frac{q_\mu}{m_b+m_c}\bra{\lc}\bar{c}\gamma^\mu\gamma_5 b\ket{\lb} \\
&=& G_0(q^2)  \frac{m_{\lb} + m_{\lc}}{m_b+m_c} \bar{u}_{\lc}\gamma_5 u_{\lb}.
\eea
In our numerical analysis, we use $m_b = 4.18(4)$ GeV, $m_c = 1.27(3)$ GeV \cite{Olive:2016xmw}. The matrix elements of the tensor currents can be written in terms of four form factors $h_+$, $h_\perp$, $\widetilde{h}_+$, $\widetilde{h}_\perp$,
\bea
&&\bra{\lc}\bar{c}i\sigma^{\mu\nu} b\ket{\lb}=\bar{u}_{\lc}\Big[2h_+(q^2)\frac{p_{\lb}^\mu p_{\lc}^{ \nu}-p_{\lb}^\nu p_{\lc}^{\mu}}{Q_+} \nonumber\\
&&+h_\perp (q^2)\Big(\frac{m_{\lb}+m_{\lc}}{q^2}(q^\mu \gamma^\nu -q^\nu \gamma^\mu)-2(\frac{1}{q^2}+\frac{1}{Q_+})(p_{\lb}^\mu p_{\lc}^{\nu}-p_{\lb}^\nu p_{\lc}^{\mu}) \Big) \nonumber\\
&&+\widetilde{h}_+ (q^2)\Big(i\sigma^{\mu \nu}-\frac{2}{Q_-}(m_{\lb}(p_{\lc}^{\mu}\gamma^\nu -p_{\lc}^{\nu}\gamma^\mu)\nonumber\\
&&-m_{\lc}(p_{\lb}^\mu \gamma^\nu -p_{\lb}^\nu \gamma^\mu)+p_{\lb}^\mu p_{\lc}^{\nu}-p_{\lb}^\nu p_{\lc}^{\mu}) \Big) \nonumber\\
&&+\widetilde{h}_\perp(q^2) \frac{m_{\lb}-m_{\lc}}{q^2 Q_-}\Big((m_{\lb}^2-m_{\lc}^2-q^2)(\gamma^\mu p_{\lb}^\nu - \gamma^\nu p_{\lb}^\mu)\nonumber\\
&&-(m_{\lb}^2-m_{\lc}^2+q^2)(\gamma^\mu p_{\lc}^{\nu}-\gamma^\nu p_{\lc}^{\mu})+2(m_{\lb}-m_{\lc})(p_{\lb}^\mu p_{\lc}^{\nu}-p_{\lb}^\nu p_{\lc}^{\mu}) \Big)
\Big]u_{\lb}. \nonumber \\ \label{eq:TFF}
\eea
The matrix elements of the current $\bar{c}i\sigma^{\mu\nu}\gamma_5b$ can be obtained from the above equation by using the identity
\bea
\sigma^{\mu \nu}\gamma_{5}=-\frac{i}{2}\epsilon^{\mu \nu \alpha \beta}\sigma_{\alpha \beta}.
\eea 

In the following, only the non-vanishing helicity amplitudes are given. The scalar and pseudo-scalar helicity amplitudes associated with the new physics scalar and pseudo-scalar interactions are
\bea
H^{SP}_{1/2,0}&=&F_0g_S \frac{\sqrt{Q_+}}{m_b-m_c}(m_{\lb}-m_{\lc})-G_0g_P\frac{\sqrt{Q_-}}{m_b+m_c}(m_{\lb}+m_{\lc}), \\
H^{SP}_{-1/2,0} &=& F_0g_S\frac{\sqrt{Q_+}}{m_b-m_c}(m_{\lb}-m_{\lc})+G_0g_P\frac{\sqrt{Q_-}}{m_b+m_c}(m_{\lb}+m_{\lc}).				
\eea
The parity-related amplitudes are
\bea
H^{S}_{\lambda_{\Lambda_{c}},\lambda_{NP}} & = & H^{S}_{-\lambda_{\Lambda_{c}},-\lambda_{NP}},\nonumber\\
H^{P}_{\lambda_{\Lambda_{c}},\lambda_{NP}} & = & -H^{P}_{-\lambda_{\Lambda_{c}},-\lambda_{NP}}.
\eea
For the vector and axial-vector helicity amplitudes, we find
\bea
H^{VA}_{1/2,0}&= &F_+ (1+g_L+g_R)\frac{\sqrt{Q_-}}{\sqrt{q^2}}(m_{\lb}+m_{\lc})\nn\\
&  &-G_+ (1+g_L-g_R)\frac{\sqrt{Q_+}}{\sqrt{q^2}}(m_{\lb}-m_{\lc}), \\
H^{VA}_{1/2,+1}&= &-F_\perp (1+g_L+g_R)\sqrt{2Q_-} +G_\perp (1+g_L-g_R)\sqrt{2Q_+}, \\		
H^{VA}_{1/2,t}&= &F_0(1+g_L+g_R)\frac{\sqrt{Q_+}}{\sqrt{q^2}}(m_{\lb}-m_{\lc})\nn\\
&  &-G_0(1+g_L-g_R)\frac{\sqrt{Q_-}}{\sqrt{q^2}}(m_{\lb}+m_{\lc}), \\
H^{VA}_{-1/2,0}&= &F_+ (1+g_L+g_R)\frac{\sqrt{Q_-}}{\sqrt{q^2}}(m_{\lb}+m_{\lc})\nn\\
&  &+G_+ (1+g_L-g_R)\frac{\sqrt{Q_+}}{\sqrt{q^2}}(m_{\lb}-m_{\lc}), \\			
H^{VA}_{-1/2,-1}&= &-F_\perp (1+g_L+g_R)\sqrt{2Q_-} -G_\perp (1+g_L-g_R)\sqrt{2Q_+}, \\		
H^{VA}_{-1/2,t}&= &F_0(1+g_L+g_R)\frac{\sqrt{Q_+}}{\sqrt{q^2}}(m_{\lb}-m_{\lc})\nn\\ 
&  &+G_0(1+g_L-g_R)\frac{\sqrt{Q_-}}{\sqrt{q^2}}(m_{\lb}+m_{\lc}).			
\eea
We also have the relations
\begin{eqnarray}
H_{\lambda_{\Lambda_{c}},\lambda_{w}}^V&=&H_{-\lambda_{\Lambda_{c}},-\lambda_{w}}^V,\nonumber\\
H_{\lambda_{\Lambda_{c}},\lambda_{w}}^A&=&-H_{-\lambda_{\Lambda_{c}},-\lambda_{w}}^A.
\end{eqnarray}
The tensor helicity amplitudes are
\bea
H^{(T)-1/2}_{-1/2,t,0}&=&-g_T\Big[-h_+\sqrt{Q_-}+\widetilde{h}_+\sqrt{Q_+}\Big], \\[10pt]
H^{(T)+1/2}_{+1/2,t,0}&=&g_T\Big[h_+\sqrt{Q_-}+\widetilde{h}_+\sqrt{Q_+}\Big],\\[10pt]
H^{(T)-1/2}_{+1/2,t,+1}&=& -g_T\frac{\sqrt{2}}{\sqrt{q^2}}\Big[h_\perp(m_{\lb}+m_{\lc})\sqrt{Q_-}+\widetilde{h}_\perp(m_{\lb}-m_{\lc})\sqrt{Q_+}\Big],\\[10pt]
H^{(T)+1/2}_{-1/2,t,-1}&=&-g_T\frac{\sqrt{2}}{\sqrt{q^2}}\Big[h_\perp(m_{\lb}+m_{\lc})\sqrt{Q_-}-\widetilde{h}_\perp(m_{\lb}-m_{\lc})\sqrt{Q_+}\Big],\\[10pt]
H^{(T)-1/2}_{+1/2,0,+1}&=&-g_T\frac{\sqrt{2}}{\sqrt{q^2}}\Big[h_\perp(m_{\lb}+m_{\lc})\sqrt{Q_-}+\widetilde{h}_\perp(m_{\lb}-m_{\lc})\sqrt{Q_+}\Big],\\[10pt]
H^{(T)+1/2}_{-1/2,0,-1}&=&g_T\frac{\sqrt{2}}{\sqrt{q^2}}\Big[h_\perp(m_{\lb}+m_{\lc})\sqrt{Q_-}-\widetilde{h}_\perp(m_{\lb}-m_{\lc})\sqrt{Q_+}\Big],\\[10pt]
H^{(T)+1/2}_{+1/2,+1,-1}&=&-g_T\Big[h_+\sqrt{Q_-}+\widetilde{h}_+\sqrt{Q_+}\Big],\\[10pt]
H^{(T)-1/2}_{-1/2,+1,-1}&=&-g_T\Big[h_+\sqrt{Q_-}-\widetilde{h}_+\sqrt{Q_+}\Big].
\eea
The other non-vanishing helicity amplitudes of tensor type are related to the above by 
\bea
H^{(T)\lambda_{\lb}}_{\lambda_{\lc},\lambda,\lambda^\prime}=-H^{(T)\lambda_{\lb}}_{\lambda_{\lc},\lambda^\prime,\lambda}.
\eea

\subsubsection{Leptonic helicity amplitudes}

In the following, we define
\beq
v=\sqrt{1-\frac{m_\tau^2}{q^2}}.
\eeq
The scalar and pseudoscalar leptonic helicity amplitudes are
\bea
L^{+1/2}=& 2\sqrt{q^2} v, \\
L^{-1/2}=& 0,
\eea
the vector and axial-vector amplitudes are
\bea
L^{+1/2}_{\pm1}&=&\pm\sqrt{2}m_{\tau} v\; \sin(\theta_\tau), \\
L^{+1/2}_{0}&=&-2m_\tau v\; \cos{(\theta_\tau)}, \\
L^{+1/2}_{t}&=&2m_\tau v, \\
L^{-1/2}_{\pm1}&=&\sqrt{2 q^2}v\; (1\pm \cos(\theta_\tau)), \\
L^{-1/2}_{0}&=&2\sqrt{q^2}v \; \sin{(\theta_\tau)}, \\
L^{-1/2}_{t}&=&0,
\eea
and the tensor amplitudes are
\bea
L^{+1/2}_{0,\pm1}&=&-\sqrt{2 q^2}v\; \sin(\theta_\tau), \\
L^{+1/2}_{\pm1,t}&=&\mp\sqrt{2 q^2}v\; \sin(\theta_\tau), \\
L^{+1/2}_{t,0}&=&L^{+1/2}_{+1,-1}=-2\sqrt{q^2}v\; \cos(\theta_\tau), \\
L^{-1/2}_{0,\pm1}&=&\mp\sqrt{2}m_\tau v\; (1\pm \cos(\theta_\tau)), \\
L^{-1/2}_{\pm1,t}&=&-\sqrt{2}m_\tau v\; (1\pm \cos(\theta_\tau)), \\
L^{-1/2}_{t,0}&=&L^{-1/2}_{+1,-1}=2m_\tau v\; \sin(\theta_\tau) .
\eea
Here we have the relation
\begin{equation}
 L^{\lambda_\tau}_{\lambda,\lambda^\prime}=-L^{\lambda_\tau}_{\lambda^\prime,\lambda}.
\end{equation}
The angle $\theta_\tau$ is defined as the angle between the momenta of the $\tau$ lepton and $\Lambda_c$ baryon in the dilepton rest frame.

\subsection{Differential decay rate and forward-backward asymmetry}

From the twofold decay distribution (\ref{eq:rate}), we obtain the following
expression for the differential decay rate by integrating over $\cos{\theta_\tau}$:
\bea
\frac{d\Gamma(\lbt)}{dq^2}&=&\frac{G_{F}^{2}|V_{cb}|^2}{384 \pi^{3}}\frac{q^2\sqrt{Q_+ Q_-}}{m_{\lb}^3}\Big(1-\frac{m_\tau^2}{q^2}\Big)^2 \Bigg[ A_1^{VA}+\frac{m_\tau^2}{2q^2}A_2^{VA}+\frac{3}{2}A_3^{SP}\nonumber\\[5pt]
&&+2\Big(1+\frac{2m_\tau^2}{q^2}\Big)A_4^{T}+\frac{3m_\tau}{\sqrt{q^2}} A_5^{VA-SP}+\frac{6m_\tau}{\sqrt{q^2}} A_6^{VA-T} \Bigg],
\eea
where
\begin{align}
A_1^{VA}=&~|H^{VA}_{1/2,1}|^2+|H^{VA}_{1/2,0}|^2+|H^{VA}_{-1/2,0}|^2+|H^{VA}_{-1/2,-1}|^2,\nonumber\\[10pt]
A_2^{VA}=&~|H^{VA}_{1/2,1}|^2+|H^{VA}_{1/2,0}|^2+|H^{VA}_{-1/2,0}|^2+|H^{VA}_{-1/2,-1}|^2+3|H^{VA}_{1/2,t}|^2+3|H^{VA}_{-1/2,t}|^2,\nonumber\\[10pt]
A_3^{SP}=&~|H^{SP}_{1/2,0}|^2+|H^{SP}_{-1/2,0}|^2,\nonumber\\[10pt]
A_4^{T}=&~|H^{(T)1/2}_{1/2,t,0}+H^{(T)1/2}_{1/2,-1,1}|^2+|H^{(T)1/2}_{-1/2,t,-1}+H^{(T)1/2}_{-1/2,-1,0}|^2+|H^{(T)-1/2}_{1/2,0,1}+H^{(T)-1/2}_{1/2,t,1}|^2\nonumber\\[10pt]
&+|H^{(T)-1/2}_{-1/2,-1,1}+H^{(T)-1/2}_{-1/2,t,0}|^2,\nonumber\\[10pt]
A_5^{VA-SP}=&~\mathrm{Re}(H^{SP*}_{1/2,0} ~H^{VA}_{1/2,t}+H^{SP*}_{-1/2,0} ~H^{VA}_{-1/2,t}),\nonumber\\[10pt]
A_6^{VA-T}=&~\mathrm{Re}[H^{VA*}_{1/2,0}(H^{(T)1/2}_{1/2,-1,1}+H^{(T)1/2}_{1/2,t,0})] + \mathrm{Re}[H^{VA*}_{1/2,1}(H^{(T)-1/2}_{1/2,0,1}+H^{(T)-1/2}_{1/2,t,1})] +\nonumber\\[10pt]
&~\mathrm{Re}[H^{VA*}_{-1/2,0}(H^{(T)-1/2}_{-1/2,-1,1}+H^{(T)-1/2}_{-1/2,t,0})] + \mathrm{Re}[H^{VA*}_{-1/2,-1}(H^{(T)1/2}_{-1/2,-1,0}+H^{(T)1/2}_{-1/2,t,-1})].
\end{align}
Here, $A_1^{VA}$ and $A_2^{VA}$ are the (axial-)vector non-spin-flip and spin-flip terms respectively, $A_3^{SP}$ and $A_4^{T}$ are the pure (pseudo-)scalar and tensor terms respectively; and $A_5^{VA-SP}$ and $A_6^{VA-T}$ are interference terms. The scalar-tensor interference term is proportional to $\cos{\theta_\tau}$ and vanishes after integration over $\cos{\theta_\tau}$. 

For the forward-backward asymmetry (\ref{eq:AFB}) we have
\bea
A_{FB}(q^2)&=&\left(\frac{d\Gamma}{dq^2}\right)^{-1}~\frac{G_F^2 V_{cb}^2}{512 \pi^3}\frac{q^2 \sqrt{Q_+Q_-}}{m_{\lb}^3}\Big(1-\frac{m_\tau^2}{q^2}\Big)^2 \Bigg[ B_1^{VA}+\frac{2m_\tau^2}{q^2}B_2^{VA}+\frac{4m_\tau^2}{q^2}
B_3^T+\nonumber\\[5pt]
&&\frac{2m_\tau}{\sqrt{q^2}} B_4^{VA-SP}+\frac{4m_\tau}{\sqrt{q^2}} B_5^{VA-T}+4B_6^{SP-T} \Bigg],
\eea
where
\begin{align}
B_1^{VA}=&~|H^{VA}_{1/2,1}|^2-|H^{VA}_{-1/2,-1}|^2,\nonumber\\[10pt]
B_2^{VA}=&~\mathrm{Re}[H^{VA*}_{1/2,t}H^{VA}_{1/2,0}+H^{VA*}_{-1/2,t}H^{VA}_{-1/2,0}],\nonumber\\[10pt]
B_3^{T}=&~|H^{(T)-1/2}_{1/2,0,1}+H^{(T)-1/2}_{1/2,t,1}|^2-|H^{(T)1/2}_{-1/2,-1,0}+H^{(T)1/2}_{-1/2,t,-1}|^2,\nonumber\\[10pt]
B_4^{VA-SP}=&~\mathrm{Re}[H^{SP*}_{1/2,0}H^{VA}_{1/2,0}+H^{SP*}_{-1/2,0}H^{VA}_{-1/2,0}],\nonumber\\[10pt]
B_5^{VA-T}=&~\mathrm{Re}[H^{VA*}_{1/2,t} (H^{(T)1/2}_{1/2,-1,1}+H^{(T)1/2}_{1/2,t,0})]+\mathrm{Re}[H^{VA*}_{1/2,1} (H^{(T)-1/2}_{1/2,0,1}+H^{(T)-1/2}_{1/2,t,1})]\nonumber\\[10pt]
&+\mathrm{Re}[H^{VA*}_{-1/2,t} (H^{(T)-1/2}_{-1/2,-1,1}+H^{(T)-1/2}_{-1/2,t,0})]-\mathrm{Re}[H^{VA*}_{-1/2,-1} (H^{(T)1/2}_{-1/2,-1,0}+H^{(T)1/2}_{-1/2,t,-1})],\nonumber\\[10pt]
B_6^{SP-T}=&~\mathrm{Re}[H^{SP*}_{1/2,0}(H^{(T)1/2}_{1/2,-1,1}+H^{(T)1/2}_{1/2,t,0})]+\mathrm{Re}[H^{SP*}_{-1/2,0}(H^{(T)-1/2}_{-1/2,-1,1}+H^{(T)-1/2}_{-1/2,t,0})].
\end{align}
There is no contribution from pure (pseudo-)scalar operators to the forward-backward asymmetry, but all possible interference terms are present.

\section{$\Lambda_b \to\Lambda_c$ tensor form factors from lattice QCD}
\label{sec:latticeqcd}

This work uses $\Lambda_b \to\Lambda_c$ form factors computed in lattice QCD. The vector and axial vector form
factors defined in Eqs.~(\ref{eq:VFF}) and (\ref{eq:AFF}) are taken from Ref.~\cite{Detmold:2015aaa}. For the purposes of the present work, one of us (SM)
extended the analysis of Ref.~\cite{Detmold:2015aaa} to include the tensor form factors defined in Eq.~(\ref{eq:TFF}). The tensor form factors were extracted from the lattice QCD correlation functions using ratios defined as in Ref.~\cite{Detmold:2016pkz}. The lattice parameters are identical to those in
Ref.~\cite{Detmold:2015aaa}, except that for the tensor form factors the ``residual matching factors'' $\rho_{T^{\mu\nu}}$ and the $\mathcal{O}(a)$-improvement coefficients were set to their tree-level values, with appropriately increased estimates for the resulting systematic uncertainties as detailed further below.
Following Ref.~\cite{Detmold:2015aaa}, two separate fits were performed
to the lattice QCD data using BCL $z$-expansions \cite{Bourrely:2008za} augmented with additional terms to describe the dependence on the lattice
spacing and quark masses. The ``nominal fit'' is used to evaluate the central values and statistical uncertainties of the form factors (and of any observables
depending on the form factors), while the ``higher-order fit'' is used in conjunction with the nominal fit to evaluate the combined systematic uncertainty associated
with the continuum extrapolation, chiral extrapolation, $z$ expansion, renormalization, scale setting, $b$-quark parameter tuning, finite volume, and missing isospin symmetry breaking/QED. The procedure for evaluating the systematic uncertainties is given in Eqs.~(82)-(84) of Ref.~\cite{Detmold:2015aaa}.
The renormalization uncertainty in the tensor form factors is dominated by the use of the tree-level values, $\rho_{T^{\mu\nu}}=1$, for the residual matching factors in the mostly nonperturbative renormalization procedure. We
estimate the systematic uncertainty in $\rho_{T^{\mu\nu}}$ to be 2 times the maximum value of $|\rho_{V^{\mu}}-1|$, $|\rho_{A^{\mu}}-1|$, which is equal to $0.0404$ \cite{Detmold:2015aaa}.
Note that the tensor form factors are scale-dependent, and our results and estimates of systematic uncertainties should be interpreted as corresponding to $\mu=m_b$ in the $\overline{\text{MS}}$ scheme.
To account for the renormalization uncertainty in the higher-order fit, we introduced nuisance parameters multiplying the form factors,
with Gaussian priors equal to $1\pm 0.0404$.

\begin{figure}
\begin{center}
 \includegraphics[width=0.9\linewidth]{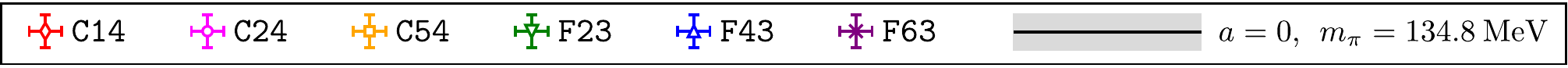} \\
 \includegraphics[width=\linewidth]{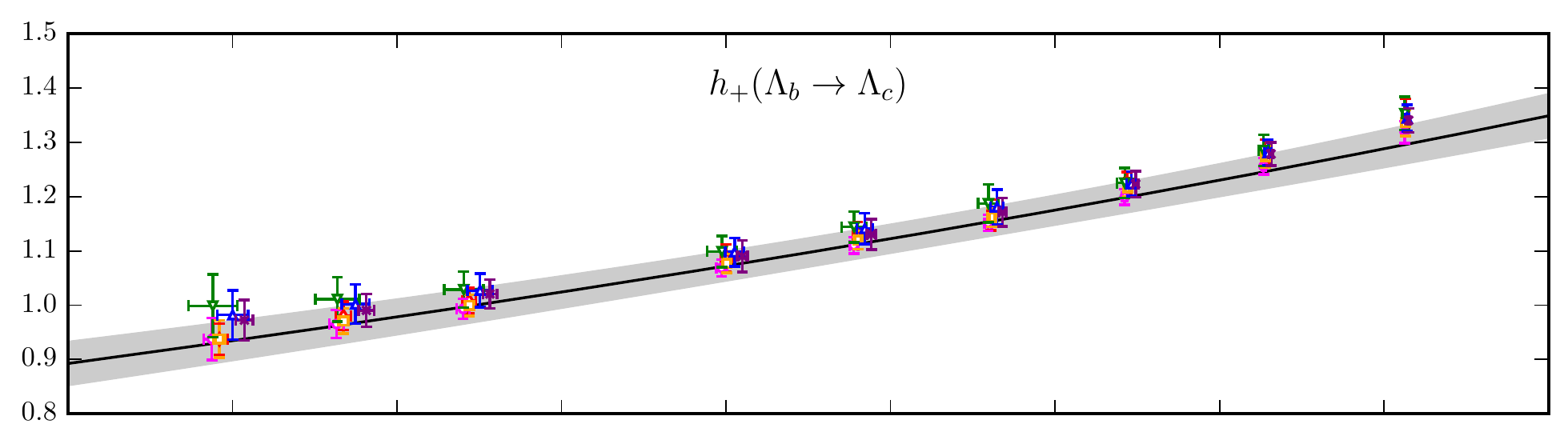} \\
 \includegraphics[width=\linewidth]{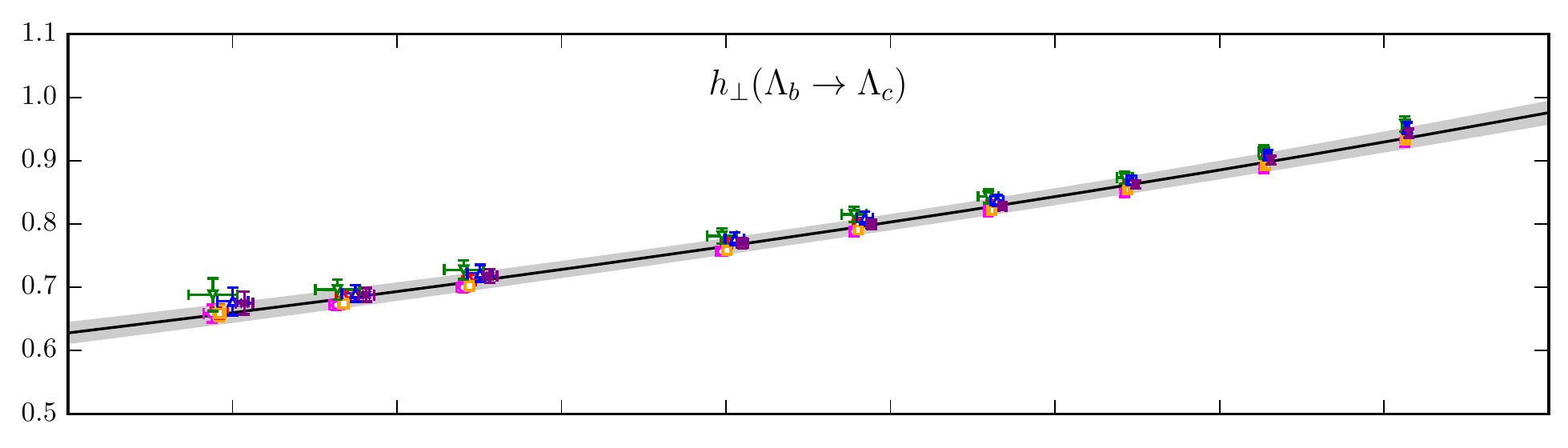} \\
 \includegraphics[width=\linewidth]{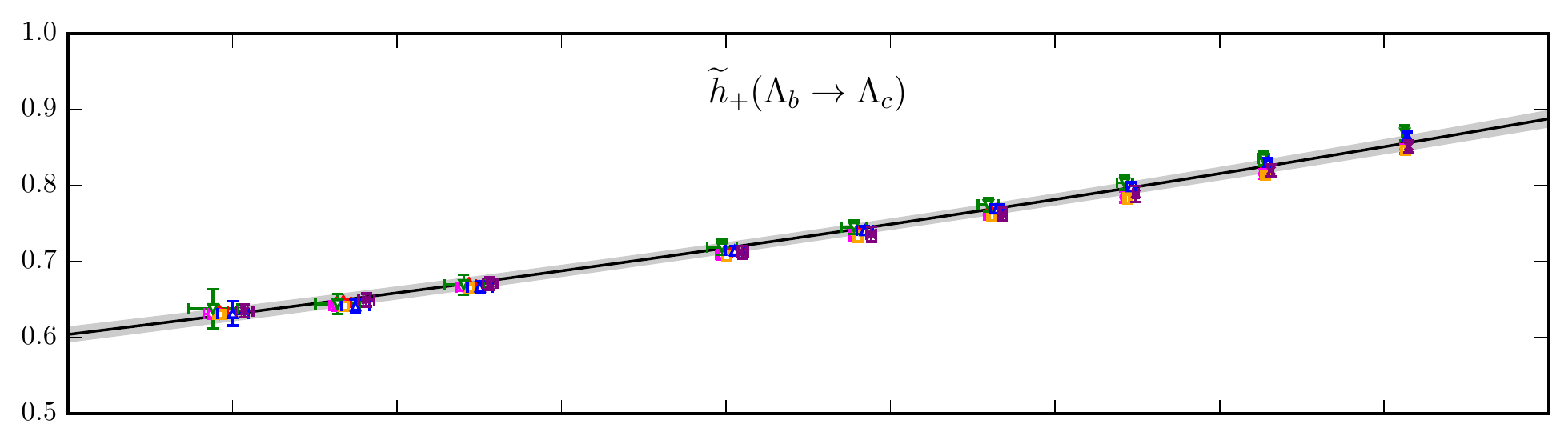} \\
 \includegraphics[width=\linewidth]{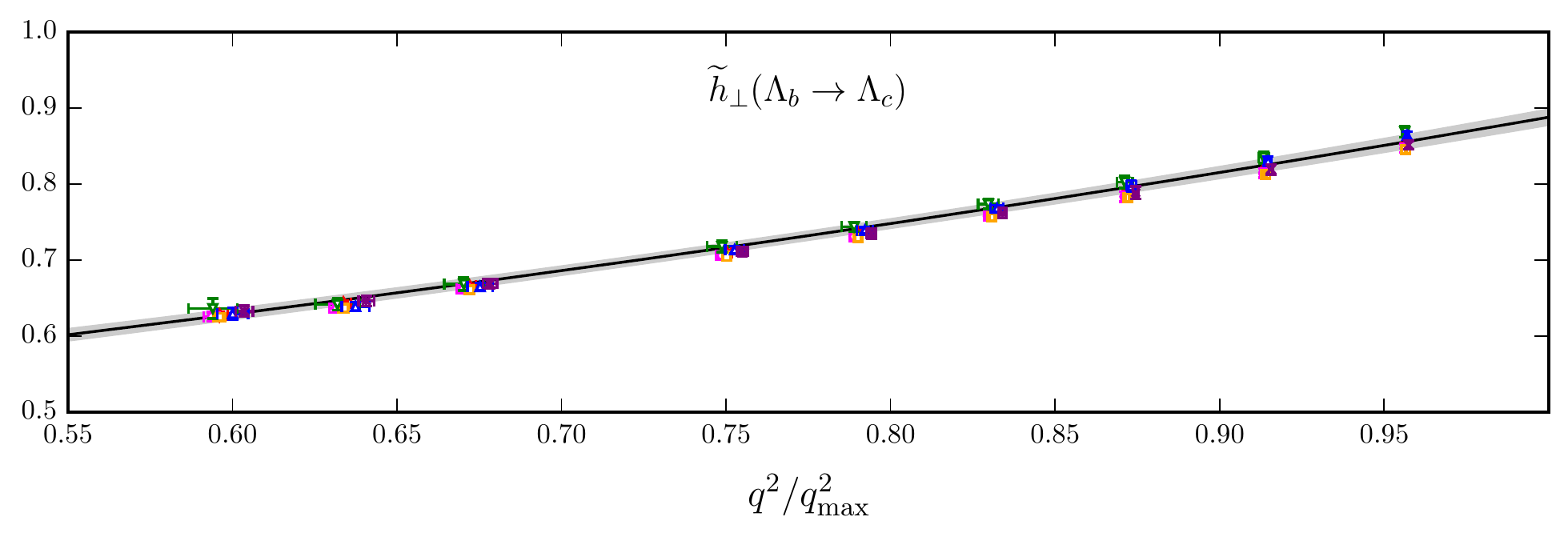}
 \caption{\label{fig:latticeFF}$\Lambda_b \to \Lambda_c$ tensor form factors in the high $q^2$-region: lattice results and extrapolation to the physical limit (nominal fit).  The bands indicate the statistical uncertainty.
 The lattice QCD data sets are labeled as in Ref.~\cite{Detmold:2015aaa}.}
\end{center}
\end{figure}

\begin{figure}
 \includegraphics[height=0.3\linewidth]{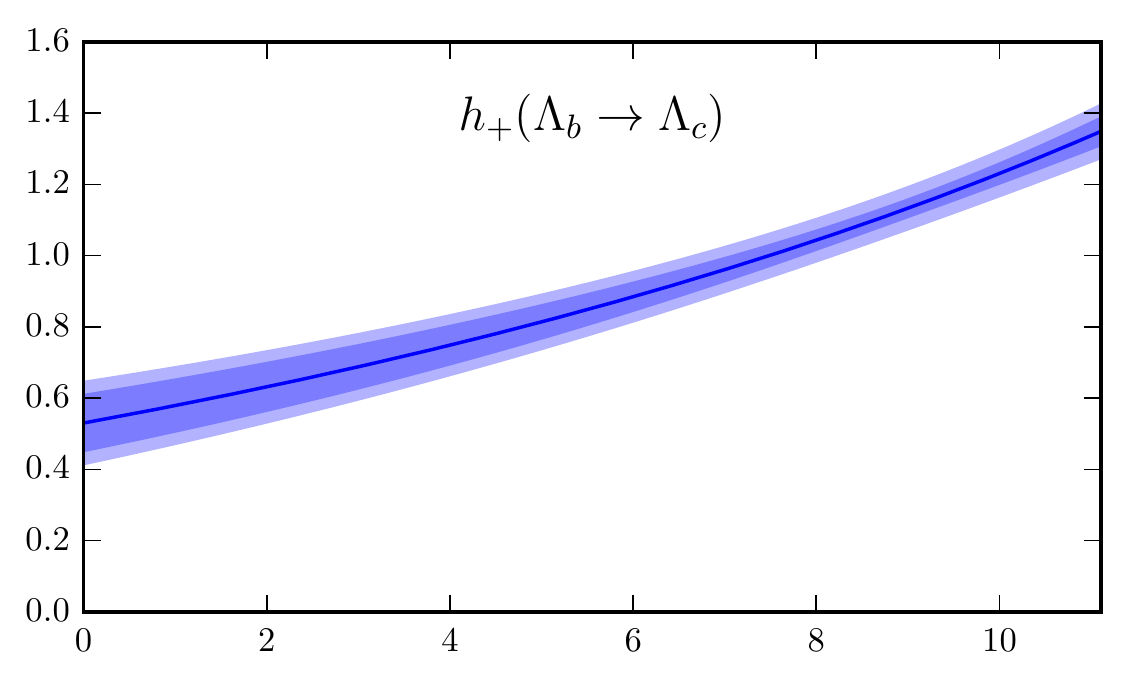} \hfill
 \includegraphics[height=0.3\linewidth]{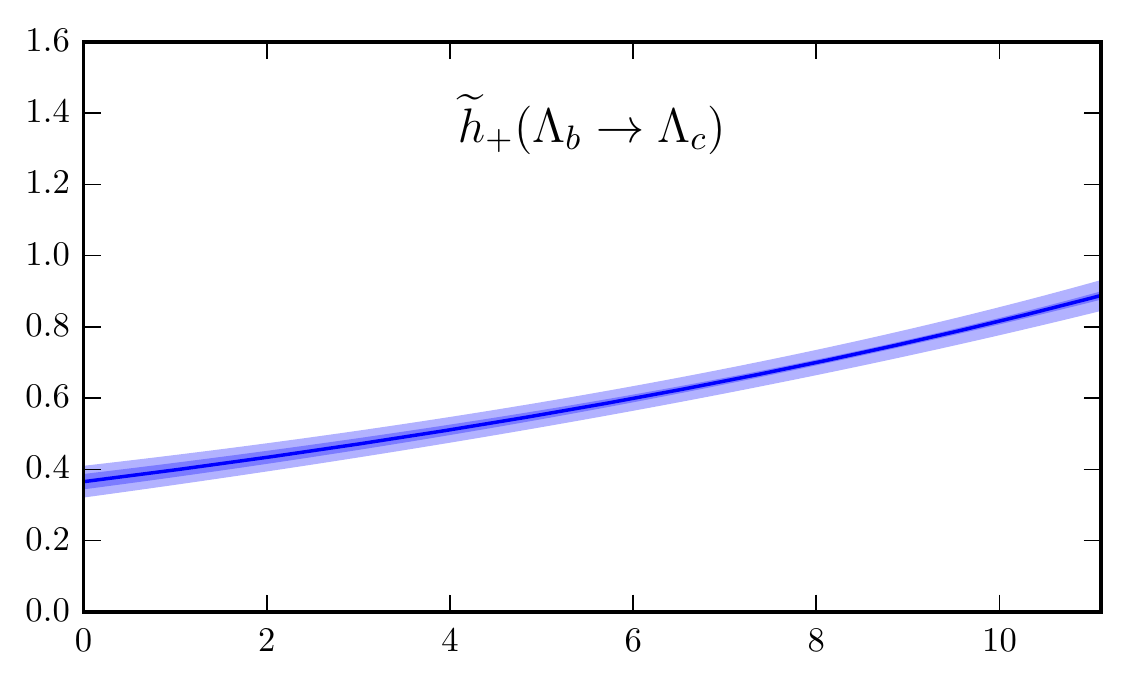} \\
 \includegraphics[height=0.345\linewidth]{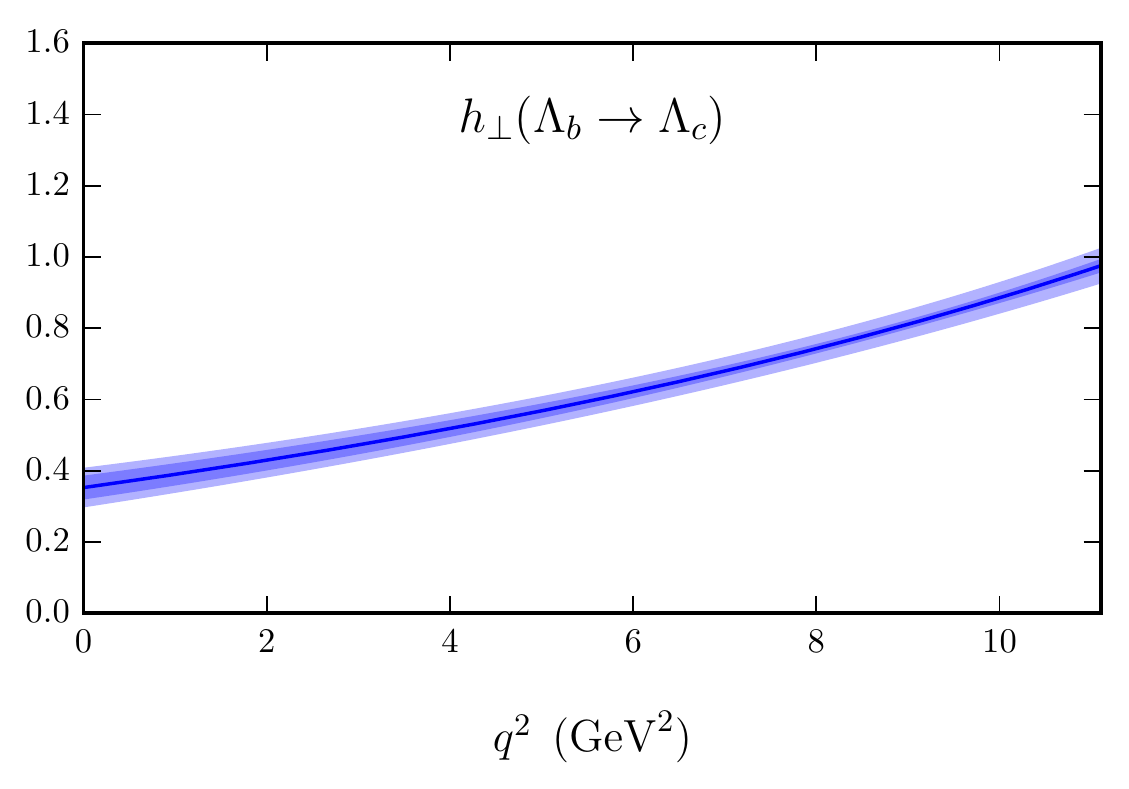} \hfill
 \includegraphics[height=0.345\linewidth]{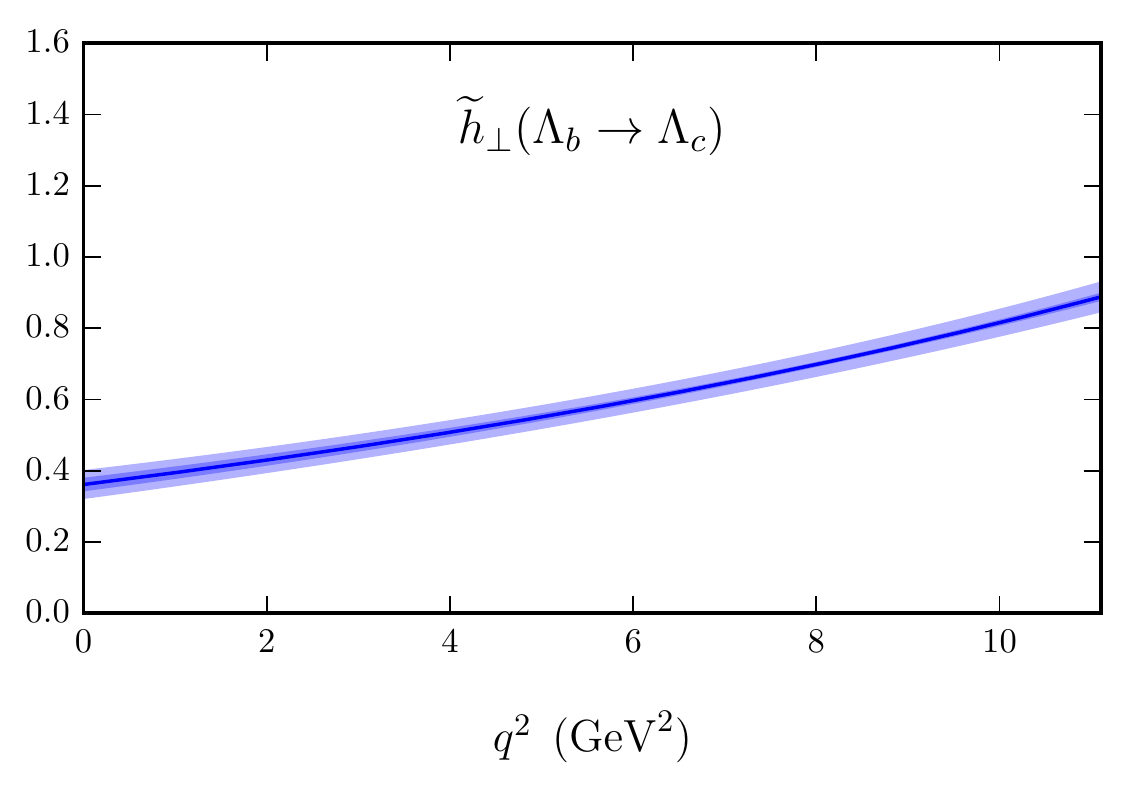} \\
 \caption{\label{fig:FF}$\Lambda_b \to \Lambda_c$ tensor form factors in the physical limit, shown in the entire kinematic range.
 The form factors are defined in the $\overline{\text{MS}}$ scheme and at $\mu=m_b$.
 The inner bands show the statistical uncertainty and the outer bands show the total (statistical plus systematic) uncertainty. The procedure for evaluating the uncertainties using the nominal and higher-order fits is given in Eqs.~(82)-(84) of Ref.~\cite{Detmold:2015aaa}.}
\end{figure}

\begin{table}
\begin{center}
\begin{tabular}{|c|c|c|}
\hline
 $f$     &  $J^P$ &  $m_{\rm pole}^f$ (GeV)   \\
\hline
$h_+$, $h_\perp$                         & $1^-$     & 6.332   \\
$\widetilde{h}_+$, $\widetilde{h}_\perp$ & $1^+$     & 6.768   \\
\hline
\end{tabular}
\caption{\label{tab:polemasses} Values of the pole masses for the tensor form factors.}
\end{center}
\end{table}

\begin{table}
\begin{center}
\begin{tabular}{|c|ll|l|}
\hline
                                      & \hspace{3ex} Nominal fit && \hspace{1ex} Higher-order fit \\
\hline
$a_0^{h_+}$                           & $\wm 0.9752\pm 0.0303$   && $\wm 0.9668\pm 0.0567$ \\ 
$a_1^{h_+}$                           &    $-5.5000\pm 1.2361$   &&    $-4.5258\pm 1.7538$ \\ 
$a_2^{h_+}$                           &                          && $\wm 2.2006\pm 10.724$ \\ 
$a_0^{h_\perp}$                       & $\wm 0.7054\pm 0.0137$   && $\wm 0.7052\pm 0.0362$ \\ 
$a_1^{h_\perp}$                       &    $-4.3578\pm 0.5114$   &&    $-4.1050\pm 0.8391$ \\ 
$a_2^{h_\perp}$                       &                          && $\wm 3.0100\pm 7.8351$ \\ 
$a_0^{\widetilde{h}_{\perp},\widetilde{h}_+}$ & $\wm 0.6728\pm 0.0088$   && $\wm 0.6763\pm 0.0328$ \\ 
$a_1^{\widetilde{h}_+}$                   &    $-4.4322\pm 0.3882$   &&    $-4.3634\pm 0.7509$ \\ 
$a_2^{\widetilde{h}_+}$                   &                          && $\wm 2.2739\pm 8.0769$ \\ 
$a_1^{\widetilde{h}_\perp}$               &    $-4.4928\pm 0.3584$   &&    $-4.5543\pm 0.7370$ \\ 
$a_2^{\widetilde{h}_\perp}$               &                          && $\wm 3.0851\pm 7.9037$ \\ 
\hline
\end{tabular}
\caption{\label{tab:FFparameters}Results for the $z$-expansion parameters describing the $\Lambda_b \to \Lambda_c$ tensor form factors in the physical limit
(in the $\overline{\text{MS}}$ scheme at the renormalization scale $\mu=m_b$). Files containing the values
and covariances of the parameters of all ten $\Lambda_b \to \Lambda_c$ form factors are provided as supplemental material.}
\end{center}
\end{table}

In the physical limit (zero lattice spacing and physical quark masses), the nominal fit function for a form factor $f$ reduces to the form
\begin{eqnarray}
f(q^2) &=& \frac{1}{1-q^2/(m_{\rm pole}^f)^2} \big[ a_0^f + a_1^f\:z^f(q^2)  \big], \label{eq:nominalfitphys}
\end{eqnarray}
while the higher-order fit function is given by
\begin{eqnarray}
f_{\rm HO}(q^2) &=& \frac{1}{1-q^2/(m_{\rm pole}^f)^2} \big[ a_{0,{\rm HO}}^f + a_{1,{\rm HO}}^f\:z^f(q^2) + a_{2,{\rm HO}}^f\:[z^f(q^2)]^2  \big]. \label{eq:HOfitphys}
\end{eqnarray}
The values of the pole masses are given in Table \ref{tab:polemasses}, and the kinematic variables $z^f$ are defined as
\begin{eqnarray}
z^f(q^2) &=& \frac{\sqrt{t_+^f-q^2}-\sqrt{t_+^f-t_0}}{\sqrt{t_+^f-q^2}+\sqrt{t_+^f-t_0}}, \\
t_0 &=& (m_{\Lambda_b} - m_{\Lambda_c})^2, \\
t_+^f &=& (m_{\rm pole}^f)^2.
\end{eqnarray}
As in Ref.~\cite{Detmold:2015aaa}, in the fits to the lattice data we evaluated the pole masses as $a m_{\rm pole}^f = a m_{B_c}^{(\rm lat)} + a \Delta^f$,
where $a m_{B_c}^{(\rm lat)}$ are the lattice QCD results for the pseudoscalar $B_c$ mass on each individual data set, and the splittings $\Delta^f$
are fixed to their physical values $\Delta^{h_+, h_\perp}=56$ MeV and $\Delta^{\widetilde{h}_+, \widetilde{h}_\perp}=492$ MeV. The form factor
results are very insensitive to the choices of $\Delta^f$ (as expected for poles far above $q^2_{\rm max}$). When varying $\Delta^f$ by $\pm 10\%$,
the $z$-expansion parameters returned from the fit are found to change in such a way that the changes in the form factors
themselves are below $0.2\%$ in the entire semileptonic region.

Plots of the lattice QCD data for the tensor form factors, along with the nominal fit functions in the physical limit, are shown
in Fig.~\ref{fig:latticeFF}. The same fit functions are plotted in the entire kinematic range in Fig.~\ref{fig:FF}, where
also the total (statistical plus systematic) uncertainties are shown. The form factor $h_+$ has larger uncertainties
than the other form factors because of larger excited-state contributions in the lattice QCD correlation functions.

The values of the nominal and higher-order fit parameters for the tensor form factors are given in Table \ref{tab:FFparameters}.
Because of the kinematic constraint
\begin{equation}
 \widetilde{h}_\perp(q^2_{\rm max}) = \widetilde{h}_+(q^2_{\rm max}),
\end{equation}
which is at the point $z=0$, the form factors $\widetilde{h}_\perp$ and $\widetilde{h}_+$
share the common parameters $a_0^{\widetilde{h}_{\perp},\widetilde{h}_+}$. To evaluate the uncertainties of the form factors and of any observables
depending on the form factors, it is essential to include the (cross-)correlations between all form factor parameters. The full covariance matrices
of the nominal and higher-order parameters of all ten $\Lambda_b\to\Lambda_c$ form factors (vector, axial vector, and tensor) are provided as supplemental files.

\section{Model-independent analysis of individual new-physics couplings}
\label{sec:modelindependent}

In this section we consider one new-physics coupling at a time. We first compute the constraints from the existing measurements
with mesons, and then study the impact of a future measurement of $R(\Lambda_c)$.

\subsection{Constraints from the existing measurements of $R(D)$, $R(D^*)$, and $\tau_{B_c}$}

We require the NP couplings to reproduce the measurements (\ref{eq:RDrexp}) and (\ref{eq:RDstrexp}) of $\RDr$ and $\RDrstar$ within the $3 \sigma$ range.
The coupling $g_S(g_P)$ only contributes to $\RDr (\RDrstar)$ while the other couplings contribute to both channels. If only $g_L$ is nonzero, the
SM contribution gets rescaled by an overall factor $| 1+ g_L|^2$, so that \cite{Bhattacharya:2014wla}
\bea
\RDr=  \RDrstar=  \Rlcr=   | 1+ g_L|^2    ,\
\label{ratio_prediction}
\eea
which is consistent with the present measurements (\ref{eq:RDrexp}) and (\ref{eq:RDstrexp}). Note that in the $g_L$-only scenario
the forward-backward asymmetry (\ref{eq:AFB}) is unmodified, $A_{FB}= A_{FB}^{SM}$.

There is also a measurement of the $\tau$ polarization by Belle \cite{Abdesselam:2016xqt}  with the result $P_{\tau} =-0.44 \pm 0.47^{+0.20}_{-0.17}$. The uncertainties of this measurement are presently too large to provide a significant additional constraint and we therefore do not include $P_{\tau}$ in our analysis.

It was recently pointed out \cite{Li:2016vvp,Alonso:2016oyd,Celis:2016azn} that the measured lifetime of the $B_c$ meson, $\tau_{B_c}=0.507(9)\:\:{\rm ps}$  \cite{Olive:2016xmw}, provides an upper bound on the $B_c \to \tau^- \bar{\nu}_\tau$ decay rate, which yields a strong constraint on the $g_P$ coupling. According to SM calculations using an operator product expansion \cite{Beneke:1996xe}, only about $5\%$ (for the central value) of the total width of the $B_c$, $\Gamma_{B_c}=1/\tau_{B_c}$, can be attributed to purely tauonic and semi-tauonic modes.
This can be relaxed as the parameters in the calculations are varied.
In our analysis, we use an upper limit of $\mathcal{B}(B_c \to \tau^- \bar{\nu}_\tau) \le 30\%$ to put constraints on the new-physics couplings.
We use $f_{B_c}=0.434(15)\:{\rm GeV}$ from lattice QCD \cite{Colquhoun:2015oha}.

\begin{figure}
\begin{adjustwidth}{-0.5cm}{-0.5cm}
\begin{center}
\includegraphics[height=0.8cm]{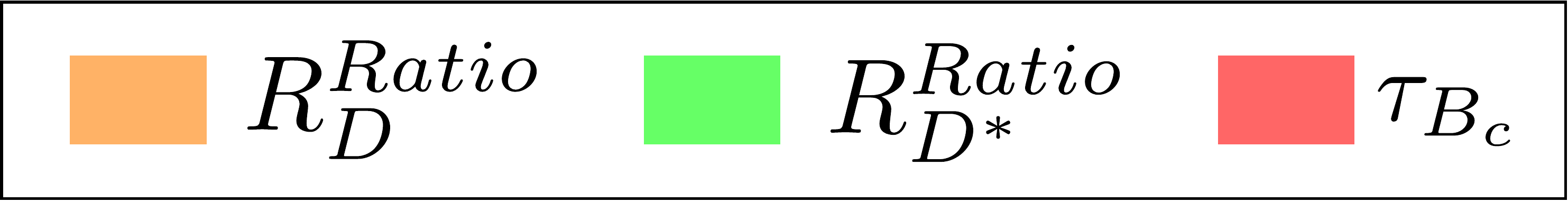}

\vspace{0.5cm}

\includegraphics[width=5cm]{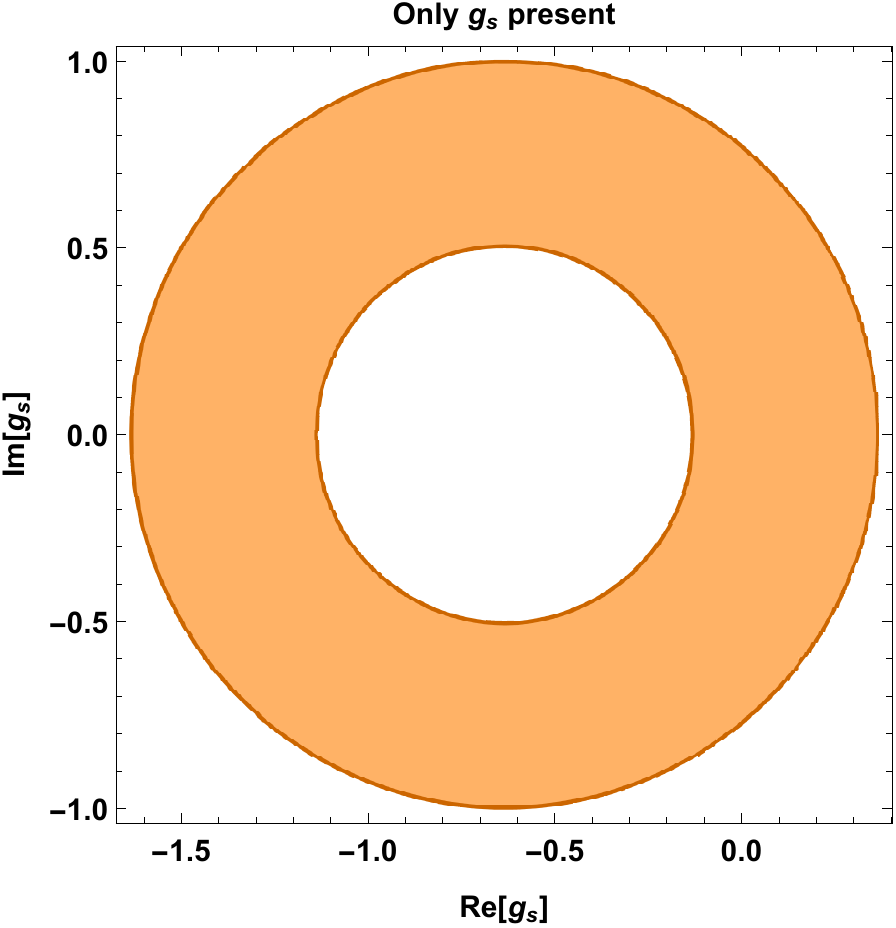}~~~
\includegraphics[width=5cm]{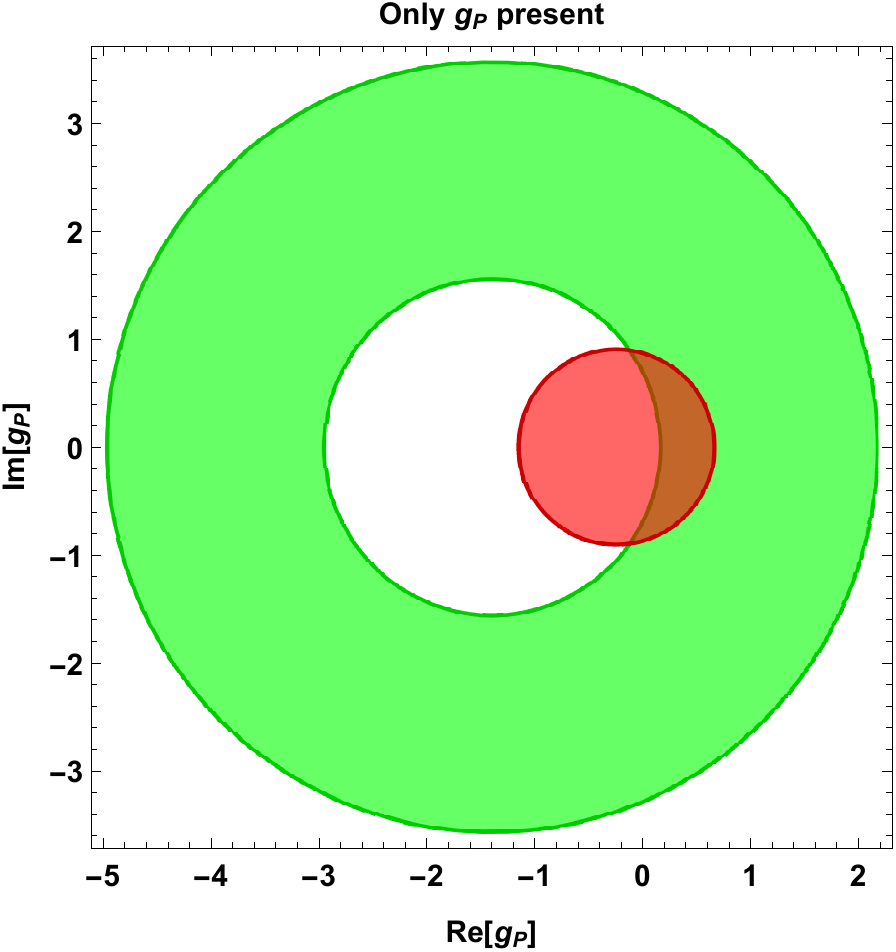}~~~\\
\includegraphics[width=5cm]{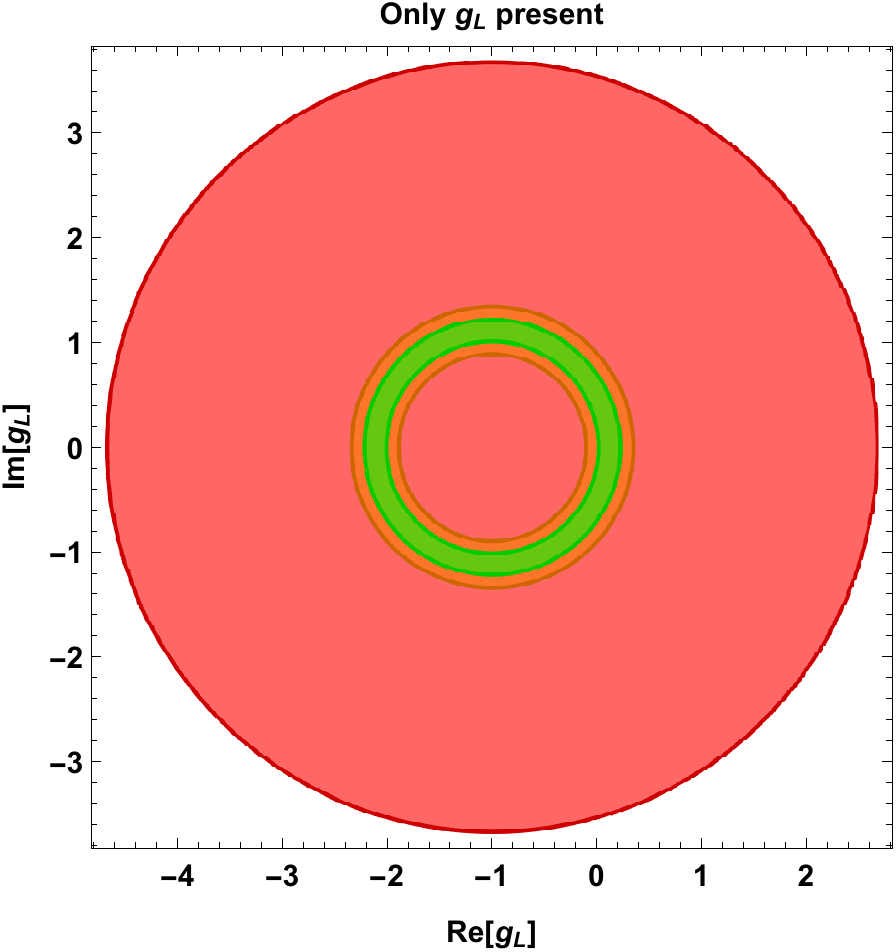}~~~
\includegraphics[width=5cm]{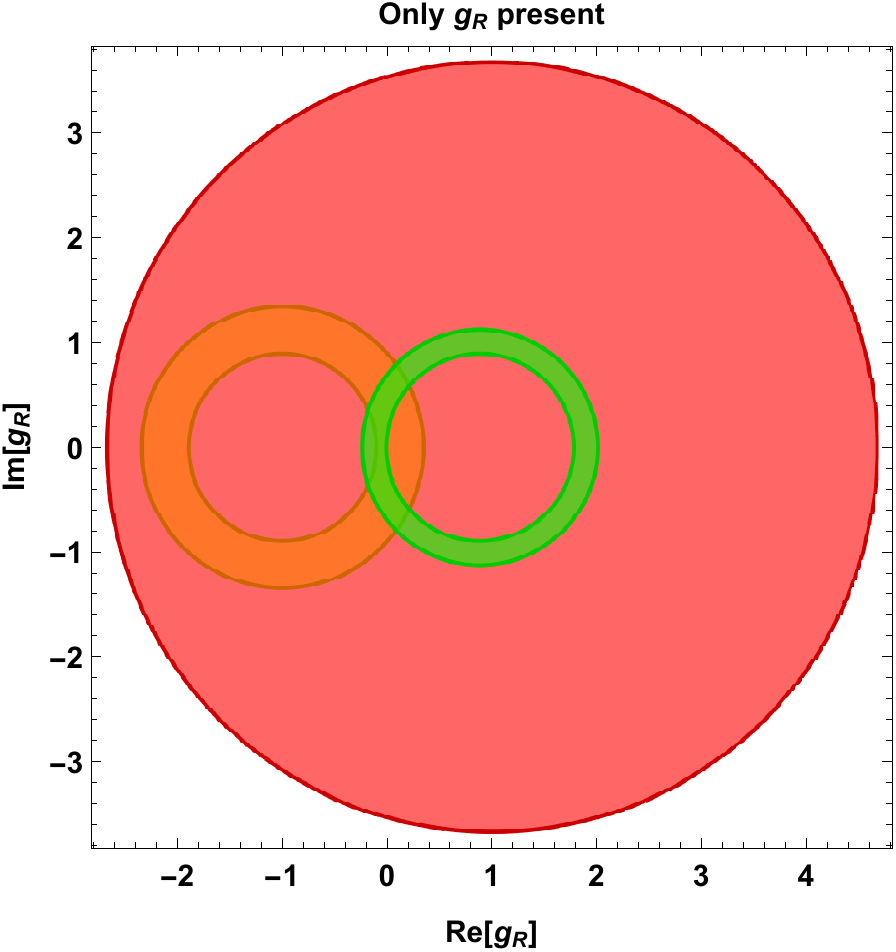}~~~
\includegraphics[width=5cm]{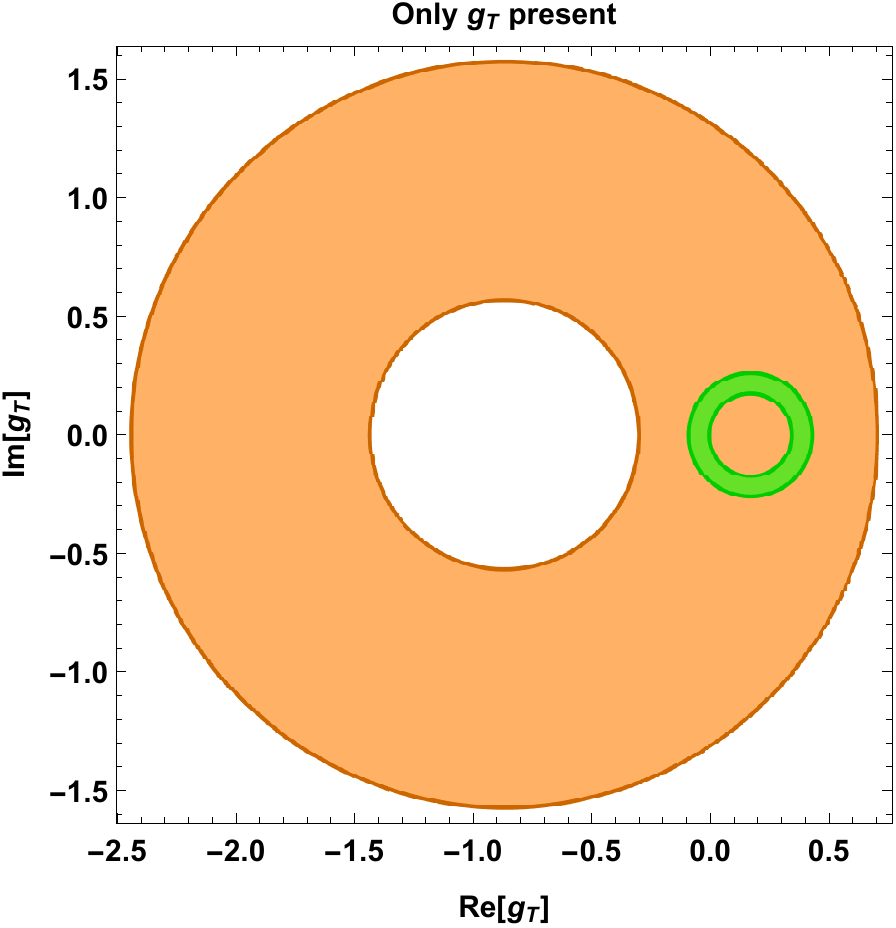}~~~
\end{center}
\end{adjustwidth}
\caption{Constraints on the individual new-physics couplings from the measurements of $\RDr$, $\RDrstar$, and $\tau_{B_c}$.
We require that the couplings reproduce the measurements of $\RDr$ and $\RDrstar$ in Eqs.~(\ref{eq:RDrexp}) and (\ref{eq:RDstrexp}) within 3$\sigma$,
and satisfy $\mathcal{B}(B_c \to \tau^- \bar{\nu}_\tau) \le 30\%$.}
\label{constraints-Ind}
\end{figure}

In Fig.~\ref{constraints-Ind}, we present the constraints on the new-physics couplings coming from the measurements of $\RDr$, $\RDrstar$, and $\tau_{B_c}$.
We see that $\tau_{B_c}$ puts a strong constraint on $g_P$, and weak constraints on $g_L$ and $g_R$.
The tensor coupling $g_T$ is strongly constrained by $\RDrstar$, and only weakly constrained by $\RDr$.
 
Example values of the ratios $\Rlc$ and $\Rlcr=\Rlc/\Rlc^{SM}$ for representative allowed values of the NP couplings are given in Table \ref{tab:RLcExamples}. The standard-model prediction for $\Rlc$ is $0.333 \pm 0.010$ \cite{Detmold:2015aaa}. We find that large deviations from this value are possible with the present mesonic constraints.  In Table \ref{TableMax}, we present the maximum and minimum allowed values of $\Rlcr=\Rlc/\Rlc^{SM}$ in the presence of each individual new-physics coupling, and the corresponding values of the coupling at which these occur.

\begin{figure}
\begin{adjustwidth}{-0.5cm}{-0.5cm}
\includegraphics[width=5cm]{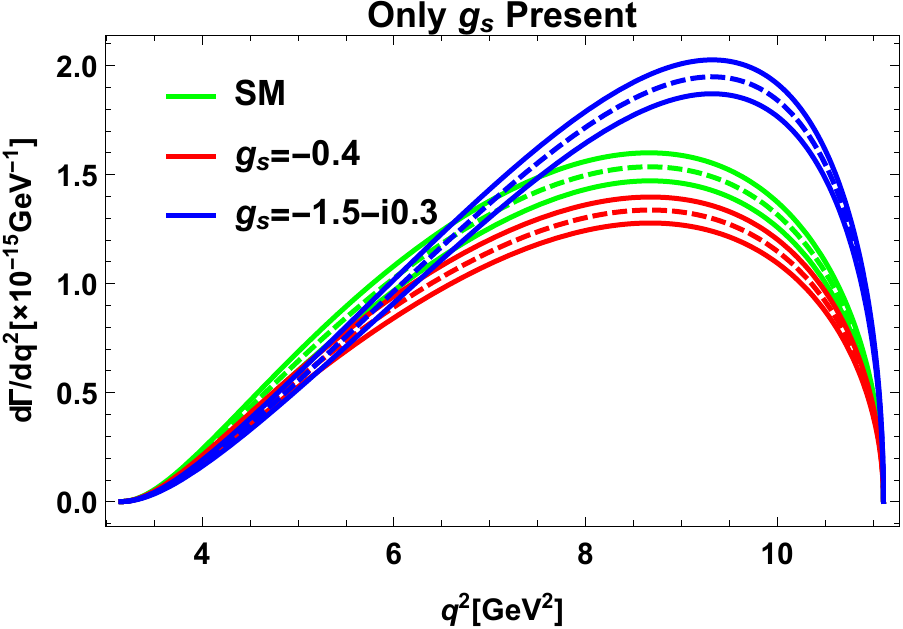}~~~
\includegraphics[width=5cm]{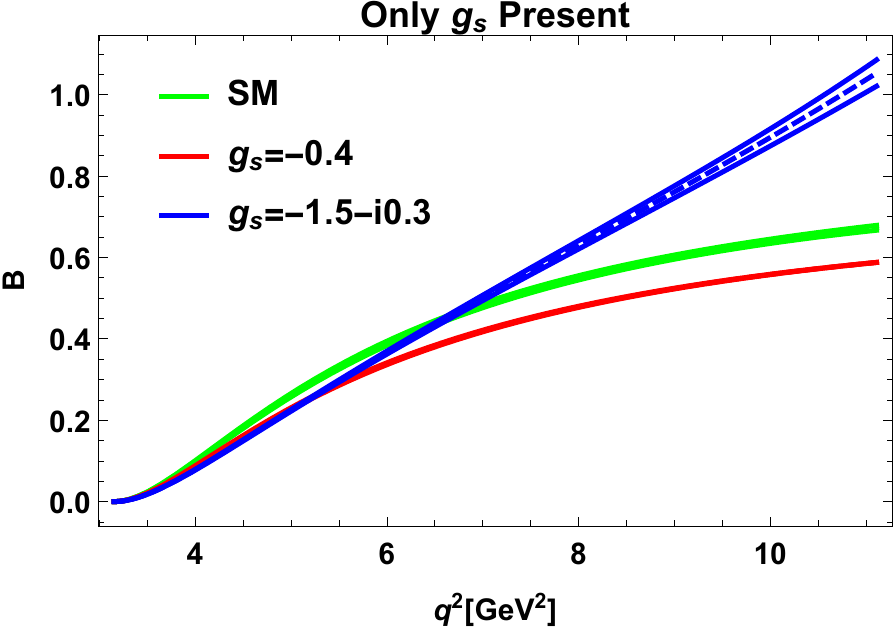}~~~
\includegraphics[width=5cm]{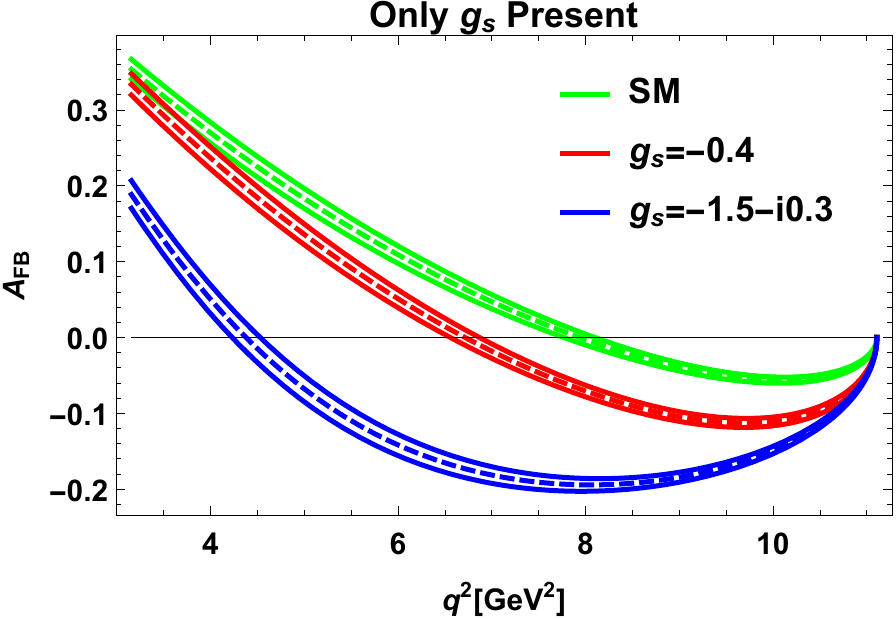}\\[0.3cm]
\includegraphics[width=5cm]{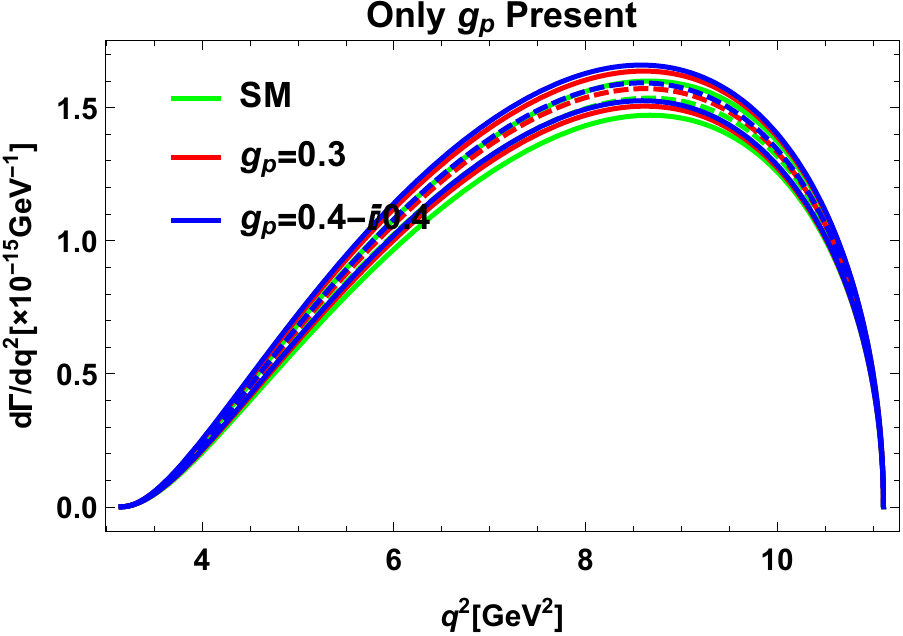}~~~
\includegraphics[width=5cm]{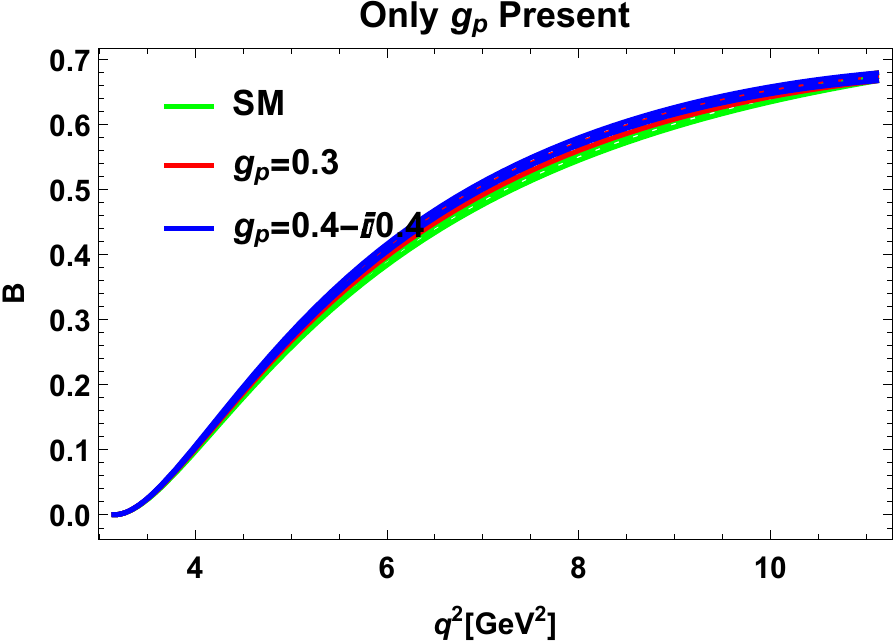}~~~
\includegraphics[width=5cm]{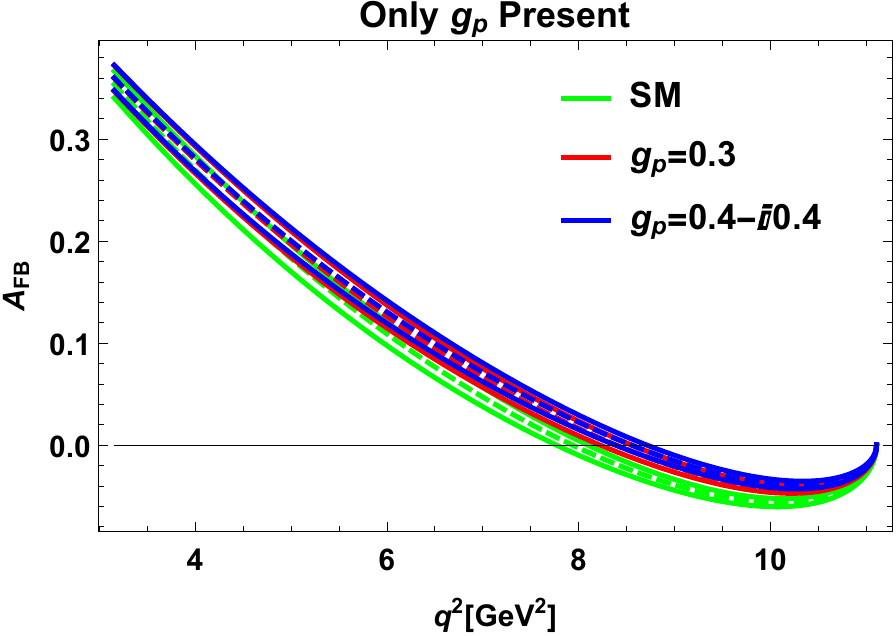}\\[0.3cm]
\includegraphics[width=5cm]{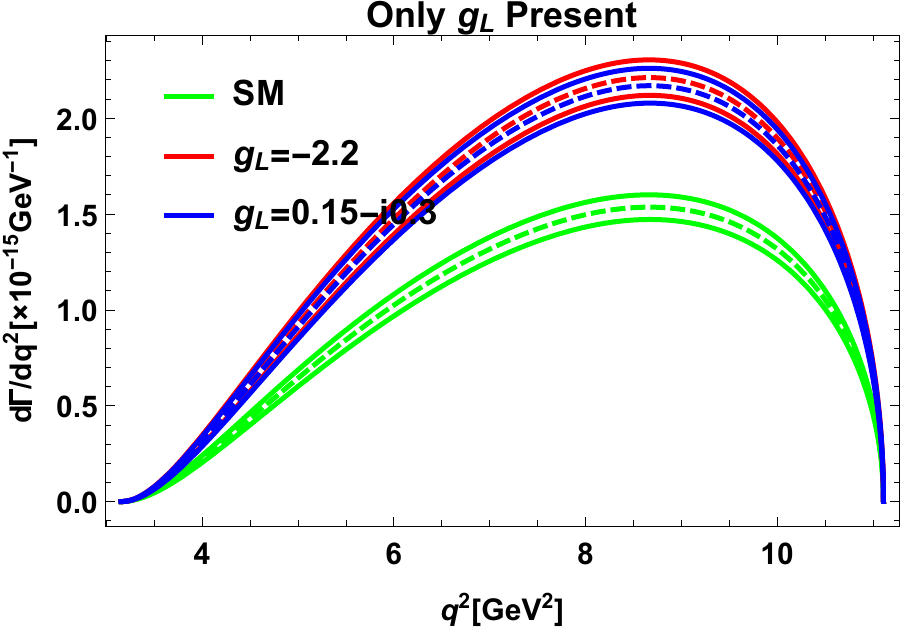}~~~
\includegraphics[width=5cm]{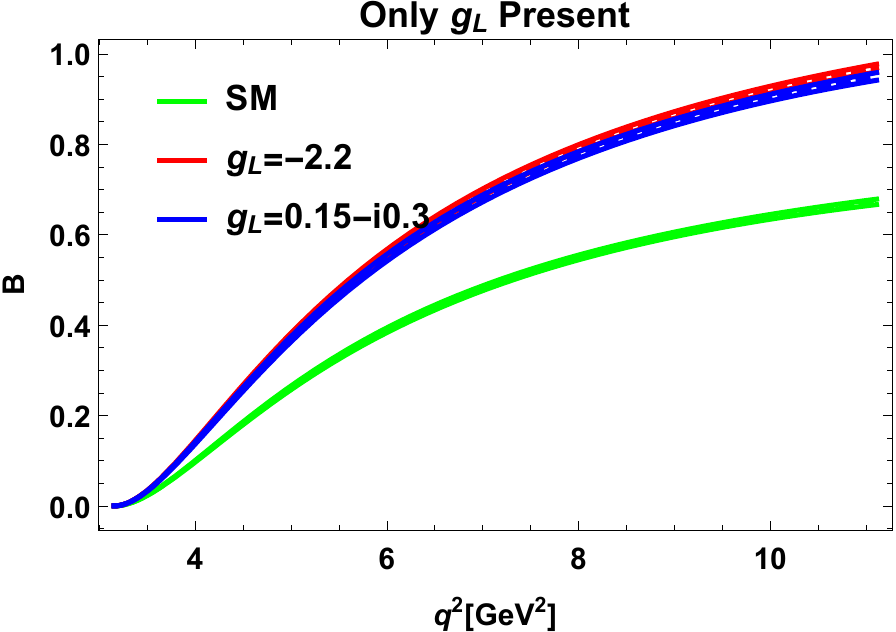}~~~
\includegraphics[width=5cm]{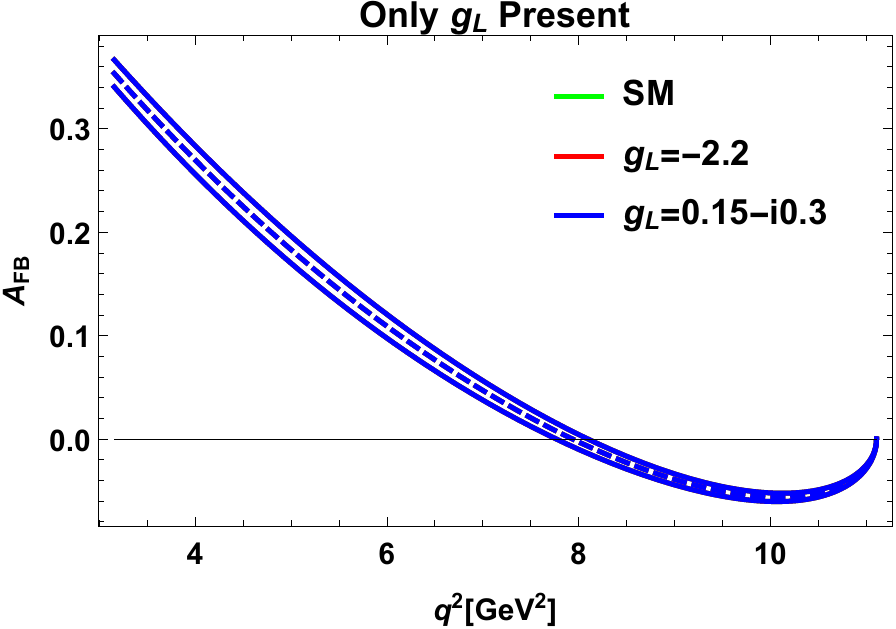}\\[0.3cm]
\includegraphics[width=5cm]{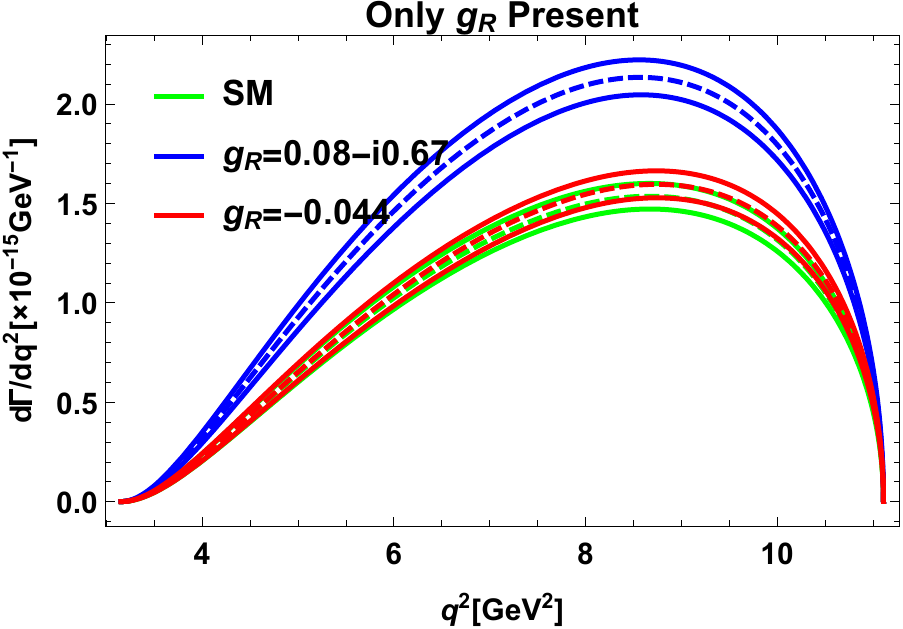}~~~
\includegraphics[width=5cm]{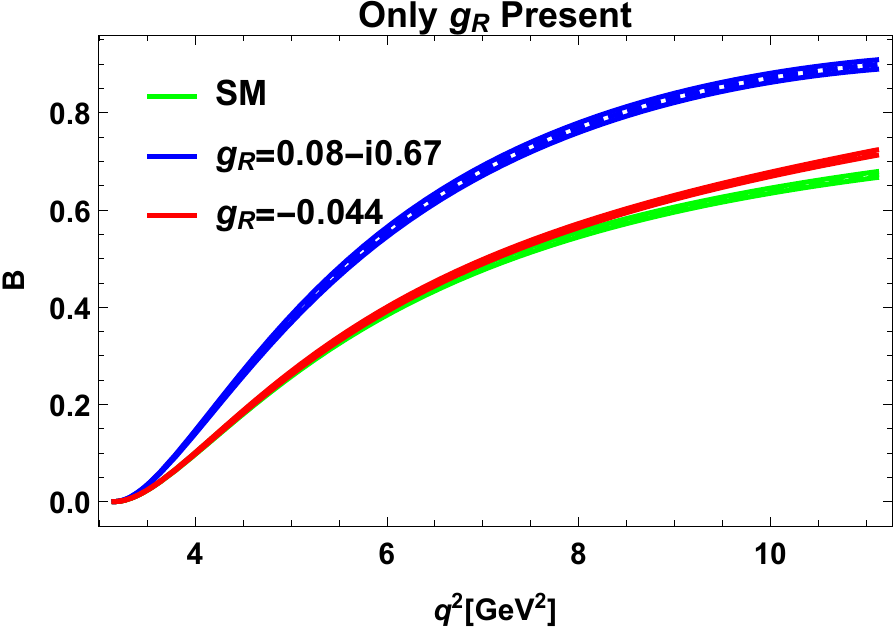}~~~
\includegraphics[width=5cm]{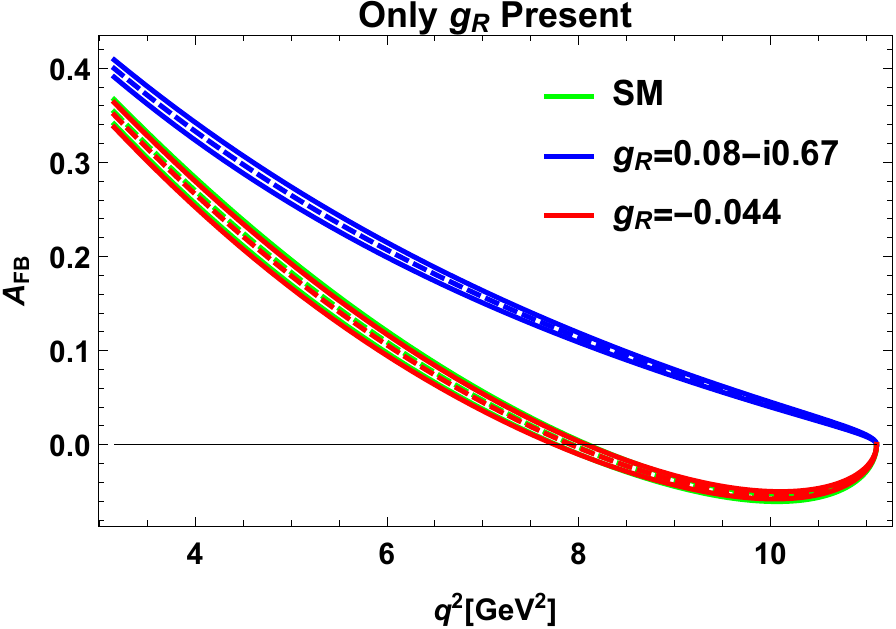}\\[0.3cm]
\includegraphics[width=5cm]{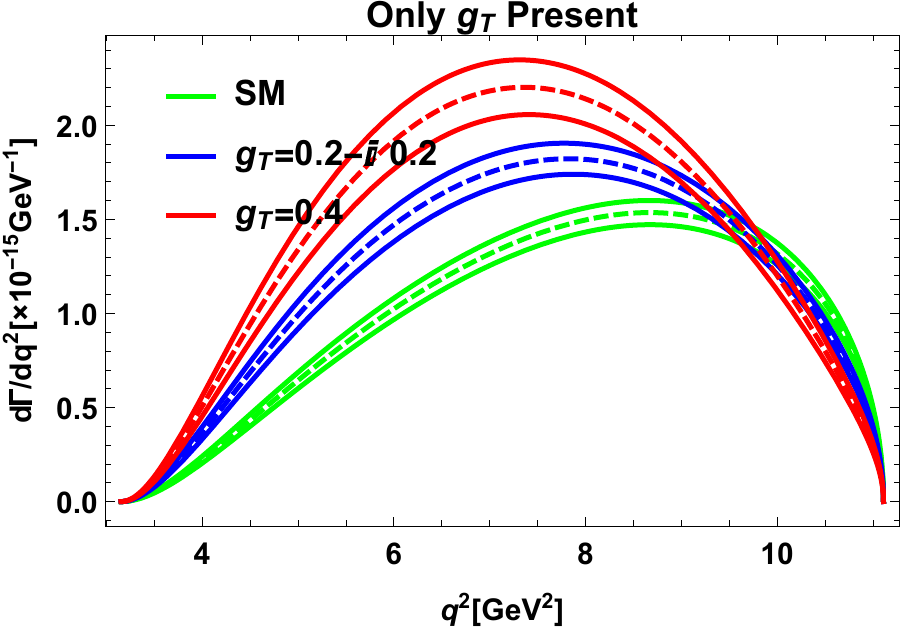}~~~
\includegraphics[width=5cm]{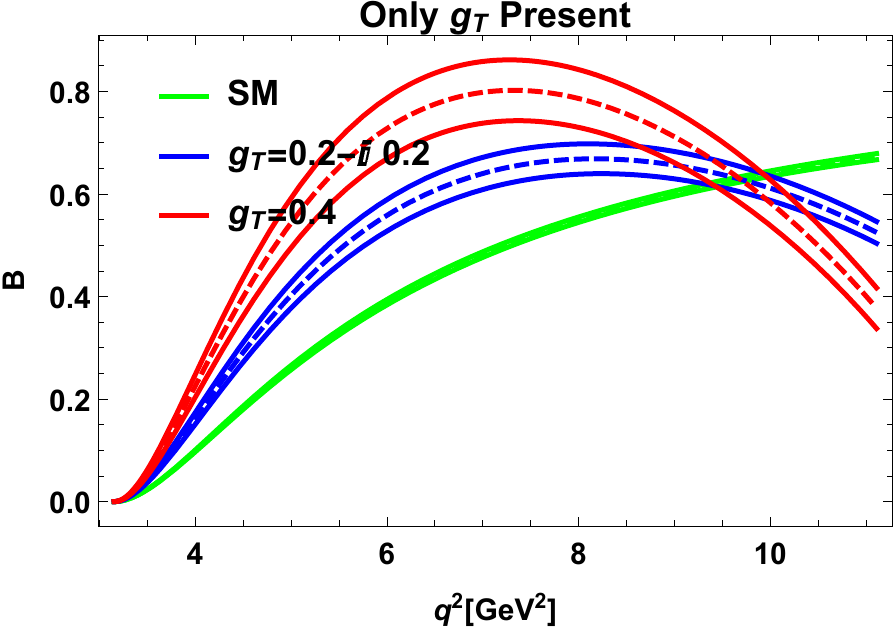}~~~
\includegraphics[width=5cm]{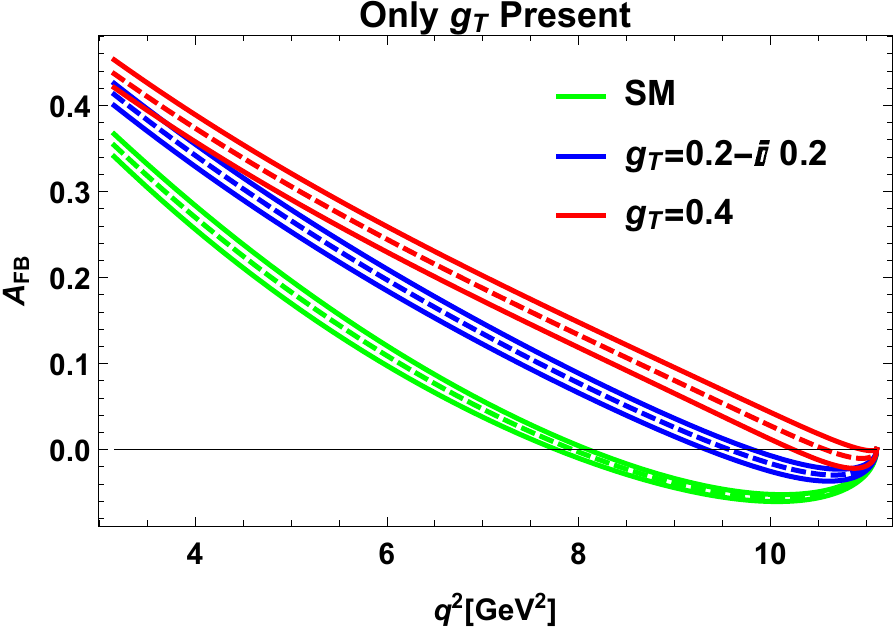}
\end{adjustwidth}
\caption{The effect of individual new-physics couplings on the $\lbt$ differential decay rate (left), the ratio of the $\lbt$ and $\lbl$ differential decay rates (middle), and the $\lbt$ forward-backward asymmetry (right). Each plot shows the observable in the Standard Model and for two representative values of the new-physics coupling (one real-valued choice and one complex-valued choice). The bands indicate the $1\sigma$ uncertainties originating from the $\Lambda_b\to\Lambda_c$ form factors.}
\label{fig:individualcouplingexamples}
\end{figure}

Figure \ref{fig:individualcouplingexamples} shows the effect of representative values of the individual NP couplings on the $\lbt$ differential decay rate (evaluated assuming $|V_{cb}|=0.041$) as well as $B_{\Lambda_c}(q^2)$ [defined in Eq.~(\ref{eq:Bqsqr})] and $A_{FB}(q^2)$. In all cases, except for the strongly constrained pure $g_P$ coupling, substantial deviations from the SM predictions are allowed. We notice that $A_{FB}$ is typically above the SM prediction
in the presence of $g_R$ or $g_T$, while it is typically below the SM prediction in the presence of $g_S$. Hence, it is possible to use $A_{FB}$ to distinguish between the different couplings.

\begin{table}
\begin{center}
 \begin{tabular}{|c | c | c | c | c | c |} 
 \hline
  & $g_S$ only & $g_P$ only & $g_L$ only & $g_R$ only & $g_T$ only  \\ [0.5ex] 
 \hline
  & $-0.4$ & $0.3$ & $-2.2$ & $-0.044$ & $0.4$  \\
  \hline
 $\Rlc$ & $0.290 \pm 0.009$ & $0.342 \pm 0.010$ & $0.479\pm 0.014$ & $0.344\pm 0.011$ & $0.475\pm 0.037$ \\
 $\Rlcr$ & $0.872 \pm 0.007$ & $1.026\pm 0.001$ & $1.44$ & $1.033\pm0.003$ & $1.426\pm 0.100$ \\
 \hline
   & $-1.5 - 0.3 i$ & $0.4 - 0.4 i$ & $0.15 - 0.3 i$ & $0.08 - 0.67 i$ & $0.2 - 0.2 i$   \\
   \hline
  $\Rlc$ & $0.384 \pm 0.013$ & $0.346\pm 0.011$ & $0.470 \pm 0.014$ & $0.465 \pm 0.014$ & $0.404\pm 0.021$ \\
  $\Rlcr$ & $1.154\pm 0.008$ & $1.040\pm 0.002$ & $1.412$ & $1.397\pm 0.005$ & $1.213\pm 0.050$ \\
  \hline
\end{tabular}
\caption{The values of $\Rlc$ and $\Rlcr$ for two example choices (real-valued and complex-valued) of the new-physics couplings. The standard-model value of $\Rlc$ is $0.333 \pm 0.010 $ \cite{Detmold:2015aaa}. The uncertainties given are due to the form factor uncertainties.}
\label{tab:RLcExamples}
\end{center}
\end{table}

\begin{table}
 \begin{center}
 \begin{tabular}{|c | c | c | c | c | c | c |} 
 \hline
 Coupling & $R(\lc)_{max}$ & $R_{\Lambda_c,max}^{Ratio}$ & coupling value &$R(\lc)_{min}$ & $R_{\Lambda_c,min}^{Ratio}$ & coupling value  \\ [0.5ex] 
 \hline
 $g_S$ only & $0.405$ & $1.217$ & $0.363 $ &$0.314$ & $0.942  $ & $-1.14$     \\ 
 $g_P$ only & $0.354$ & $1.062$ & $0.658 $ & $0.337$ & $1.014 $ & $0.168$     \\
 $g_L$ only & $0.495 $ & $1.486 $ & $0.094 + 0.538i$ & $0.340$ & $1.022$ & $-0.070 + 0.395i$    \\
  $g_R$ only & $0.525$ & $1.576 $ & $0.085 + 0.793i$ & $0.336$ & $1.009 $ & $-0.012$    \\ 
  $g_T$ only & $0.526$ & $1.581$ & $0.428 $ & $0.338$ & $1.015 $ & $-0.005$   \\
  \hline
\end{tabular}
\caption{The maximum and minimum values of  $\Rlc$ and $\Rlcr$ allowed by the mesonic constraints for each new-physics coupling, and the coupling
values at which these extrema are reached.}
\label{TableMax}
\end{center}
\end{table}

\FloatBarrier
\subsection{Impact of a future $R(\Lambda_c)$ measurement}

\begin{figure}
\begin{adjustwidth}{-0.5cm}{-0.5cm}
\begin{center}
\includegraphics[height=0.8cm]{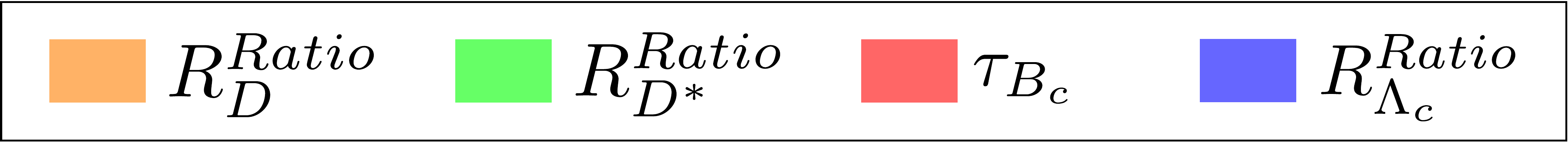}

\vspace{0.5cm}

\includegraphics[width=5cm]{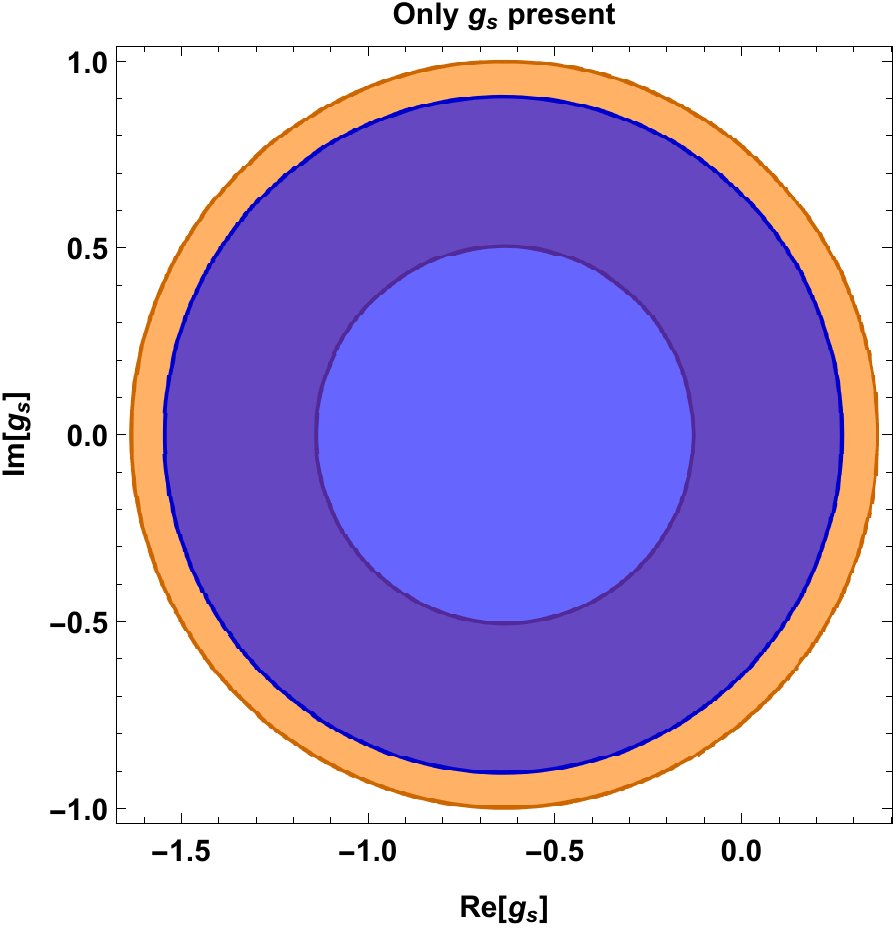}~~~
\includegraphics[width=5cm]{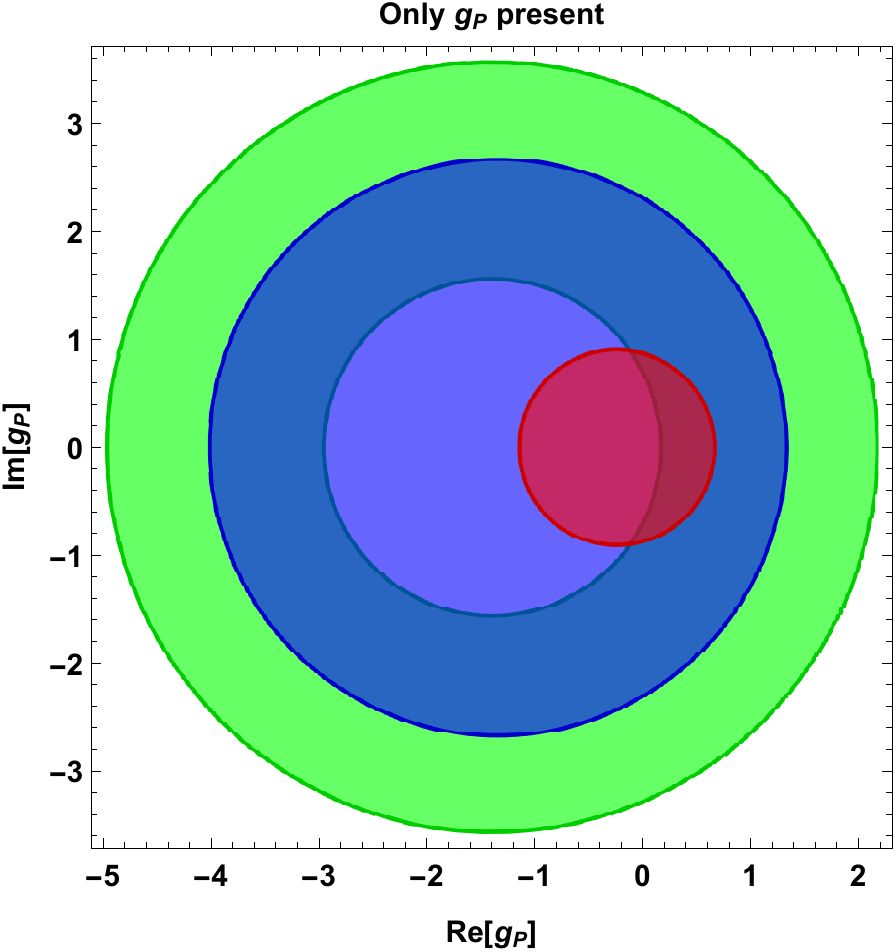}~~~\\
\includegraphics[width=5cm]{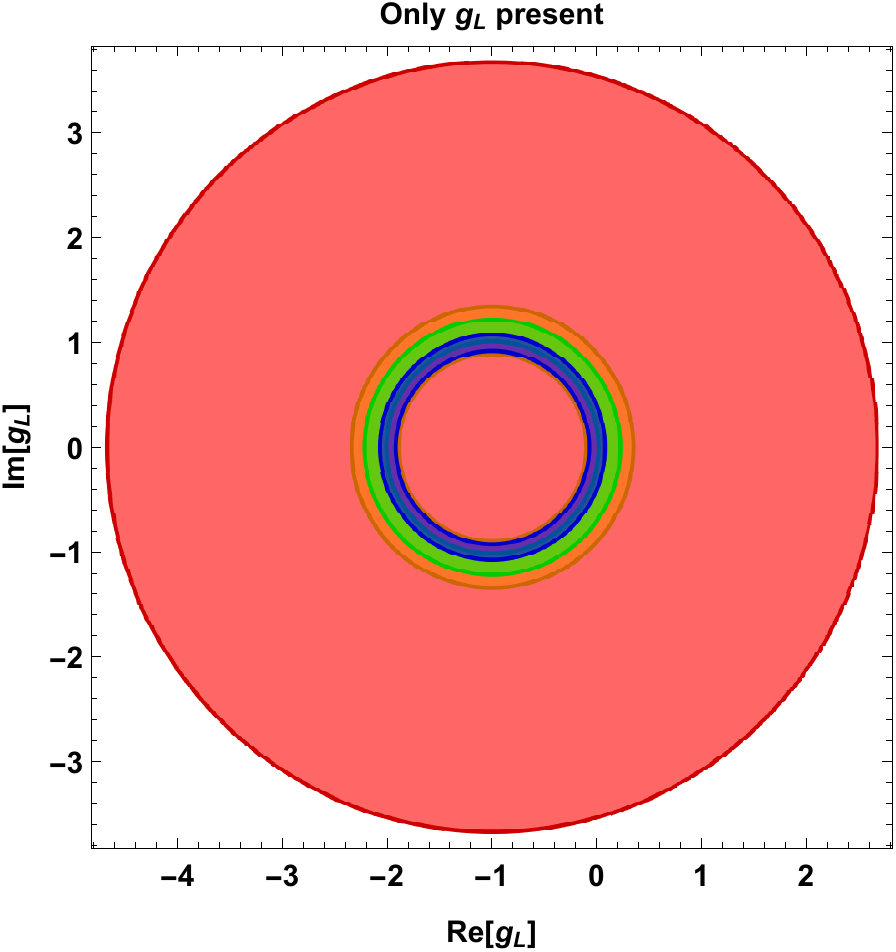}~~~
\includegraphics[width=5cm]{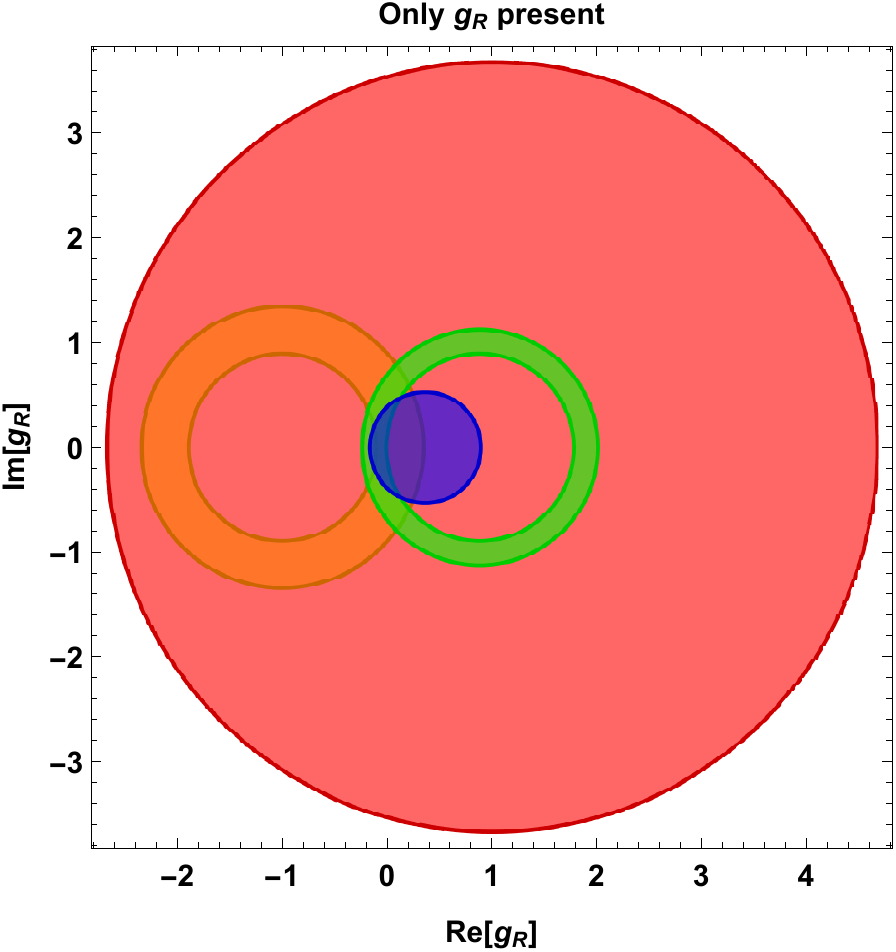}~~~
\includegraphics[width=5cm]{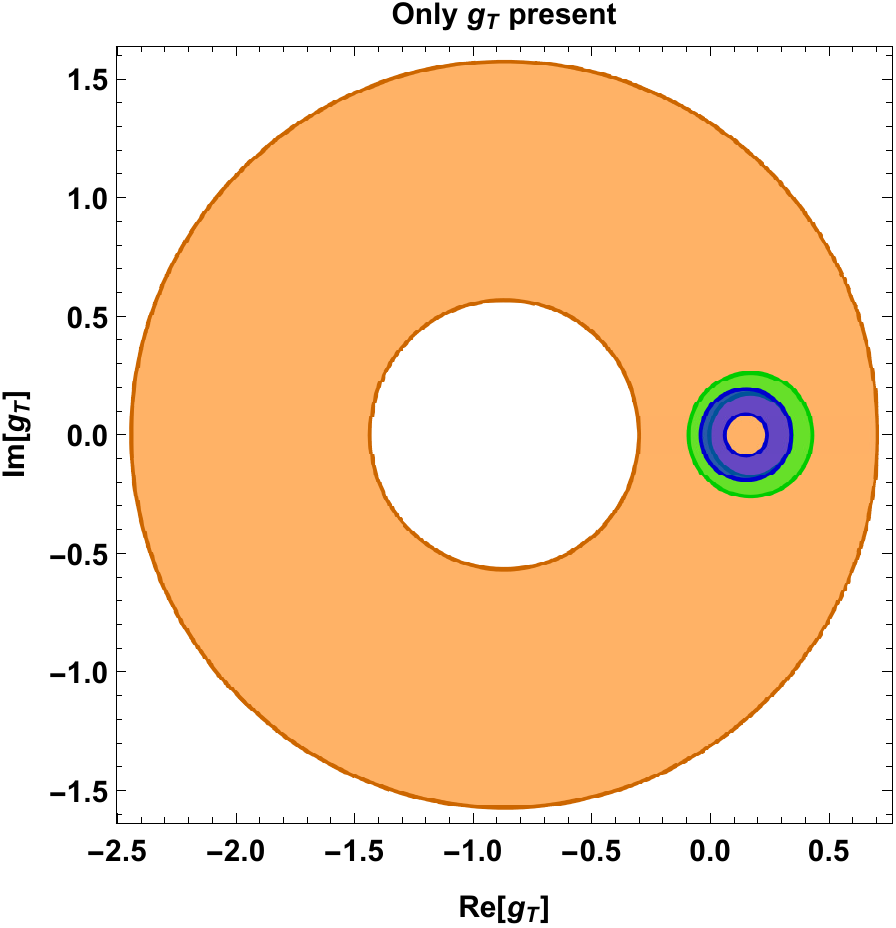}~~~
\end{center}
\end{adjustwidth}
\caption{Constraints on individual new-physics couplings from a possible $\Rlc$ measurement (shown in blue), assuming that $\Rlcr=1\pm 3\times 0.05$ where the $1\sigma$ uncertainty is $0.05$. Also shown are the mesonic constraints as in Fig.~\protect\ref{constraints-Ind}.}
\label{constraints-Ind-NearSM}
\end{figure}

\begin{figure}
\begin{adjustwidth}{-0.5cm}{-0.5cm}
\begin{center}
\includegraphics[height=0.8cm]{legend_constraints-Ind_with_RLc.pdf}

\vspace{0.5cm}

\includegraphics[width=5cm]{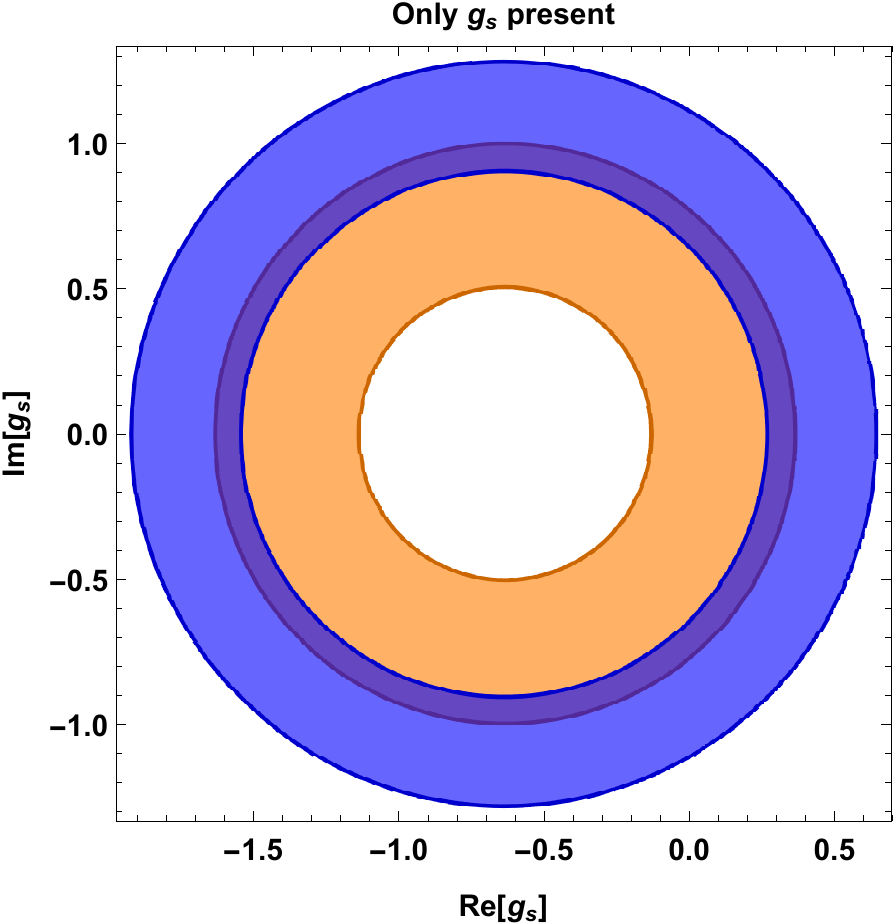}~~~
\includegraphics[width=5cm]{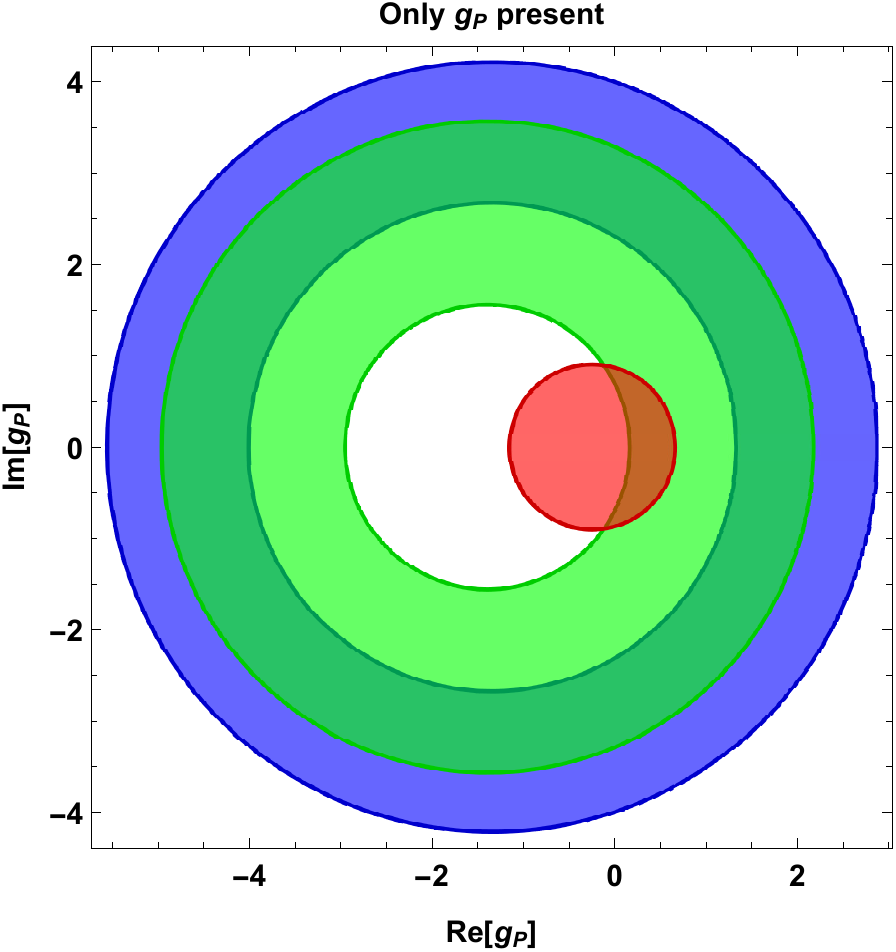}~~~\\
\includegraphics[width=5cm]{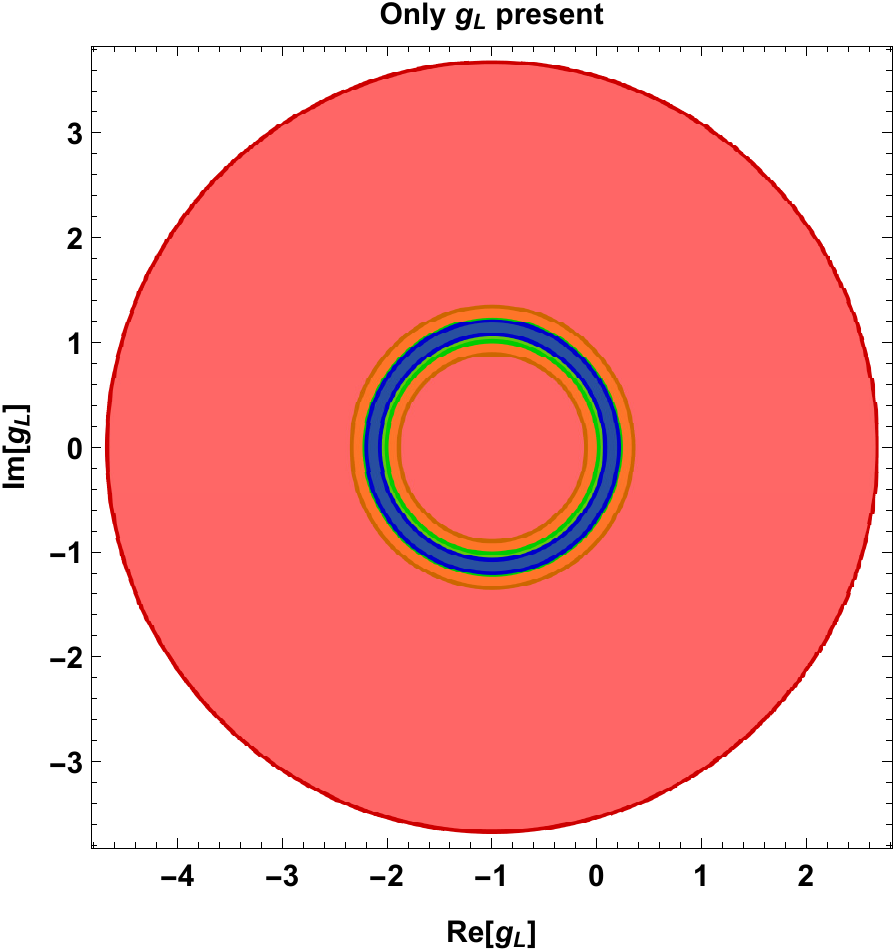}~~~
\includegraphics[width=5cm]{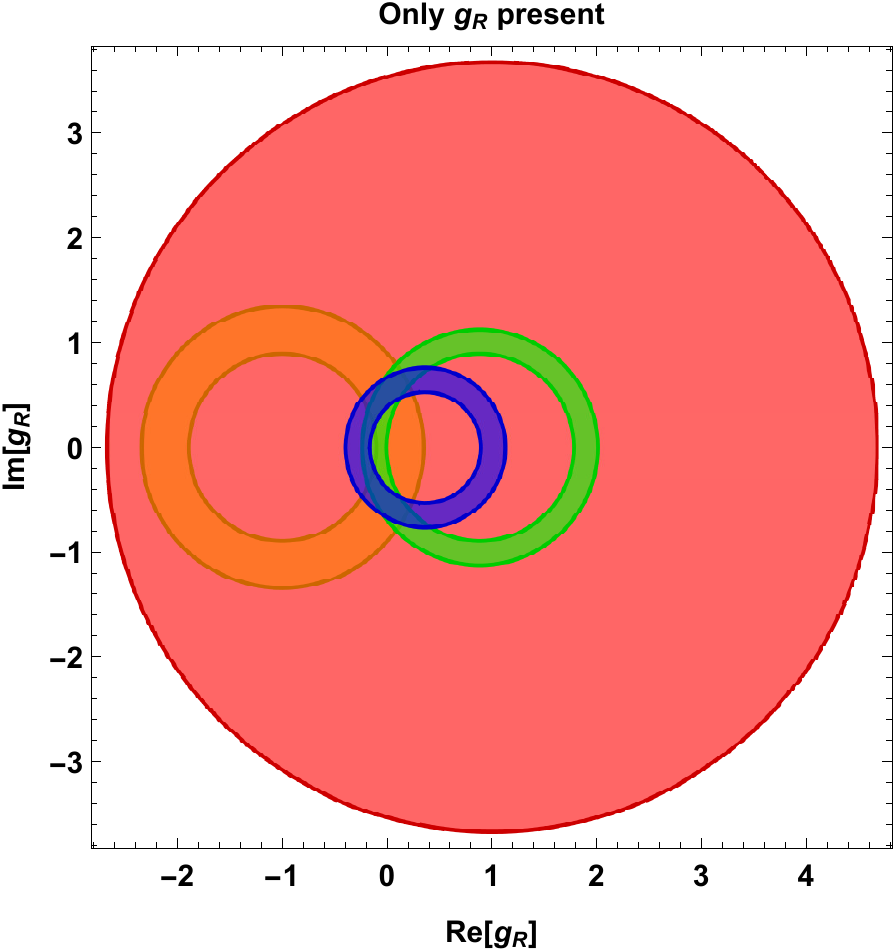}~~~
\includegraphics[width=5cm]{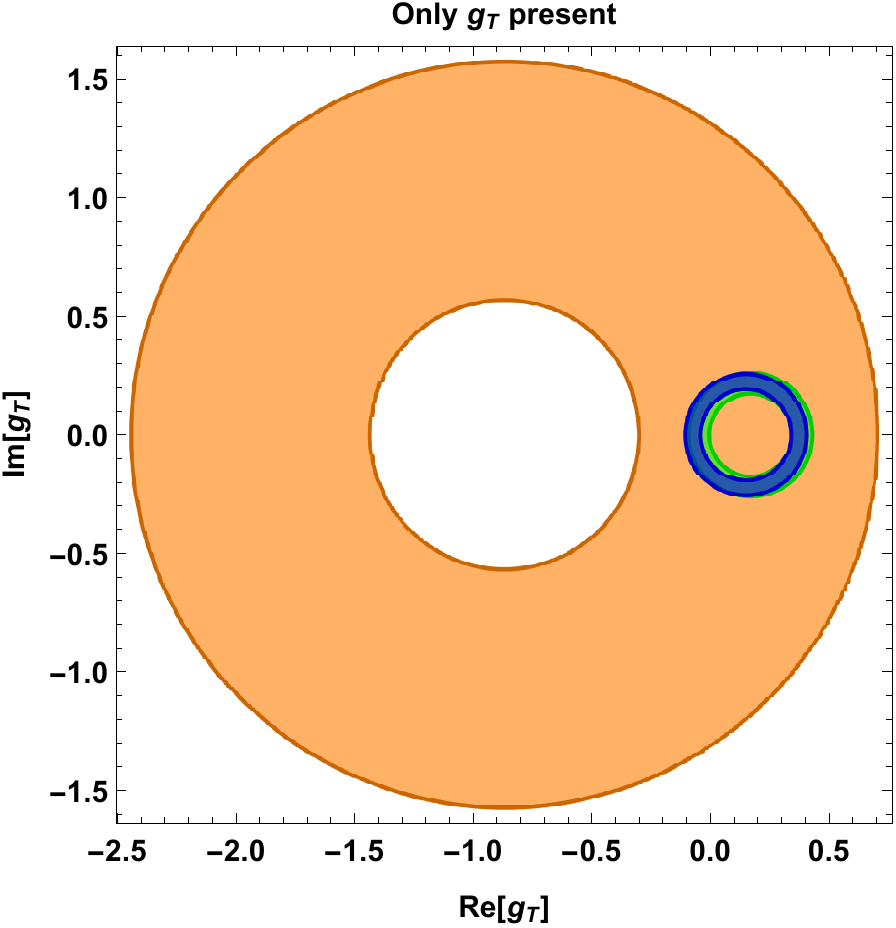}~~~
\end{center}
\end{adjustwidth}
\caption{Constraints on individual new-physics couplings from a possible $\Rlc$ measurement (shown in blue), assuming that $\Rlcr=1.3\pm 3\times 0.05$ where the $1\sigma$ uncertainty is $0.05$. Also shown are the mesonic constraints as in Fig.~\protect\ref{constraints-Ind}.}
\label{constraints-Ind-FarSM}
\end{figure}

In this subsection we present the effect of possible future measurements of $R(\Lambda_c)$ on the NP couplings constraints. We consider two cases, one in which the measured value is near the SM prediction and one with measured value far from SM. For the first case we take $\Rlcr=1\pm 3\times 0.05$, and for the second case  $\Rlcr=1.3\pm 3\times 0.05$ (the same central values as $\RDr$). Note that we take the $1\sigma$ uncertainty as $0.05$. Figures \ref{constraints-Ind-NearSM} and \ref{constraints-Ind-FarSM} show the allowed regions of the parameter space for the first and second case, respectively. We observe the following
when adding the $\Rlcr$ constraints to the mesonic constraints:
\begin{itemize}
 \item For $R(\Lambda_c)$ near the SM (Fig.~\ref{constraints-Ind-NearSM}), the allowed regions for $(g_L,\; g_R,\; g_T)$ are reduced significantly, the allowed region for $g_S$ shrinks only slightly, and the allowed region for $g_P$ remains the same (as it is dominantly constrained by $\tau_{B_c}$).
 \item For $R(\Lambda_c)$ far from the SM (Fig.~\ref{constraints-Ind-FarSM}), most of the previously allowed region for $g_S$ becomes excluded by $R(\Lambda_c)$. Even more importantly, the $g_P$-only scenario becomes ruled out. In this case, $R(\Lambda_c)$ also provides strong constraints on $(g_L,\; g_R,\; g_T)$, but these constraints still overlap with the mesonic constraints.
\end{itemize}

\FloatBarrier
\section{Explicit models}
\label{sec:models}

In this section we will discuss explicit models that can generate the couplings in the effective Hamiltonian (\ref{eq1:Lag}). We will consider three categories: Two-Higgs-doublet models which generate ($g_S$, $g_P$), $SU(2)$ models which generate $g_L$, and leptoquark models which generate ($g_S$, $g_P$, $g_L$, $g_T$). We do not consider models that generate $g_R$, as in the standard-model-effective-theory picture it is difficult to have a $g_R$ coupling that leads to lepton universality violation effects \cite{Alonso:2015sja}. 


\subsection{Two-Higgs-doublet models}

The simplest scalar extensions of the SM are the two-Higgs-doublet models (2HDM). The 2HDM of type II is disfavored by experiment \cite{Lees:2013uzd}. 
We will consider the Aligned Two-Higgs-Doublet Model (A2HDM) from Ref.~\cite{Celis:2012dk}. The Lagrangian of the model is
 \begin{equation}\label{lagrangian}
 \mathcal L_Y^{H^\pm} =  - \frac{\sqrt{2}}{v}\, H^+ \left\{ \bar{u} \left[ \xi_d\, V M_d \mathcal P_R - \xi_u\, M_u V \mathcal P_L \right] d\, + \, \xi_l\, \bar{\nu} M_l \mathcal P_R l \right\}
 \,+\,\mathrm{h.c.}, \
\end{equation}
where $u$, $d$, and $l$ denote all three generations of up-type quarks, down-type quarks, and charged leptons, $M_u$ and $M_d$ are the quark mass matrices,
and $V$ is the CKM matrix. Above, $\xi_f$~($f=u,d,l$) are the proportionality parameters in the so-called ``Higgs basis'', in which only one scalar doublet acquires a nonzero vacuum expectation value. The cases $\xi_d = \xi_l = -1/\xi_u = - \tan \beta$ and $\xi_u = \xi_d = \xi_l = \cot \beta$ correspond to the Type-II and Type-I models, respectively.
The general effective couplings in Eq.~(\ref{eq1:Lag}) read
\bea
g_S^{q_{u}q_{d}l} &=& g_R^{q_{u}q_{d}l}+g_L^{q_{u}q_{d}l},\nonumber\\
g_P^{q_{u}q_{d}l} &=& g_R^{q_{u}q_{d}l}-g_L^{q_{u}q_{d}l},
\eea
where
\bea \label{eq:hy}
g_L^{q_{u}q_{d}l}\; =\; \xi_u\xi_l^*\; \frac{m_{q_u} m_l}{M_{H^\pm}^2}\, ,
\qquad\qquad
g_R^{q_{u}q_{d}l}\; =\; -\xi_d\xi_l^*\; \frac{m_{q_d} m_l}{M_{H^\pm}^2}\, .
\label{scalar_coup}
\eea
The scenario in which the $\xi_{u,d,l}$ parameters are universal for all three generations is ruled out \cite{Celis:2012dk}. We therefore assume that Eq.~(\ref{scalar_coup}) only gives the couplings for processes involving the $b$ quark, while the couplings for the first two generations are considered independently. 
In this model we find significant deviation from the standard model contribution to the decay $\lbt$, but for a more complete analysis RGE evolution should be considered.
The RGE evolution of the couplings of the A2HDM  has been discussed in Ref.~\cite{Ferreira:2010xe}. The alignment condition, which guarantees the absence of tree-level FCNC processes, is preserved by the RGE only in the case of the standard type-I, II, X, and Y models which are discussed in \cite{Pich:2009sp}. However, our framework requires non-universal flavor dependent couplings and the RGE evolution has not been worked out and is not included in the analysis.
Keeping in mind that RGE effects could change the phenomenology of the model, the discussion of the full numerical analysis of the model is not included in this work.

\subsection{$SU(2)$ and Leptoquark models}

The analysis of the $\RD$ and $R_K$ anomalies could  favor  the left-handed operator $g_L$.
In Ref.~\cite{Bhattacharya:2014wla}, it was pointed out that, assuming that the scale of NP
is much higher than the weak scale, the $g_L$ operator should be invariant under the full
$SU(3)_C \times SU(2)_L \times U(1)_Y$ gauge group. There are two
possibilities:
\bea
{\cal O}_1^{NP} &=& \frac{G_1}{\Lambda_\NP^2} ({\bar Q}'_L \gamma_\mu Q'_L) ({\bar L}'_L \gamma^\mu L'_L) ~, \nn\\
{\cal O}_2^{NP} &=& \frac{G_2}{\Lambda_\NP^2} ({\bar Q}'_L \gamma_\mu \sigma^I Q'_L) ({\bar L}'_L \gamma^\mu \sigma^I L'_L) \nn\\
&=& \frac{G_2}{\Lambda_\NP^2} \left[
2 ({\bar Q}'^{i}_L \gamma_\mu Q'^{j}_L) ({\bar L}'^{j}_L \gamma^\mu L'^{i}_L)
- ({\bar Q}'_L \gamma_\mu Q'_L) ({\bar L}'_L \gamma^\mu L'_L) \right] ~,
\label{NPoperators}
\eea
where $G_1$ and $G_2$ are both $O(1)$, and the $\sigma^I$ are the
Pauli matrices. Here $Q' \equiv (t',b')^T$ and $L' \equiv
(\nu'_\tau,\tau')^T$. The key point is that ${\cal O}_2^{NP}$ contains
both neutral-current (NC) and charged-current (CC) interactions. The
NC and CC pieces can be used to respectively explain the $R_K$ and
$R(D^{(*)})$ puzzles. In the following, we briefly review the literature
on models of this type.

In Ref.~\cite{Calibbi:2015kma}, UV completions
that can give rise to ${\cal O}_{1,2}^{NP}$ [Eq.~(\ref{NPoperators})],
were discussed. 
One among the  four possibilities for the underlying NP model is 
 a vector boson (VB) that transforms as $({\bf 1},{\bf 3},0)$
under $SU(3)_C \times SU(2)_L \times U(1)_Y$, as in the SM.

Concrete VB models were discussed in Ref.~\cite{Greljo:2015mma, Boucenna:2016wpr}
 and the simplest VB model was considered in Ref.~\cite{Bhattacharya:2016mcc}.
We refer to the VBs as $V = W'$, $Z'$. In the gauge basis, the Lagrangian describing the couplings of the
VBs to left-handed third-generation fermions is
\bea
\De\cL^{}_{V} &=& g^{33}_{qV}\(\oQ^{\prime}_{L3}~\ga^\mu\si^I~Q^{\prime}_{L3}\)V^{I}_{\mu}
~+~ g^{33}_{\ell V}\(\oL'_{L3}~\ga^\mu\si^I~L'_{L3}\)V^{I}_{\mu}~,~~
\eea
where $\sigma^I$ ($I=1,2,3$) are the Pauli matrices. Once the heavy
VB is integrated out, one obtains the following effective Lagrangian,
relevant for $\bsll$, $\bctaunu$ and $\bsnunubar$ decays:
\beq
\cL^\eff_{V} = - \frac{g^{33}_{qV}g^{33}_{\ell V}}{m^2_{V}}\(\oQ'_{L3}\ga^\mu
\si^I~Q'_{L3}\)\(\oL'_{L3}\ga^{}_\mu\si^I L'_{L3}\) ~.
\eeq
One can study the phenomenology of the model with an ansatz for the mixing matrices.  The
assumption of Ref.~\cite{Calibbi:2015kma, Bhattacharya:2016mcc} is that the transformations $D$
and $L$ involve only the second and third generations.
The key observation in Ref.~\cite{Bhattacharya:2016mcc} is the $Z'$ interaction also contributes to $B_s$ mixing and the model becomes highly constrained. If fact only a few percent deviation from the SM is allowed in the $\RD$ observables. 
  For this reason, we do not present a detailed numerical analysis of the $SU(2)$ models for the $\lbt$ decay.


We next move to leptoquark models. 
In Ref.~\cite{Dumont:2016xpj}, several leptoquark models are considered that generate scalar, vector, and tensor operators.
The $SU(3)\times SU(2)\times U(1)$ quantum numbers of these models are summarized in Table~\ref{LQ_numbers}.
We can group the leptoquarks as vector or scalar leptoquarks. These leptoquarks can in turn be $SU(2)$ singlets, doublets, or triplets.

\begin{table}[t]
 \begin{center}\begin{tabular}{|c|c|c|c|c|c|}
 \hline
    		& \,\,spin\,\, 	& \,\,$SU(3)_c$\,\, 	& \,\,$SU(2)_L$\,\, 	& \,\,$U(1)_{Y=Q-T_3}$\,\, \\
 \hline
 $S_1$ 	& $0$	 		& $3^*$ 		& $1$			& $1/3$        \\
 ${\bm S}_3$ 	& $0$			& $3^*$		& $3$			& $1/3$        \\     
 $R_2$ 	& $0$			& $3$		& $2$			& $7/6$        \\
 $V_2$ 	& $1$			& $3^*$		& $2$		 	& $5/6$        \\
 $U_1$ 	& $1$				& $3$		& $1$			& $2/3$        \\
 ${\bm U}_3$ 	& $1$					& $3$		& $3$			& $2/3$        \\
 \hline
 \end{tabular}
 \caption{Quantum numbers of scalar and vector leptoquarks.}
 \label{LQ_numbers}
 \end{center}
\end{table}

The Lagrangians for the various leptoquarks are
\begin{align}
      & \mathcal L^{\rm LQ} = \mathcal L_{V}^{\rm LQ} + \mathcal L_{S}^{\rm LQ} \,, \label{EQ:LagLQ1} \\[1em]
      & \mathcal L_{V}^{\rm LQ}
      = \left( {h_{1L}^{ij}}\,\bar Q_L^i \gamma_\mu L_{L}^j + {h_{1R}^{ij}} \,\bar d_R^i \gamma_\mu \ell_{R}^j \right)U_{1}^\mu + {h_{3L}^{ij}} \,\bar Q_{L}^i {\bm\sigma}\gamma_\mu L_{L}^j {\bm U}_3^\mu \notag \\
      &\hspace{3em} +        \left( {g_{2L}^{ij}} \,\bar d_{R}^{c,i} \gamma_\mu L_{L}^j + {g_{2R}^{ij}} \,\bar Q_{L}^{c,i} \gamma_\mu \ell_{R}^j \right)V_{2}^\mu +\text{h.c.} \\
           & \mathcal L_{S}^{\rm LQ}
      = \left( {g_{1L}^{ij}} \,\bar Q_{L}^{c,j} i\sigma_2 L_{L}^j + {g_{1R}^{ij}}\,\bar u_{R}^{c,i} \ell_{R}^j \right)S_1 + {g_{3L}^{ij}} \,\bar Q_{L}^{c,i} i\sigma_2{\bm\sigma} L_{L}^j {\bm S}_3 \notag \\
      &\hspace{3em}+  
      \left( { h_{2L}^{ij}} \,\bar u_{R}^i L_{L}^j + {h_{2R}^{ij}} \,\bar Q_{L}^i i\sigma_2 \ell_{R}^j \right)R_2 +\text{h.c.}, \
      \label{EQ:LagLQ2}
\end{align}
where $h^{ij}$ and $g^{ij}$ are dimensionless couplings,
$S_1$, ${\bm S}_3$, and $R_2$ are the scalar leptoquark bosons,  
$U_{1}^\mu$, ${\bm U}_3^\mu$, and $V_2^\mu$ are the vector leptoquark bosons, and
the index $i$ ($j$) indicates the generation of quarks (leptons).

The leptoquark Lagrangian generates the following couplings in Eq.~(\ref{eq1:Lag}):
\bea
g_S(\mu_b) &=& \frac{\sqrt{2}}{4G_F V_{cb}}\left(C_{\mathcal S_{1}}(\mu_b)+C_{\mathcal S_{2}}(\mu_b) \right),\\  
g_P(\mu_b) &=& \frac{\sqrt{2}}{4G_F V_{cb}}\left(C_{\mathcal S_{1}}(\mu_b)-C_{\mathcal S_{2}}(\mu_b) \right),\\
g_L &=& \frac{\sqrt{2}}{4G_F V_{cb}}C_{\mathcal V_1}^l,\\
g_R &=& \frac{\sqrt{2}}{4G_F V_{cb}}C_{\mathcal V_2}^l,\\
g_T(\mu_b) &=& \frac{\sqrt{2}}{4G_F V_{cb}}C_{\mathcal T}(\mu_b) ,
\eea
where the Wilson coefficients in the leptoquark models are given by 
\begin{align}
      & C_\text{SM} = 2 \sqrt 2 G_F V_{cb} \,, \label{EQ:CVsm} \\
      & C_{\mathcal V_1}^l =  \sum_{k=1}^3 V_{k3} 
      \left[ 
      {g_{1L}^{kl}g_{1L}^{23*} \over 2M_{S_1}^2} - {g_{3L}^{kl}g_{3L}^{23*} \over 2M_{{\bm S}_3}^2} + {h_{1L}^{2l}h_{1L}^{k3*} \over M_{U_1}^2} - {h_{3L}^{2l}h_{3L}^{k3*} \over M_{{\bm U}_3}^2}
      \right] \,,  \label{EQ:CV1} \\
      & C_{\mathcal V_2}^l = 0 \,, \\
      & C_{\mathcal S_1}^l = \sum_{k=1}^3 V_{k3} 
      \left[ 
      -{2g_{2L}^{kl}g_{2R}^{23*} \over M_{V_2}^2} - {2h_{1L}^{2l}h_{1R}^{k3*} \over M_{U_1}^2} 
      \right] \,, \\
      & C_{\mathcal S_2}^l = \sum_{k=1}^3 V_{k3} 
      \left[ 
      -{g_{1L}^{kl}g_{1R}^{23*} \over 2M_{S_1}^2} - {h_{2L}^{2l}h_{2R}^{k3*} \over 2M_{R_2}^2} 
      \right] \,, \label{EQ:CS2} \\
      & C_{\mathcal T}^l = \sum_{k=1}^3 V_{k3} 
      \left[ 
      {g_{1L}^{kl}g_{1R}^{23*} \over 8M_{S_1}^2} - {h_{2L}^{2l}h_{2R}^{k3*} \over 8M_{R_2}^2} 
      \right] \,. \label{EQ:CT}
\end{align}
These Wilson coefficients are defined at the energy scale $\mu = M_X$, where $X$ represents a leptoquark. Above, $V_{k3}$ denotes the relevant CKM matrix element,
where the $3$ corresponds to the bottom quark. In the following, we neglect the CKM-suppressed contributions from $k=1$ and $k=2$ in the sums.
 Because the neutrino is not observed, we have $l=1,2,3$. Note that there is a Standard-Model contribution for $l=3$ but not for $l=1,2$; hence, the constraints
for different $l$ will be different.

The renormalization-group running of the scalar and tensor Wilson coefficients from $\mu=M_X$ to $\mu=\mu_b$, 
where $\mu_{b}$ is the mass scale of the bottom quark, is given by
\begin{align}
  C_{\mathcal S_{1,2}}(\mu_b) &= \left[ \alpha_s(m_t) \over \alpha_s(\mu_b) \right]^{-\frac{12}{23}} \left[ \alpha_s(m_{\rm LQ}) \over \alpha_s(m_t) \right]^{-\frac{4}{7}}\, C_{\mathcal S_{1,2}} (m_{\rm LQ}) \,, \\ 
  C_{\mathcal T}(\mu_b) &=\left[ \alpha_s(m_t) \over \alpha_s(\mu_b) \right]^{\frac{4}{23}} \left[ \alpha_s(m_{\rm LQ}) \over \alpha_s(m_t) \right]^{\frac{4}{21}}\, C_{\mathcal T} (m_{\rm LQ}) \,, 
\end{align}
where $\alpha_s (\mu)$ is the QCD coupling at scale $\mu$. Because the anomalous dimensions of the vector and axial-vector currents are zero, the Wilson coefficients for $\mathcal V_{1,2}$ are scale-independent.

The different leptoquarks produce different effective operators as summarized below:
\begin{itemize}
 \item The $S_1$ leptoquark with nonzero $(g_{1L},g_{1R}^*)$  generates  $C_{\mathcal V_1}^l$, $C_{\mathcal S_2}^l$, and $C_{\mathcal T}^l$, with the relation $C_{\mathcal S_2}^l =-4C_{\mathcal T}^l$. 
  \item The $R_2$ leptoquark with $(h_{2L},h_{2R}^*)$  generates $C_{\mathcal S_2}^l$ and  $C_{\mathcal T}^l$ with the relation $C_{\mathcal S_2}^l =4C_{\mathcal T}^l$.
  \item The $V_2$ leptoquark generates  $C_{\mathcal S_1}^l$  and is tightly constrained, so we do not consider this model.
  \item The $U_1$ leptoquark with nonzero $(g_{2L},g_{2R}^*)$ generates $C_{\mathcal S_1}^l$ and  $C_{\mathcal V_1}^l$.
 \item The ${\bm S}_3$ and ${\bm U}_3$ leptoquarks with nonzero values of $(g_{3L},g_{3L}^*)$ and  $(h_{3L},h_{3L}^*)$  generate $C_{\mathcal V_1}^l$.
\end{itemize}
The leptoquark couplings can also be constrained using $b\to s\nu\bar\nu$ decays. As pointed out in Ref.~\cite{Bhattacharya:2016mcc},
the exclusive decays $\bar B \to K \nu\bar\nu$ and $\bar B \to K^* \nu\bar\nu$ provide more stringent bounds than the inclusive
mode $B \to X_s \nu\bar\nu$.
The $U_1$ and $R_2$ leptoquarks do not contribute to $b\to s\nu\bar\nu$, while the left-handed couplings of $S_1$, ${\bm S}_3$, and ${\bm U}_3$ do.
(The $V_2$ leptoquark also contributes to $b\to s\nu\bar\nu$, but we do not consider this model.)
The BaBar
and Belle Collaborations give the following 90\% C.L. upper limits
\cite{Lees:2013kla,Lutz:2013ftz}:

\bea
\mathcal{B}(B^+ \to K^+ \nu\bar\nu) & \le & 1.7 \times 10^{-5} ~, \nn\\
\mathcal{B}( B^+ \to K^{*+} \nu\bar\nu) & \le & 4.0 \times 10^{-5} ~, \nn\\
\mathcal{B}( B^0 \to K^{*0} \nu\bar\nu) & \le & 5.5 \times 10^{-5} ~.
\eea

In Ref.~\cite{Buras:2014fpa}, these are compared with the SM
predictions

\bea
\mathcal{B}_K^{\rm SM} \equiv \mathcal{B}(B \to K \nu\bar\nu)_{\rm SM} = (3.98 \pm 0.43 \pm 0.19) \times 10^{-6} ~, \nn\\
\mathcal{B}_{K^*}^{\rm SM} \equiv \mathcal{B}(B \to K^{*} \nu\bar\nu)_{\rm SM} = (9.19 \pm 0.86 \pm 0.50) \times 10^{-6} ~.
\eea

Taking into account the theoretical uncertainties
\cite{Buras:2014fpa}, the 90\% C.L. upper bounds on the NP
contributions are

\beq
\frac{\mathcal B_K^{{\rm SM} + {\rm NP}}}{\mathcal B_K^{\rm SM}} \le 4.8 ~,~~~~
\frac{\mathcal B_{K^*}^{{\rm SM} + {\rm NP}}}{\mathcal B_{K^*}^{\rm SM}} \le 4.9 ~.
\label{pk}
\eeq

Following Ref.~\cite{Sakaki:2013bfa}, the $b\to s \nu_j \bar{\nu}_i$ process can be described by the effective Hamiltonian
\begin{equation}
   H_{eff} = {4G_F \over \sqrt{2}} V_{tb} V_{ts}^* \left[ \left(\delta_{ij}C_L^{(\rm SM)} + C_L^{ij}\right)O_L^{ij} + C_R^{ij}O_R^{ij} \right] \,,
\end{equation}
where the left-handed and right-handed operators are defined as
\begin{equation}
   \begin{split}
      O_L^{ij} =& (\bar{s}_L \gamma^\mu b_L)(\bar{\nu}_{jL} \gamma_\mu \nu_{iL}) \,, \\
      O_R^{ij} =& (\bar{s}_R \gamma^\mu b_R)(\bar{\nu}_{jL} \gamma_\mu \nu_{iL}) \,.
   \end{split}
\end{equation}
The SM Wilson coefficient $C_L^{(\rm SM)}$ receives contributions from box and $Z$-penguin diagrams, which yield
\begin{equation}
   C_L^{(\rm SM)} = {\alpha \over 2\pi\sin^2\theta_W}X(m_t^2/M_W^2) \,,
\end{equation}
where the loop function $X(x_t)$ can be found e.g.~in Ref.~\cite{Buras:1998raa}.
The leptoquarks that we consider produce contributions to $C_L^{ij}$ which, to leading order, are equal to \cite{Sakaki:2013bfa}
\begin{subequations}
   \label{eq:LQ_coeff_Xsnunu}
   \begin{align}
      C_L^{ij} =& -{1 \over 2\sqrt2 G_F V_{tb} V_{ts}^* } 
      \left[ {g_{1L}^{3i}g_{1L}^{2j*} \over 2M_{S_1^{1/3}}^2} + {g_{3L}^{3i}g_{3L}^{2j*} \over 2M_{S_3^{1/3}}^2} - {2h_{3L}^{2i}h_{3L}^{3j*} \over M_{U_3^{-1/3}}^2} \right] \,.
   \end{align}
\end{subequations}

We obtain common coefficients for $b \to c \tau \bar{\nu}_l$ and $b \to s \nu_\tau \bar{\nu}_l$ processes,

\begin{subequations}
   \begin{align}
      C_L^{l3} =& -{1 \over 2\sqrt2 G_F V_{tb} V_{ts}^* } 
      \left[ {g_{1L}^{3l}g_{1L}^{23*} \over 2M_{S_1^{1/3}}^2} + {g_{3L}^{3l}g_{3L}^{23*} \over 2M_{S_3^{1/3}}^2} - {2h_{3L}^{2l}h_{3L}^{33*} \over M_{U_3^{-1/3}}^2} \right] \,.
   \end{align}
\end{subequations}

Hence, for $l=3$ we obtain
\bea
\frac{\mathcal B_K^{{\rm SM} + {\rm NP}}}{\mathcal B_K^{\rm SM}} = \frac{\mathcal B_{K^*}^{{\rm SM} + {\rm NP}}}{\mathcal B_{K^*}^{\rm SM}}  & = & \left | \frac{ 3C_L^{(\rm SM)}  + C_L^{33}}
{3C_L^{(\rm SM)}} \right | ^2 ,\
\eea

while for $l=1,2$ we have

\bea
\frac{\mathcal B_K^{{\rm SM} + {\rm NP}}}{\mathcal B_K^{\rm SM}} = \frac{\mathcal B_{K^*}^{{\rm SM} + {\rm NP}}}{\mathcal B_{K^*}^{\rm SM}}  & = & \left | \frac{  C_L^{l3}}
{3C_L^{(\rm SM)}} \right |^2 .\
\eea

\begin{figure}
\begin{center}
\includegraphics[height=0.8cm]{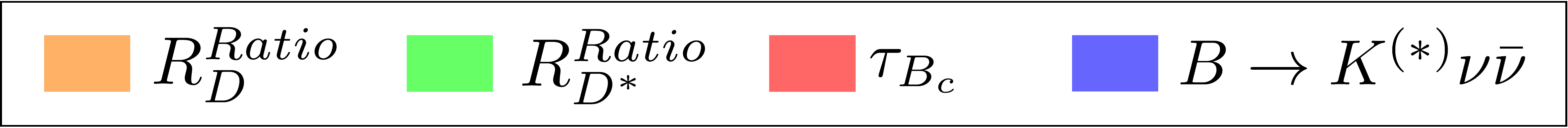}

\vspace{0.25cm}

\includegraphics[width=5.5cm]{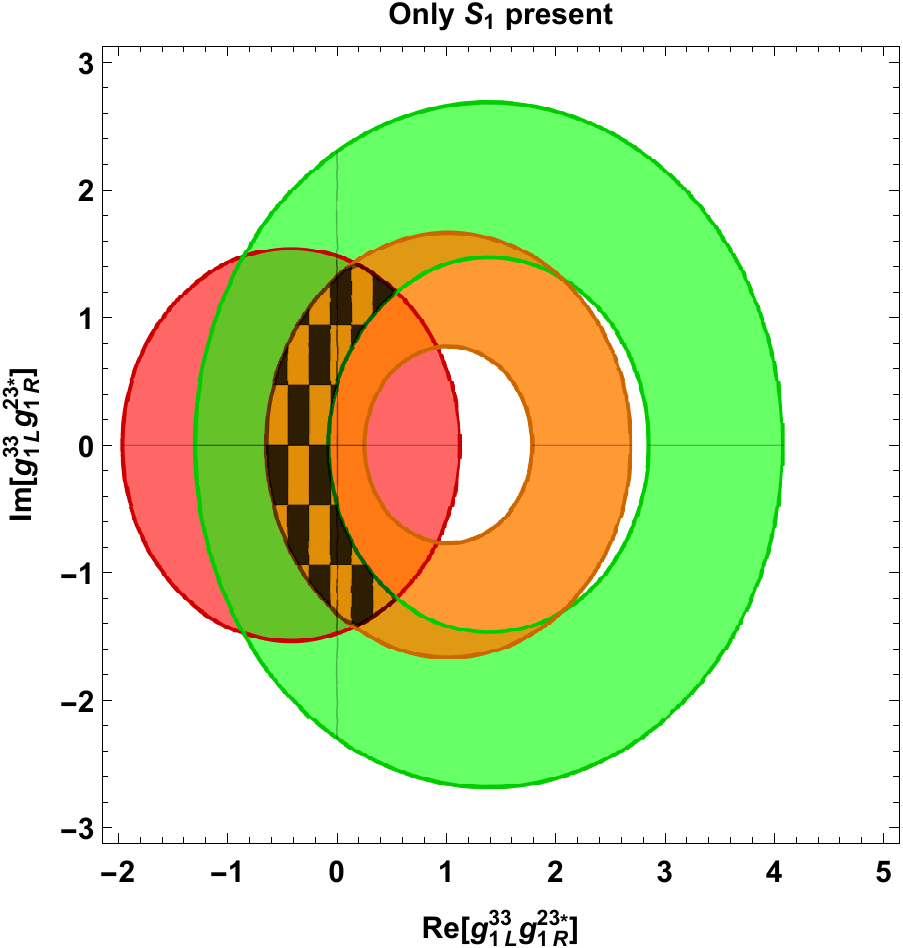} \hspace{4ex}
\includegraphics[width=5.5cm]{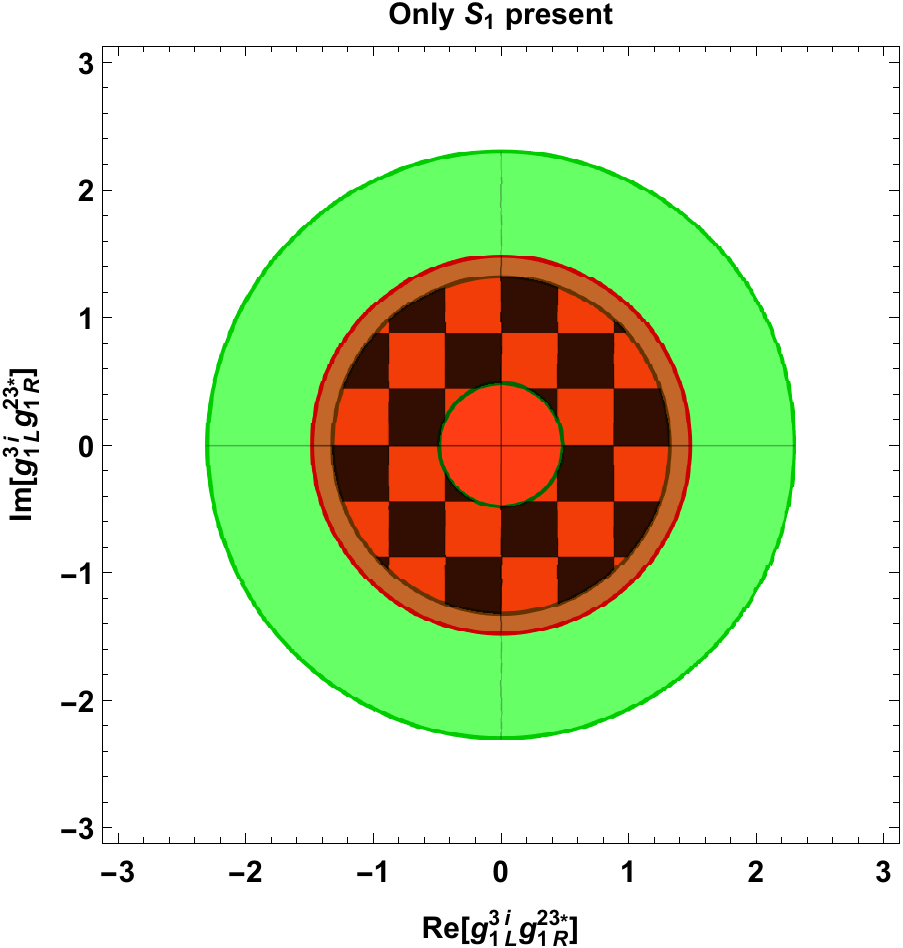}\\

\vspace{3ex}

\includegraphics[width=5.8cm]{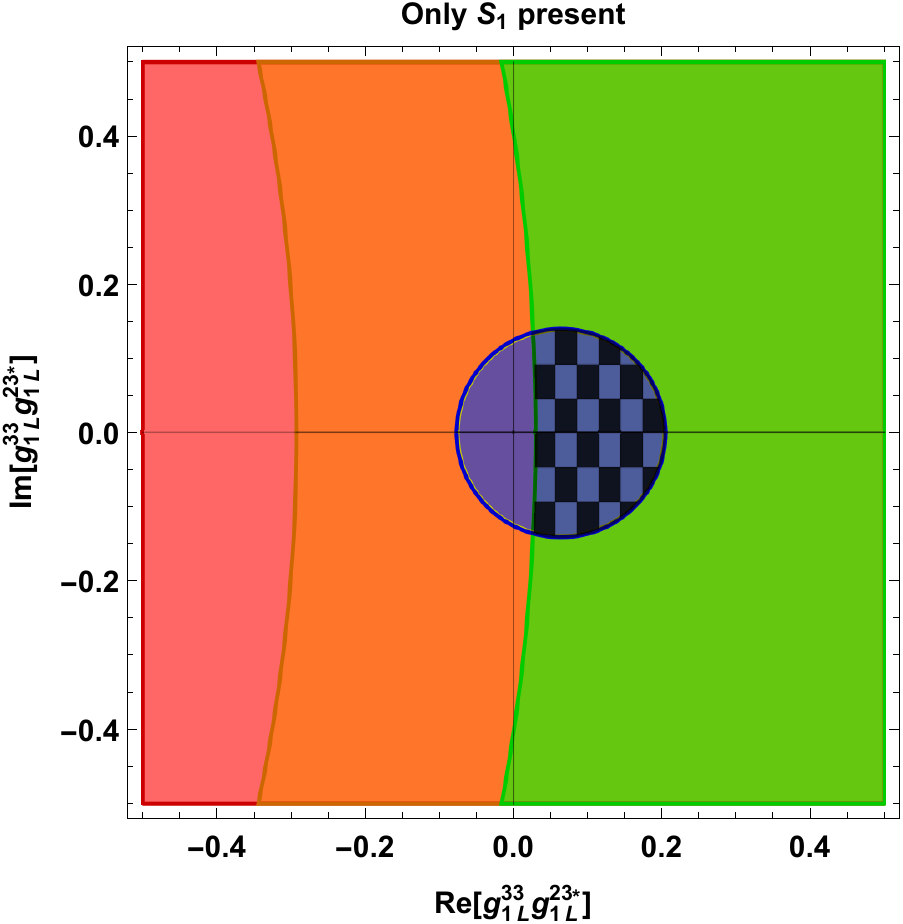}  \hspace{4ex}
\includegraphics[width=5.8cm]{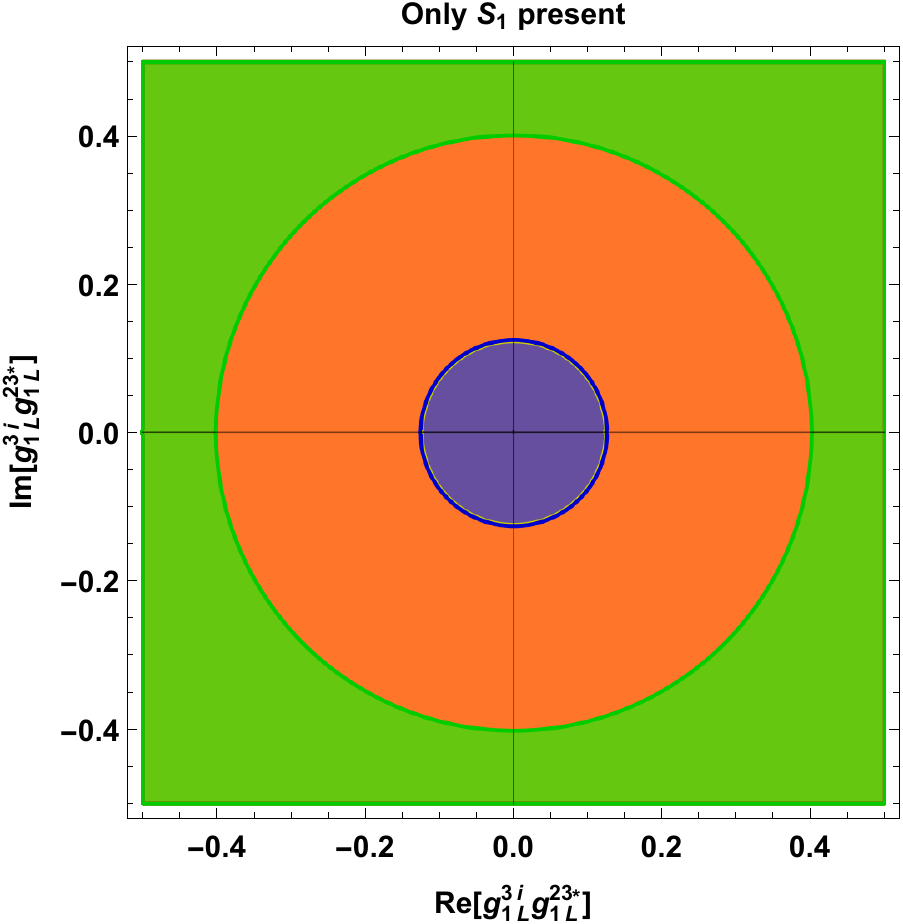}\\

\vspace{3ex}

\includegraphics[width=5.5cm]{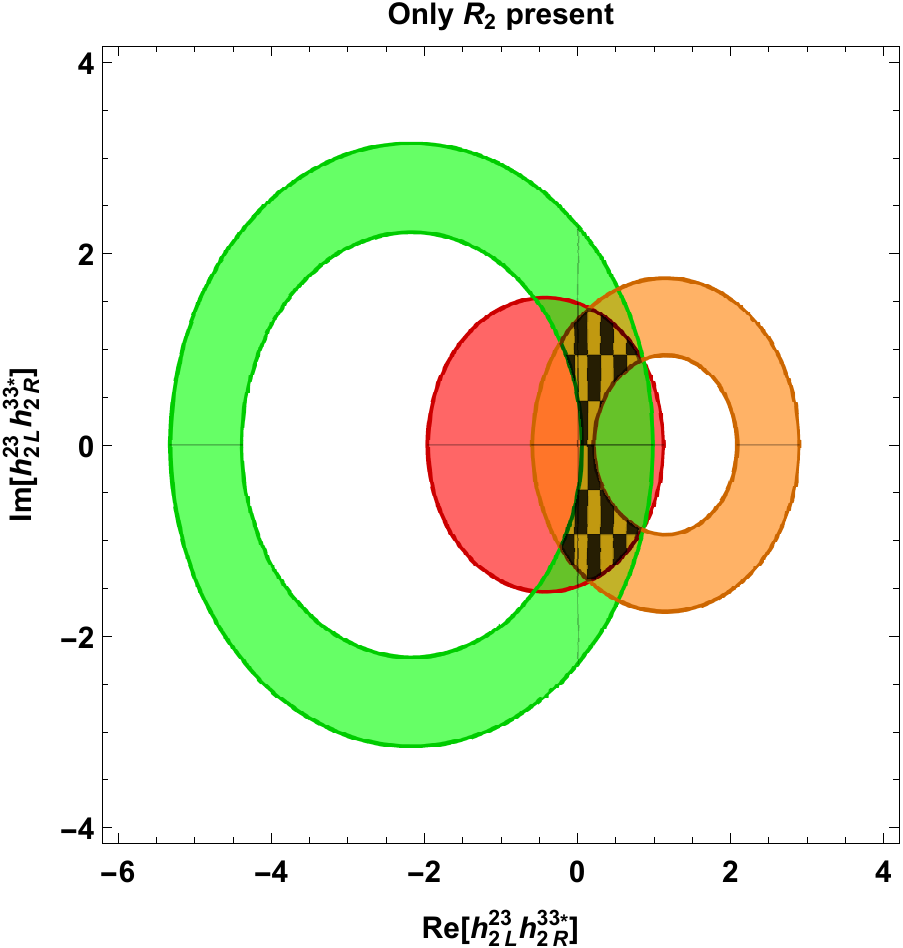}  \hspace{4ex}
\includegraphics[width=5.5cm]{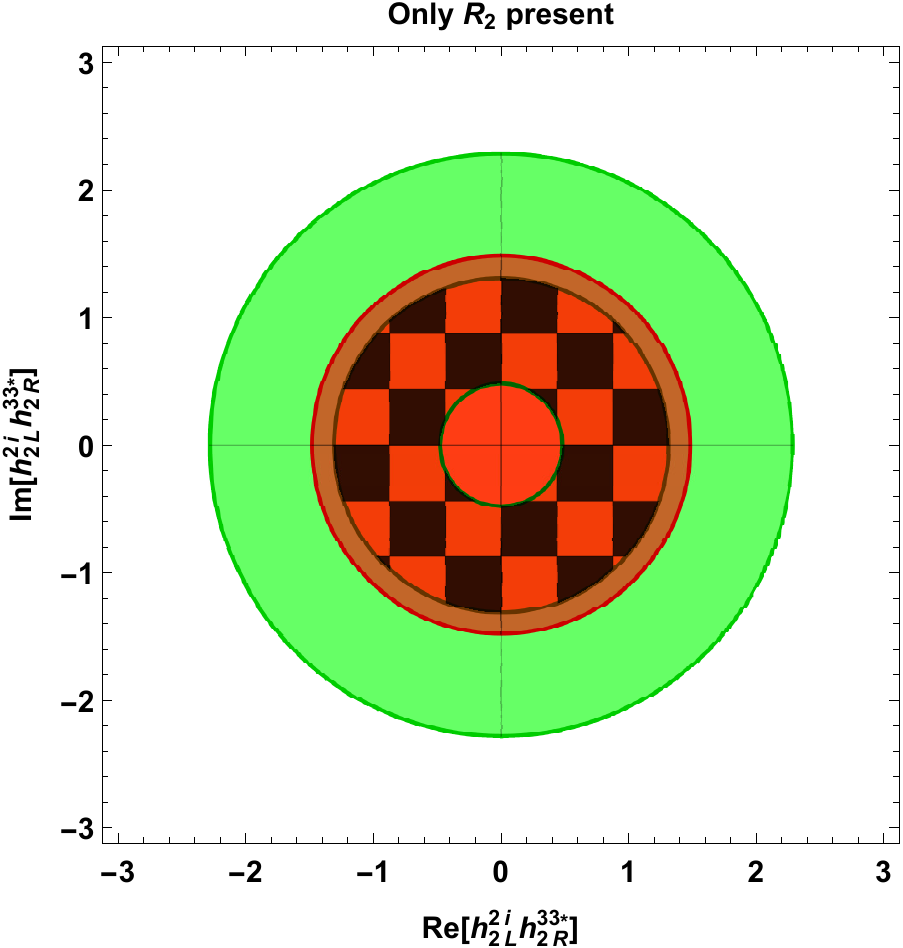}\\

\end{center}
\caption{Constraints on the $S_1$ and $R_2$ leptoquark models when considering one coupling at a time. Here, $i=1,2$ denotes the electron and muon neutrinos.
We require that the couplings reproduce the measurements of $\RDr$ and $\RDrstar$ in Eqs.~(\ref{eq:RDrexp}) and (\ref{eq:RDstrexp}) within 3$\sigma$,
satisfy $\mathcal{B}(B_c \to \tau^- \bar{\nu}_\tau) \le 30\%$, and are consistent with the upper bounds on $\mathcal{B}(B \to K^{(*)} \nu\bar\nu)$ at 90\% C.L. The allowed regions of the parameter space when combining all constraints are highlighted with a black mesh.}
\label{LQ-S1-R2-Ind}
\end{figure}

\begin{figure}
\begin{center}
\includegraphics[height=0.8cm]{legend_constraints-Ind.pdf}

\vspace{0.25cm}

\includegraphics[width=6.5cm]{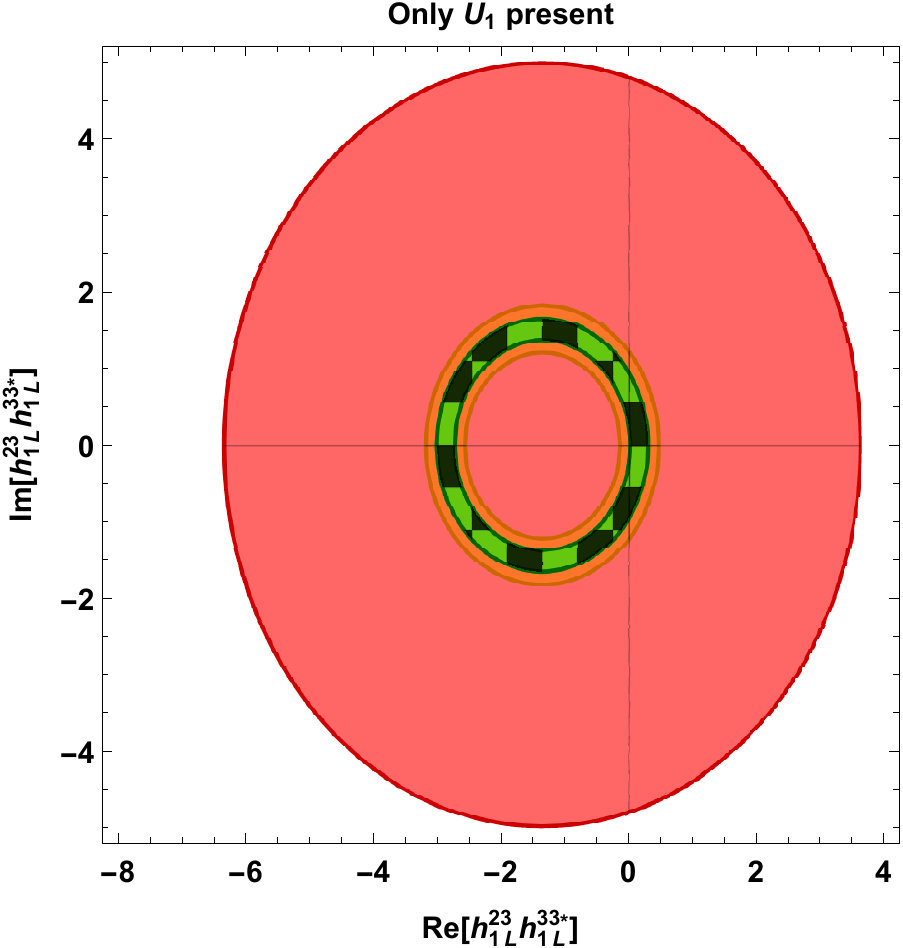} \hspace{4ex}
\includegraphics[width=6.5cm]{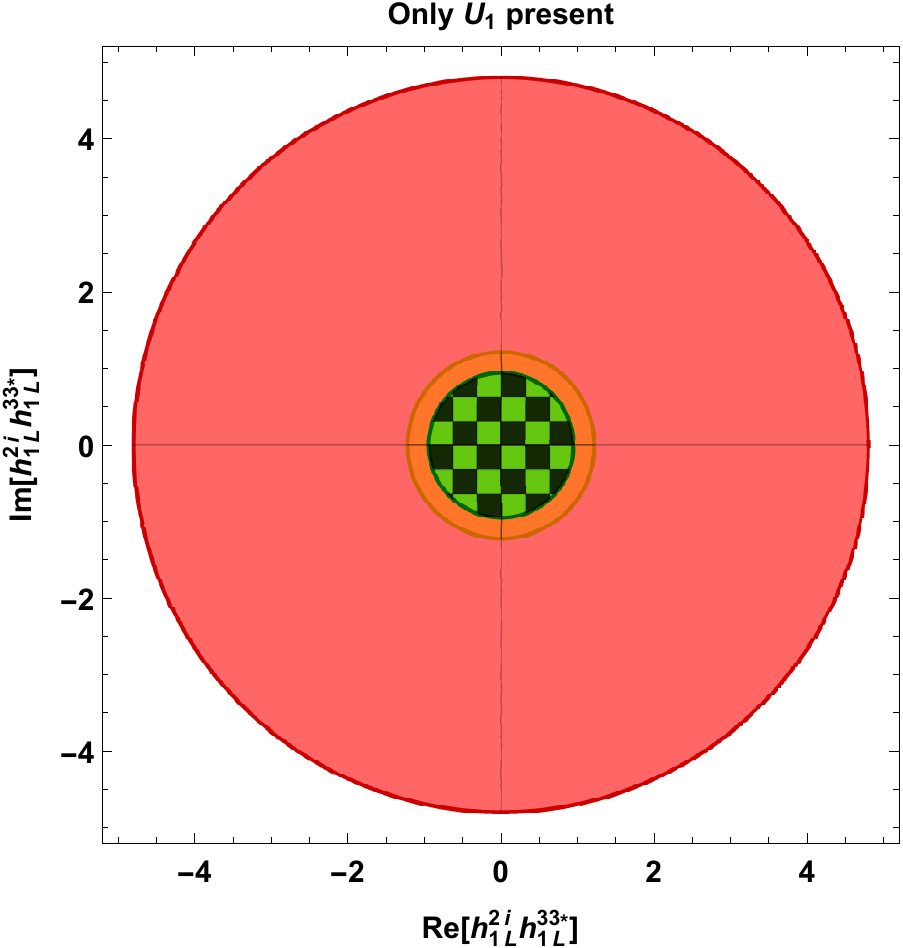}\\

\vspace{3ex}

\includegraphics[width=7cm]{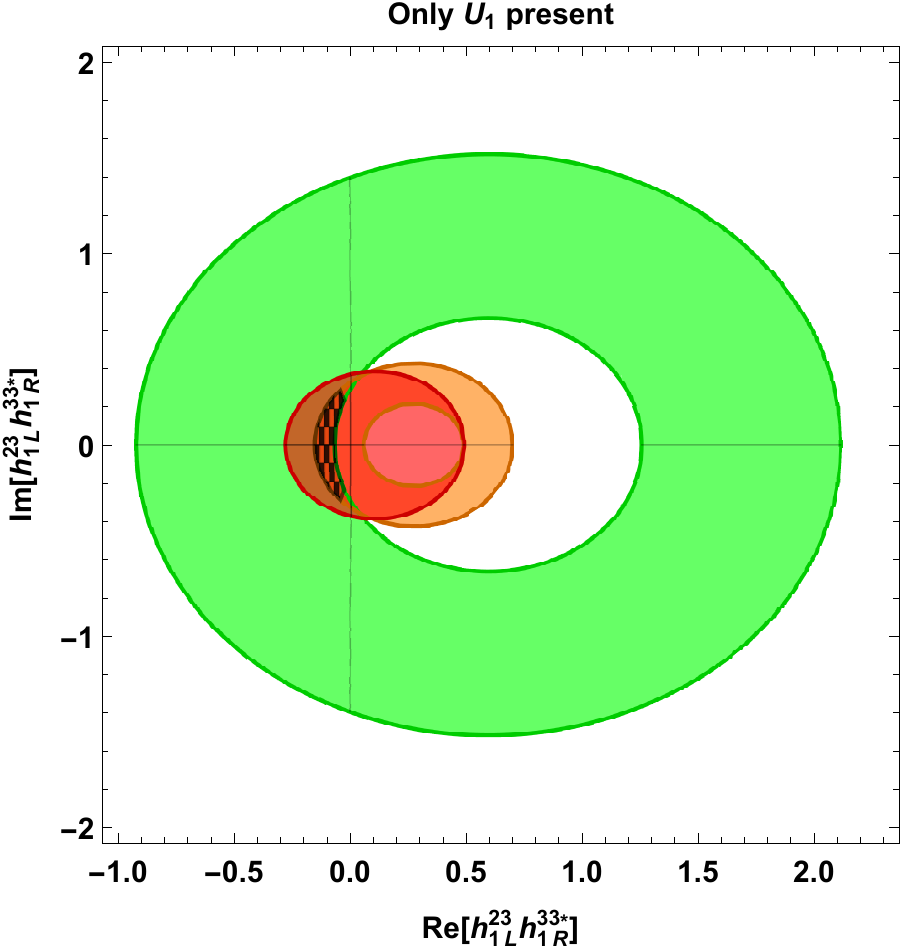}  \hspace{4ex}
\includegraphics[width=7cm]{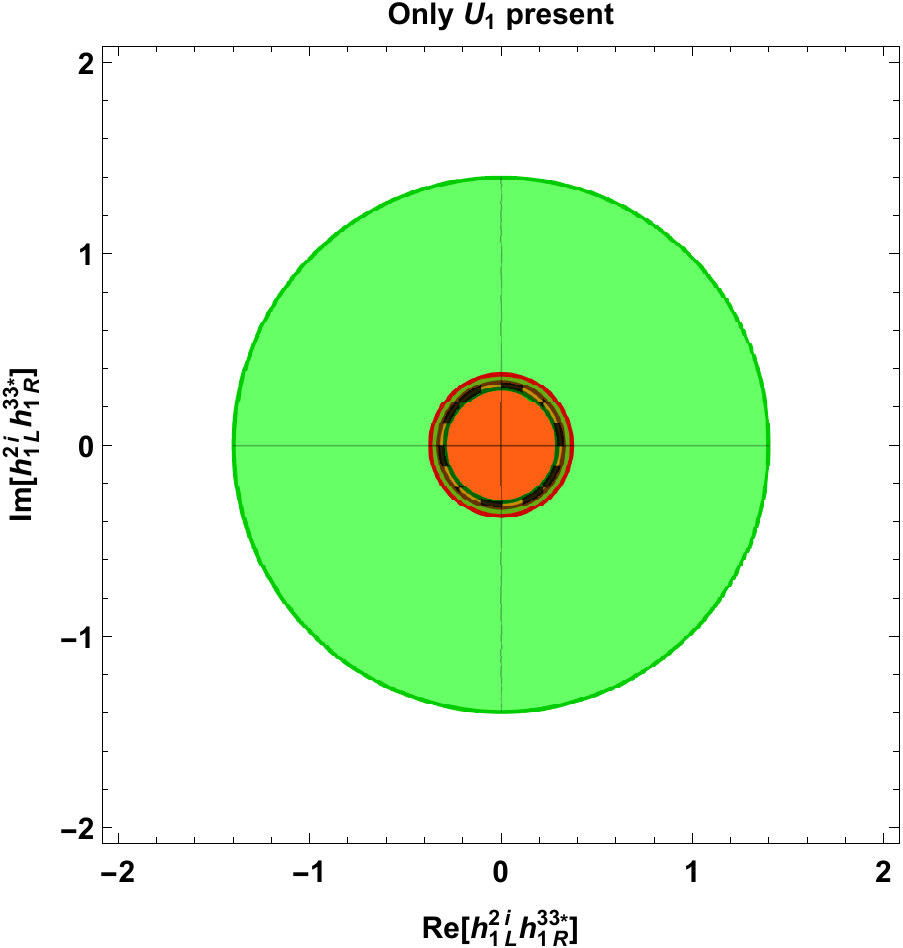}\\

\end{center}
\caption{Constraints on the $U_1$ leptoquark model when considering one coupling at a time. Here, $i=1,2$ denotes the electron and muon neutrinos.
We require that the couplings reproduce the measurements of $\RDr$ and $\RDrstar$ in Eqs.~(\ref{eq:RDrexp}) and (\ref{eq:RDstrexp}) within 3$\sigma$
and satisfy $\mathcal{B}(B_c \to \tau^- \bar{\nu}_\tau) \le 30\%$. The allowed regions of the parameter space when combining all constraints are highlighted with a black mesh.}
\label{LQ-U1-Ind}
\end{figure}

\begin{figure}
\begin{center}
\includegraphics[height=0.8cm]{legend_constraints-Ind_with_BKnunu.pdf}

\vspace{0.25cm}

\includegraphics[width=7cm]{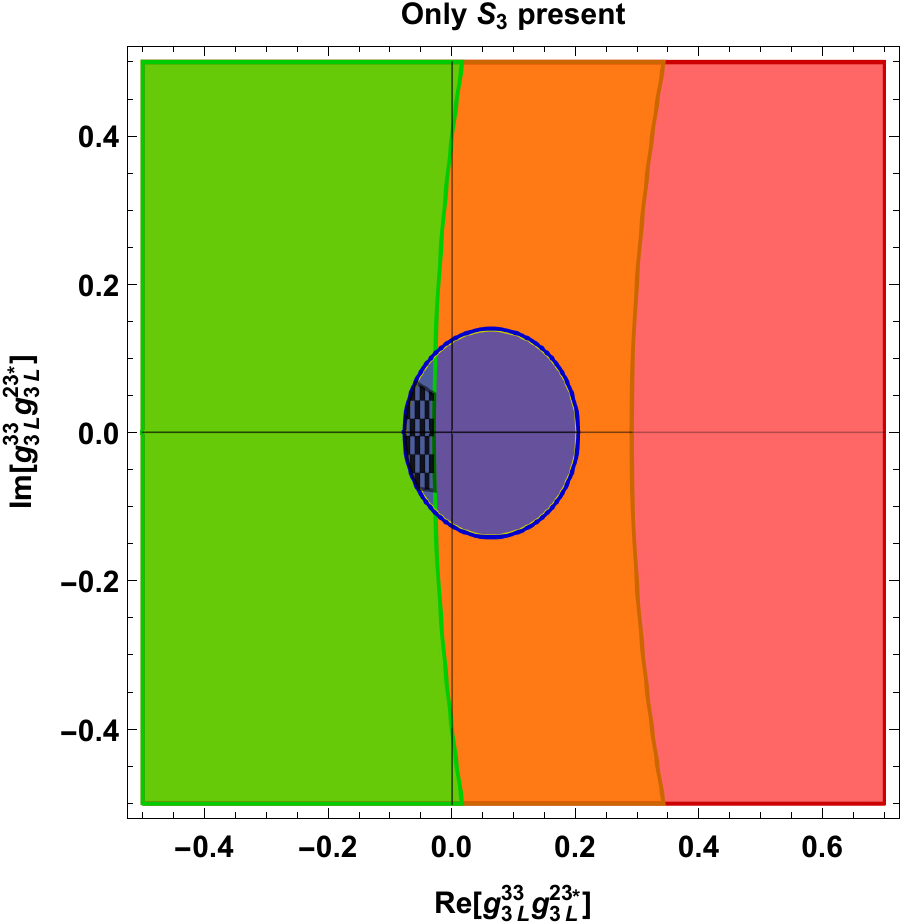}  \hspace{4ex}
\includegraphics[width=7cm]{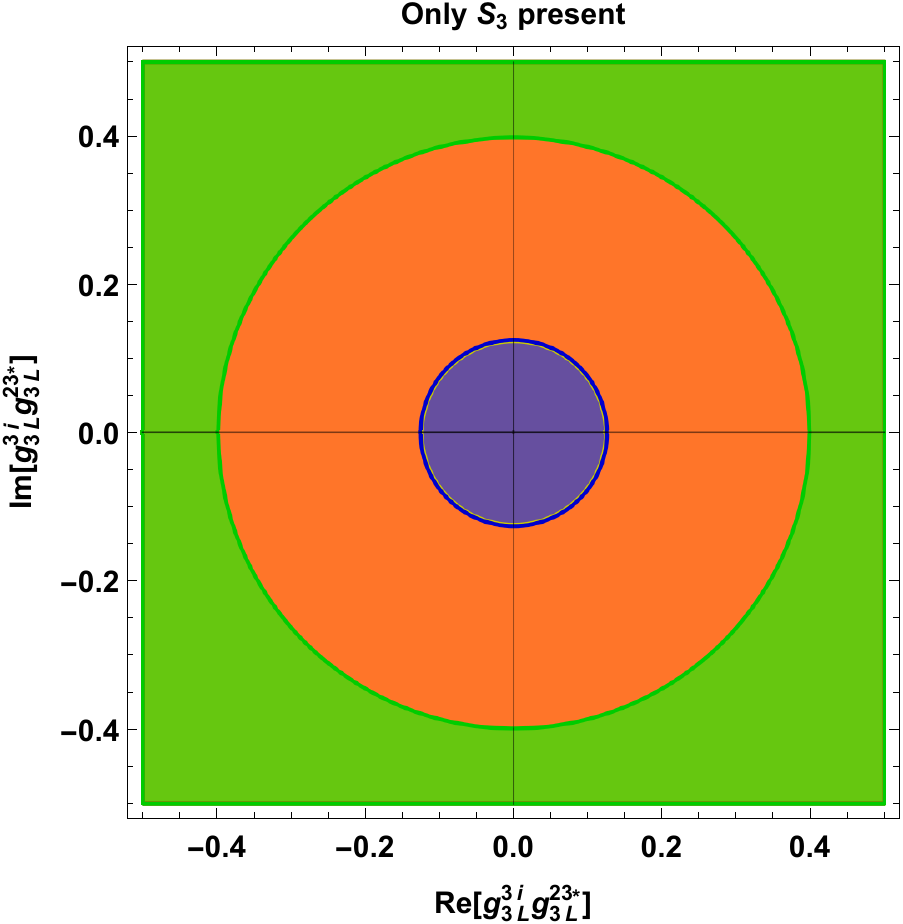}\\

\vspace{3ex}

\includegraphics[width=7cm]{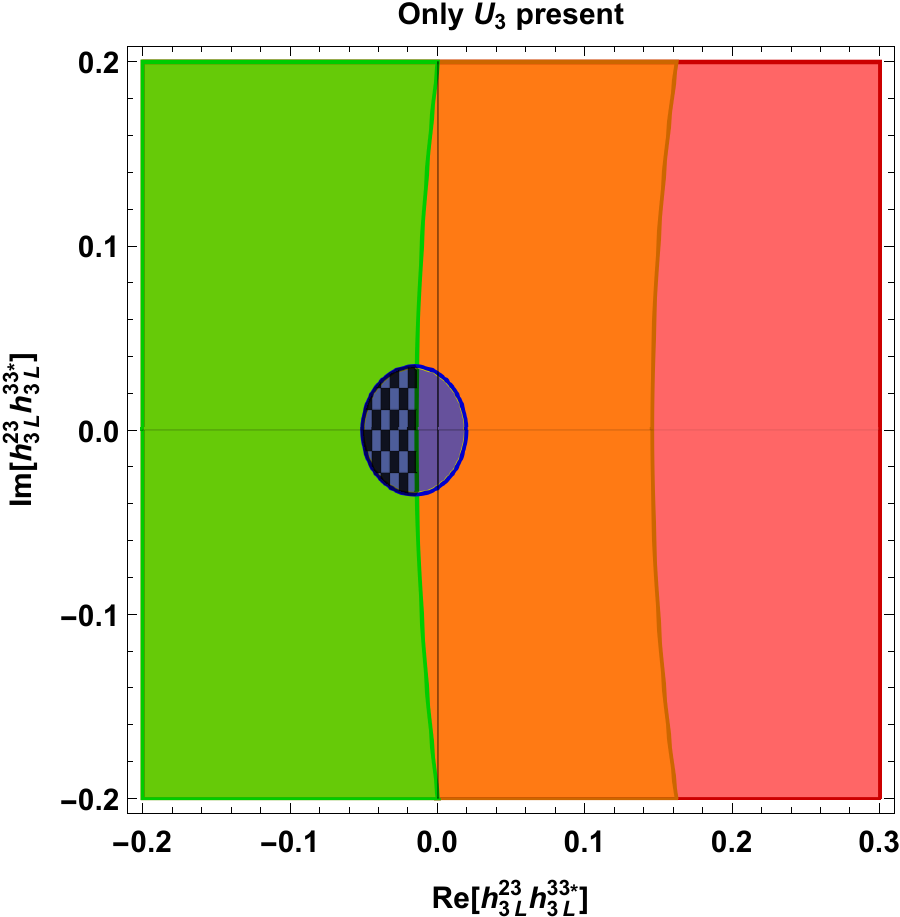} \hspace{4ex}
\includegraphics[width=7cm]{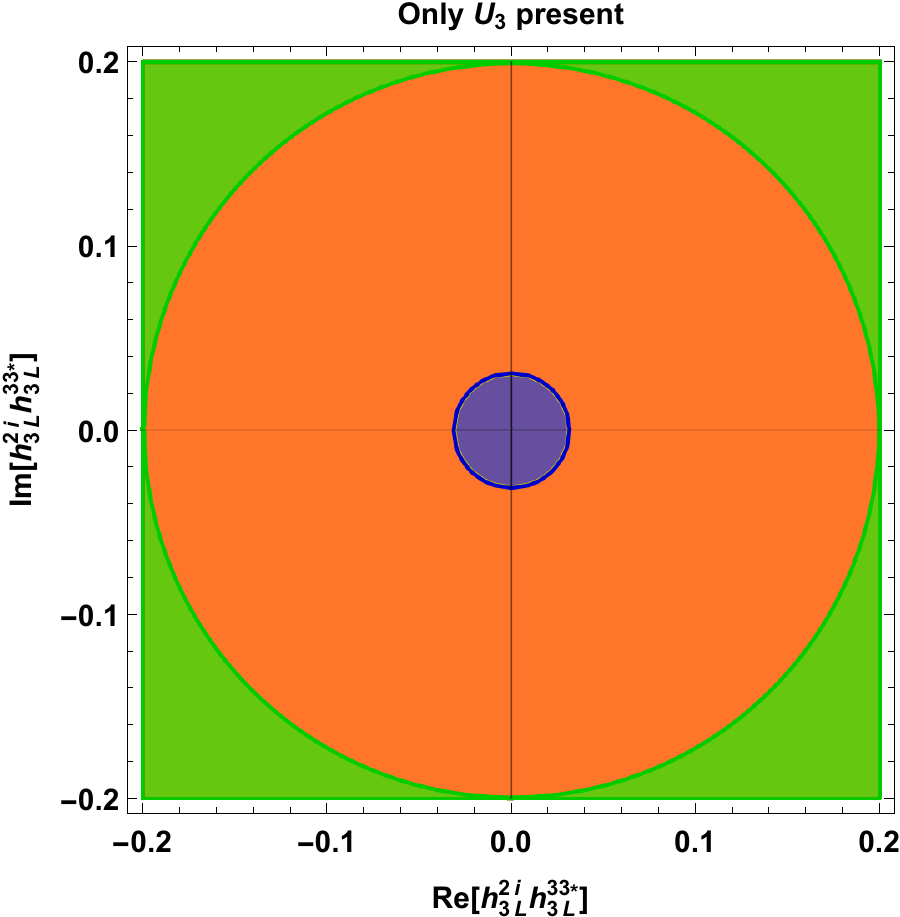}\\

\end{center}
\caption{Constraints on the ${\bm S}_3$ and ${\bm U}_3$ leptoquark models when considering one coupling at a time. Here, $i=1,2$ denotes the electron and muon neutrinos.
We require that the couplings reproduce the measurements of $\RDr$ and $\RDrstar$ in Eqs.~(\ref{eq:RDrexp}) and (\ref{eq:RDstrexp}) within 3$\sigma$,
satisfy $\mathcal{B}(B_c \to \tau^- \bar{\nu}_\tau) \le 30\%$, and are consistent with the upper bounds on $\mathcal{B}(B \to K^{(*)} \nu\bar\nu)$ at 90\% C.L. The allowed regions of the parameter space when combining all constraints are highlighted with a black mesh.}
\label{LQ-S3-U3-Ind}
\end{figure}

When considering nonzero values only for one coupling at a time ($l=1,2,3$), the experimental measurements
of $\RDr$, $\RDrstar$, $\tau_{B_c}$, and ${\cal B}(B \to K^{(*)} \nu\bar\nu)$ yield the constraints shown in
Figures \ref{LQ-S1-R2-Ind}, \ref{LQ-U1-Ind}, and \ref{LQ-S3-U3-Ind}. The cases with $g_{3L}^{3i}g_{3L}^{23*}$ in the ${\bm S}_3$ model,
$g_{1L}^{3i}g_{1L}^{23*}$ in the $S_1$ model, and $h_{3L}^{2i}h_{3L}^{23*}$ in the ${\bm U}_3$ model are ruled out for $i=1,2$.

\begin{figure}
\begin{adjustwidth}{-0.5cm}{-0.5cm}
\begin{center}
\includegraphics[width=5cm]{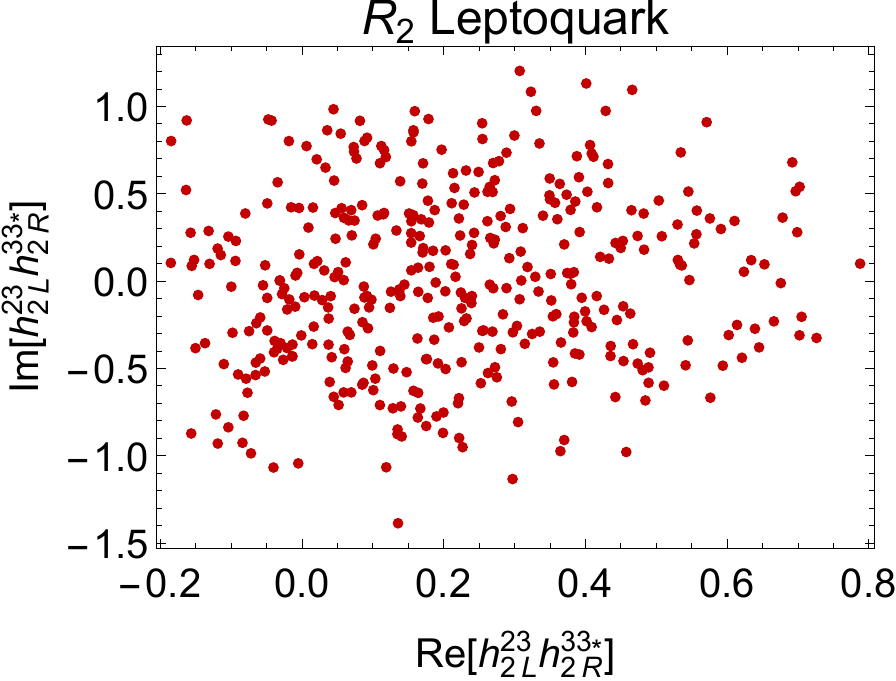}~~~
\includegraphics[width=5cm]{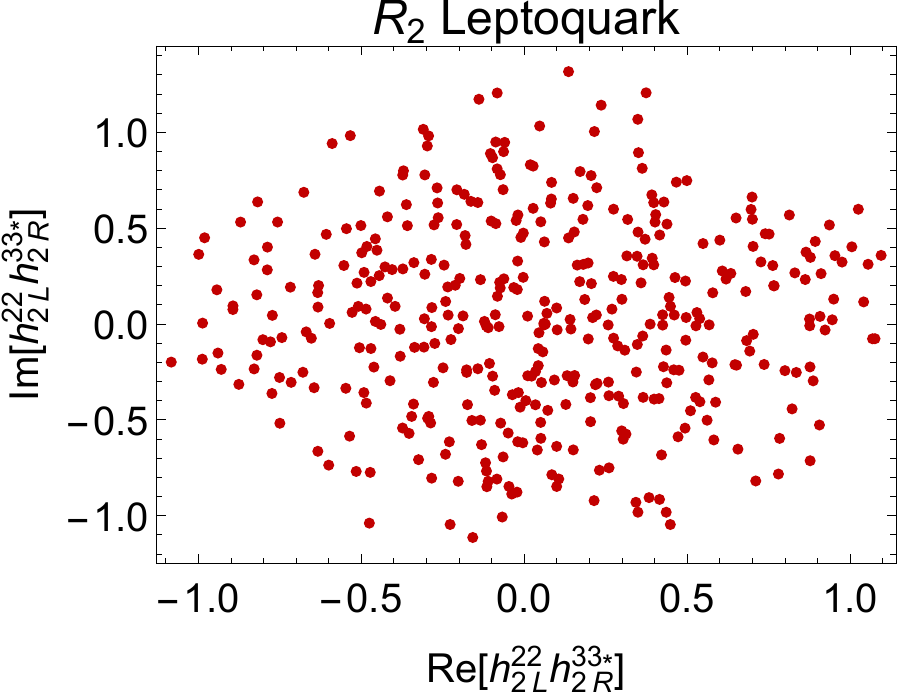}~~~
\includegraphics[width=5cm]{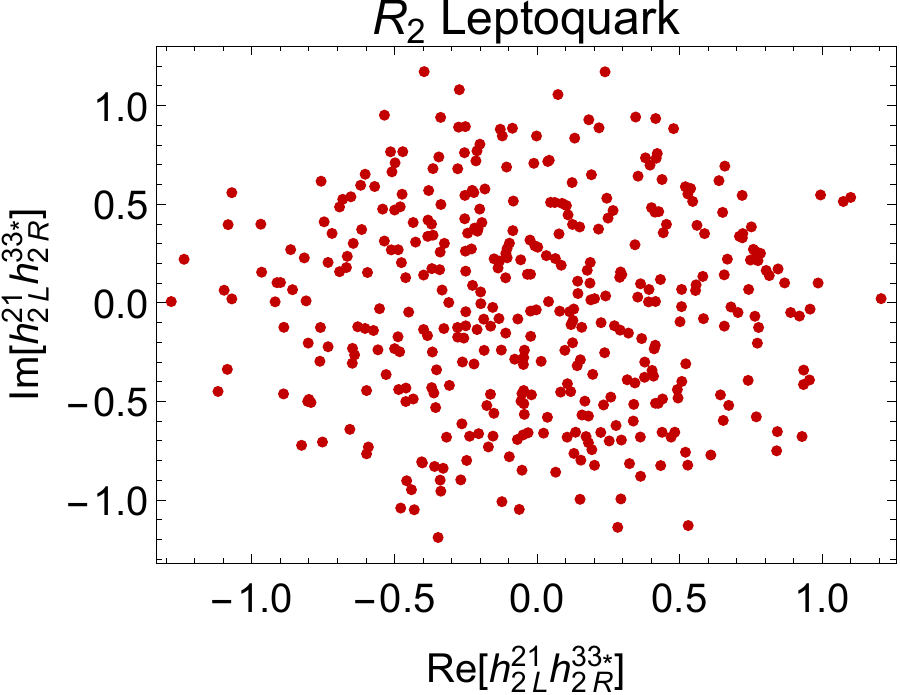}~~~\\

\vspace{2ex}

\includegraphics[width=5cm]{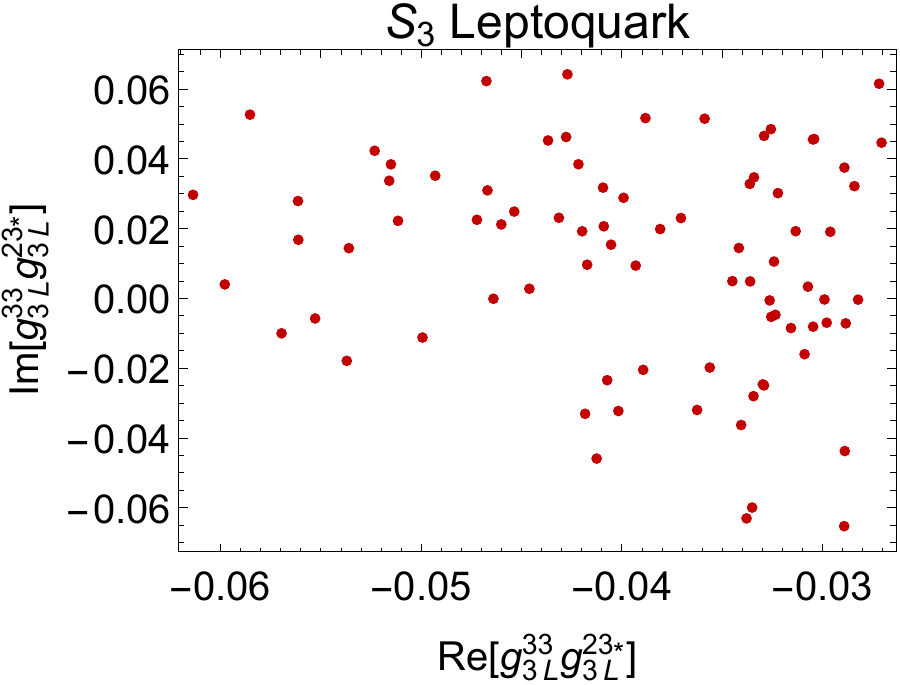}~~~
\includegraphics[width=5cm]{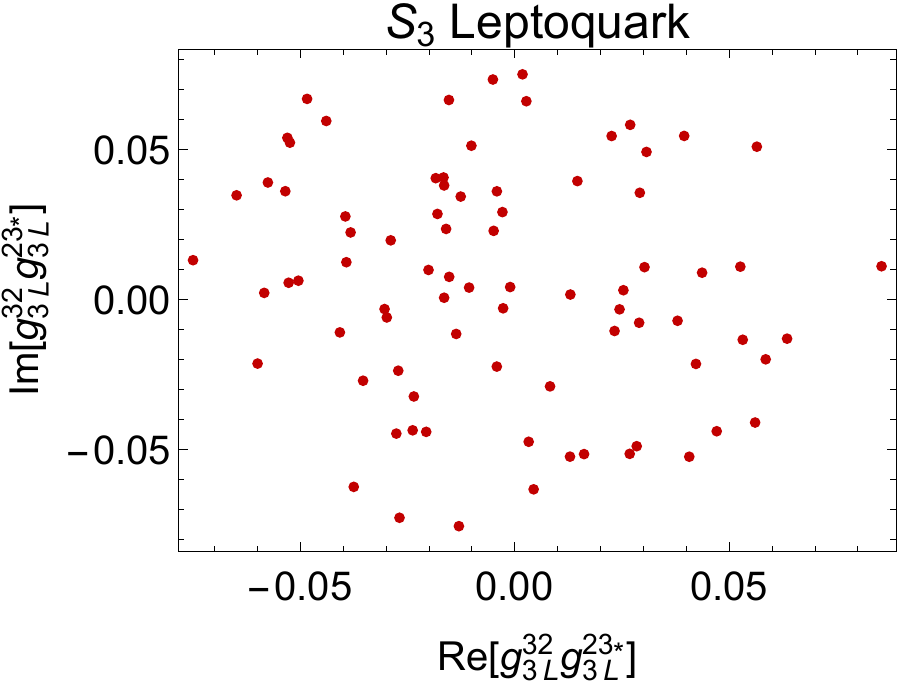}~~~
\includegraphics[width=5cm]{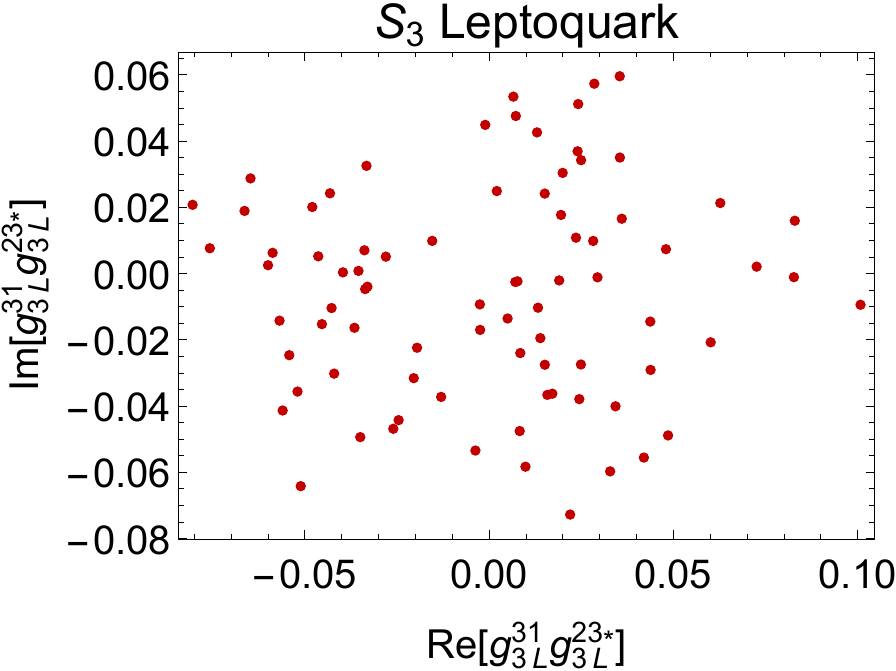}~~~\\

\vspace{2ex}

\includegraphics[width=5cm]{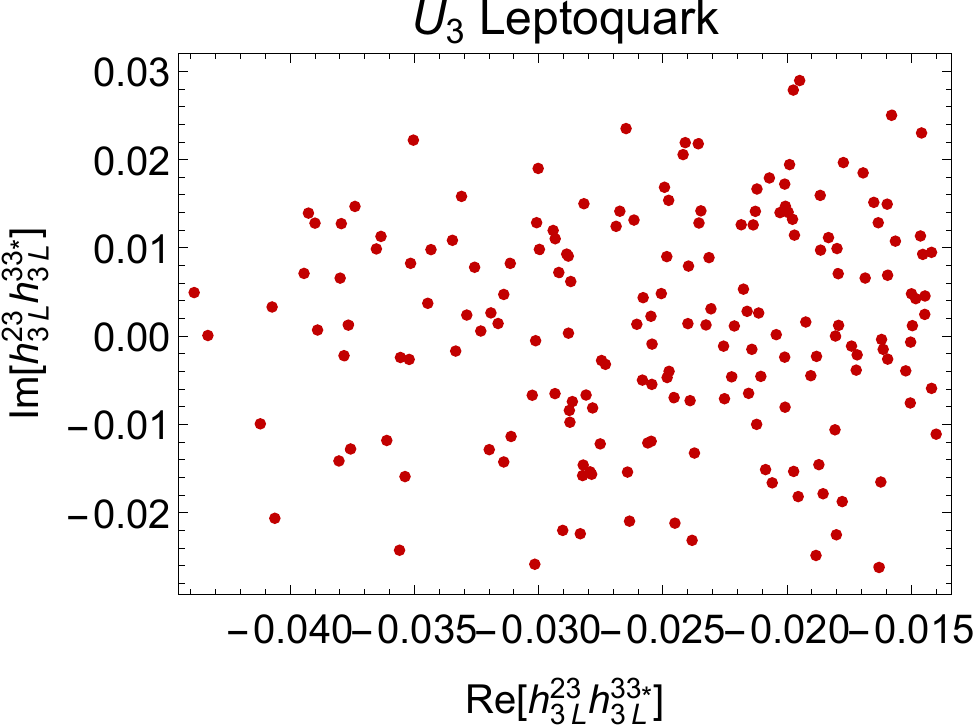}~~~
\includegraphics[width=5cm]{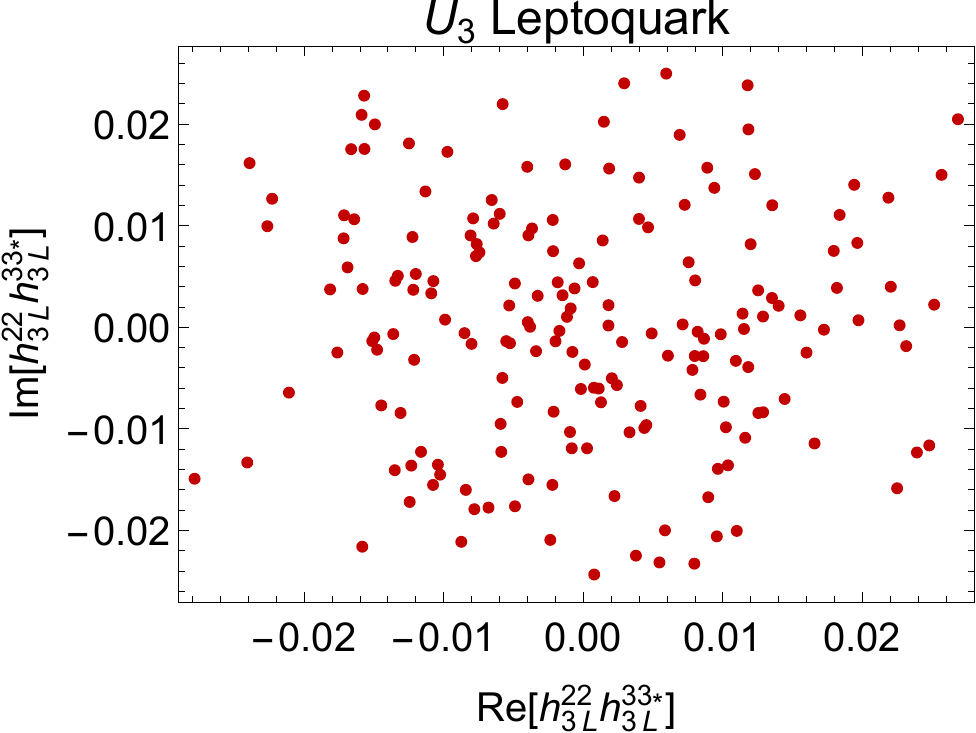}~~~
\includegraphics[width=5cm]{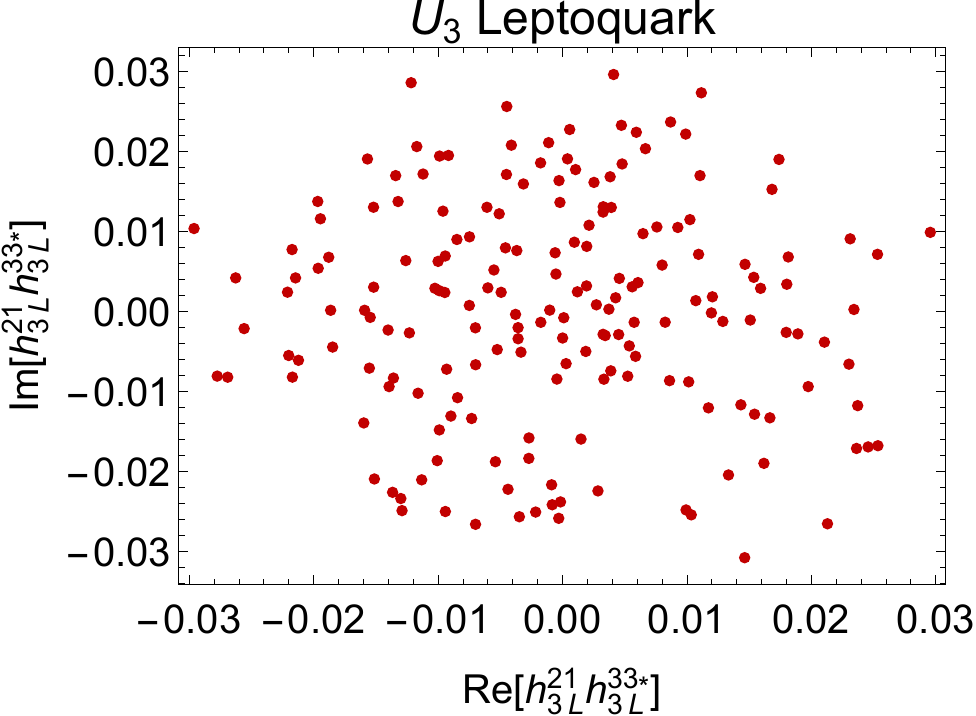}~~~\\

\end{center}
\end{adjustwidth}
\caption{Allowed regions for the couplings of the $ R_2$, ${\bm S}_3$, and ${\bm U}_3$ leptoquark models in the case that all relevant couplings in each model are included simultaneously. We require that the couplings reproduce the measurements of $\RDr$ and $\RDrstar$ in Eqs.~(\ref{eq:RDrexp}) and (\ref{eq:RDstrexp}) within 3$\sigma$, satisfy $\mathcal{B}(B_c \to \tau^- \bar{\nu}_\tau) \le 30\%$, and are consistent with the upper bounds on $\mathcal{B}(B \to K^{(*)} \nu\bar\nu)$ at 90\% C.L (the latter is only relevant for the left-handed couplings in the ${\bm S}_3$ and ${\bm U}_3$ models).}
\label{LQ-R2-S3_U3}
\end{figure}

\begin{figure}
\begin{adjustwidth}{-0.5cm}{-0.5cm}
\begin{center}
\includegraphics[width=5cm]{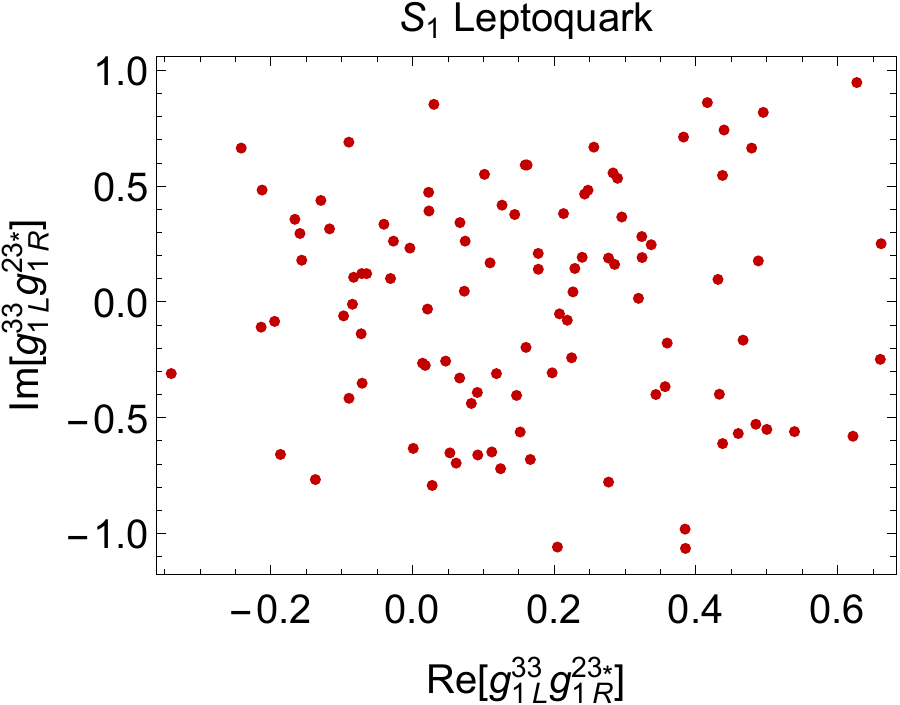}~~~
\includegraphics[width=5cm]{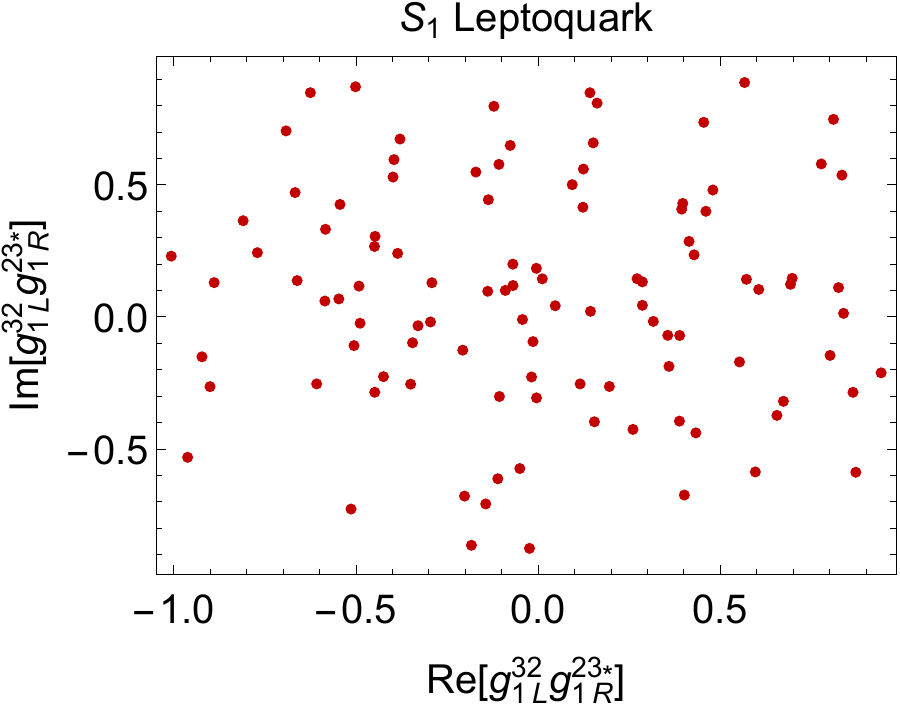}~~~
\includegraphics[width=5cm]{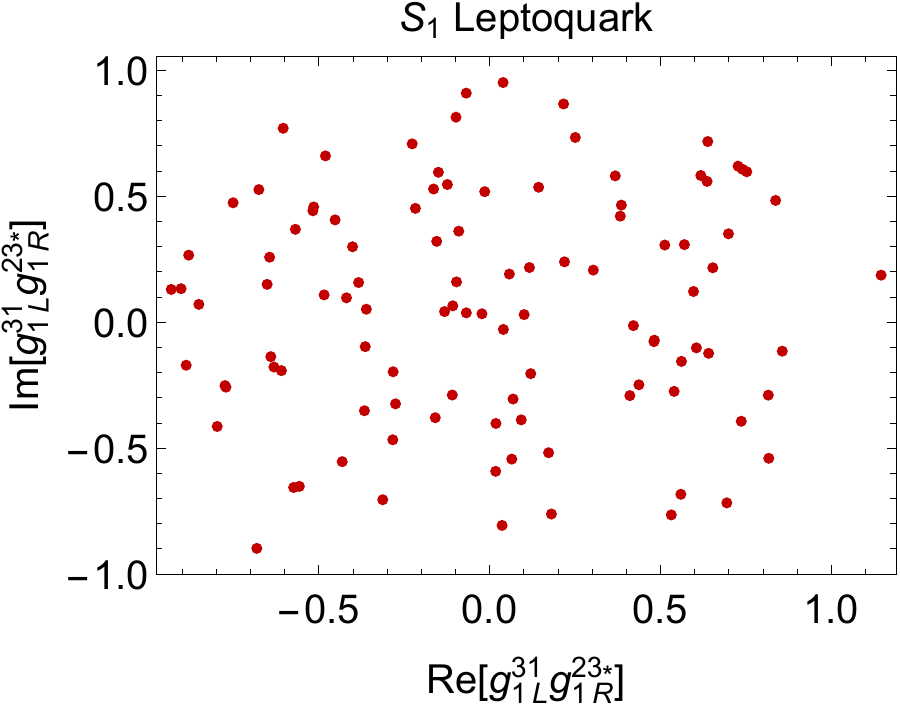}~~~\\

\vspace{2ex}

\includegraphics[width=5cm]{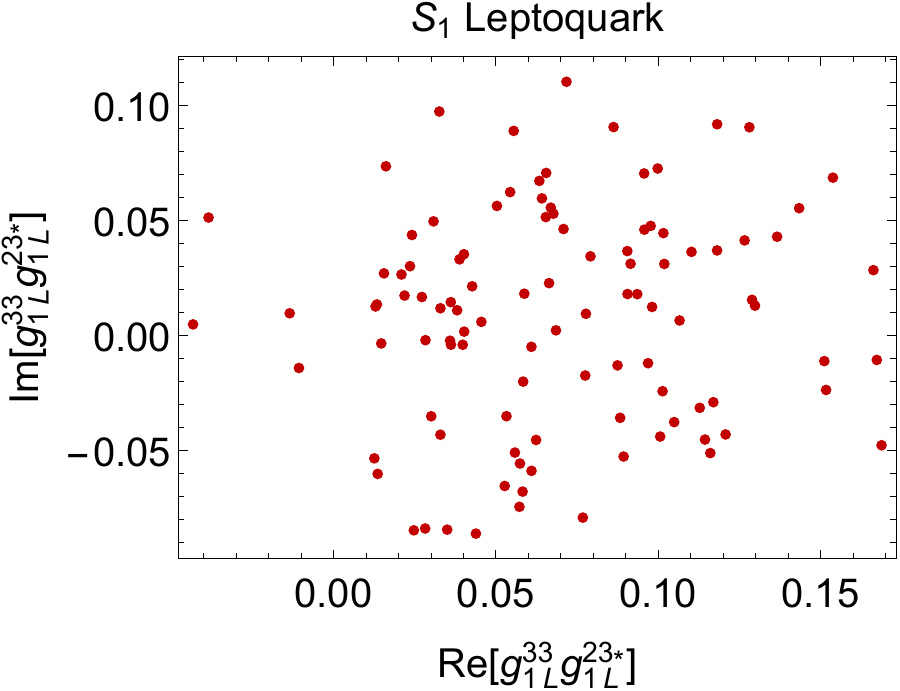}~~~
\includegraphics[width=5cm]{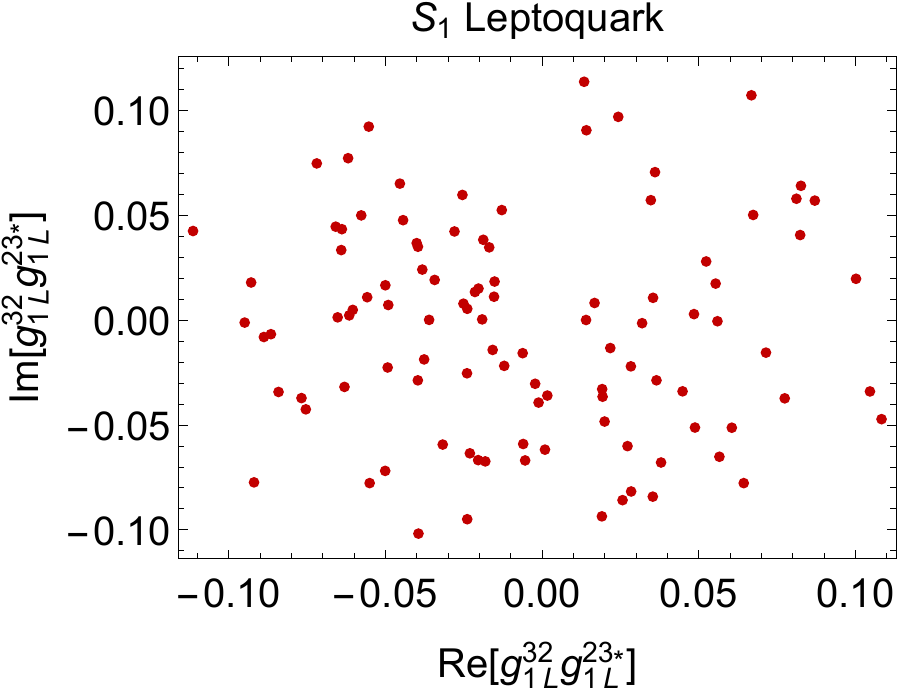}~~~
\includegraphics[width=5cm]{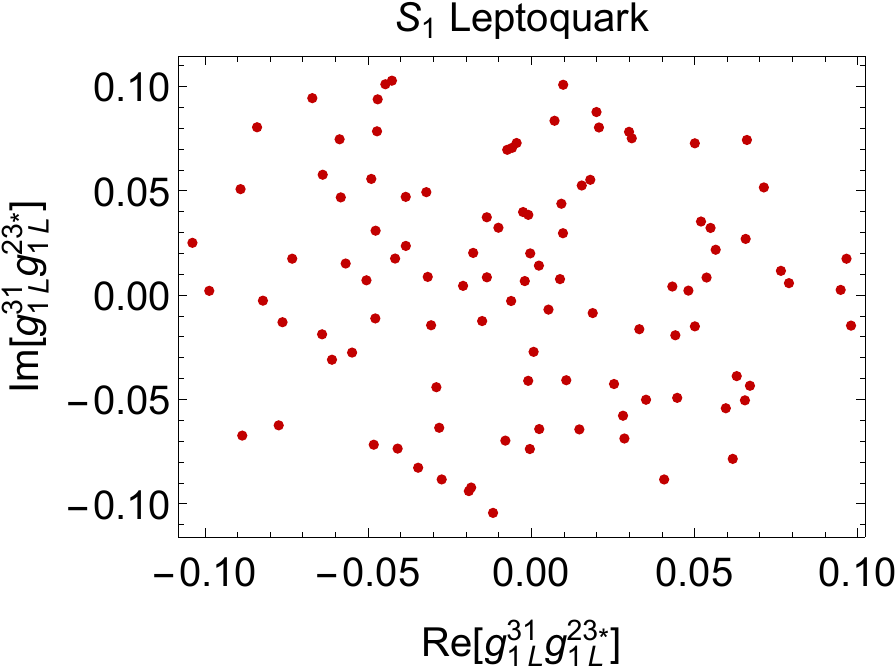}~~~\\

\vspace{2ex}

\includegraphics[width=5cm]{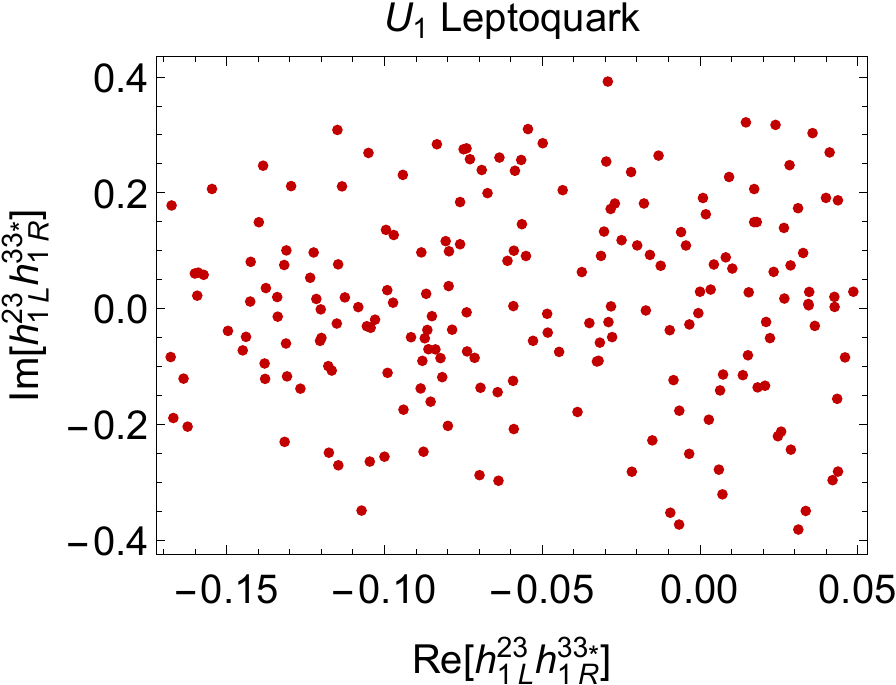}~~~
\includegraphics[width=5cm]{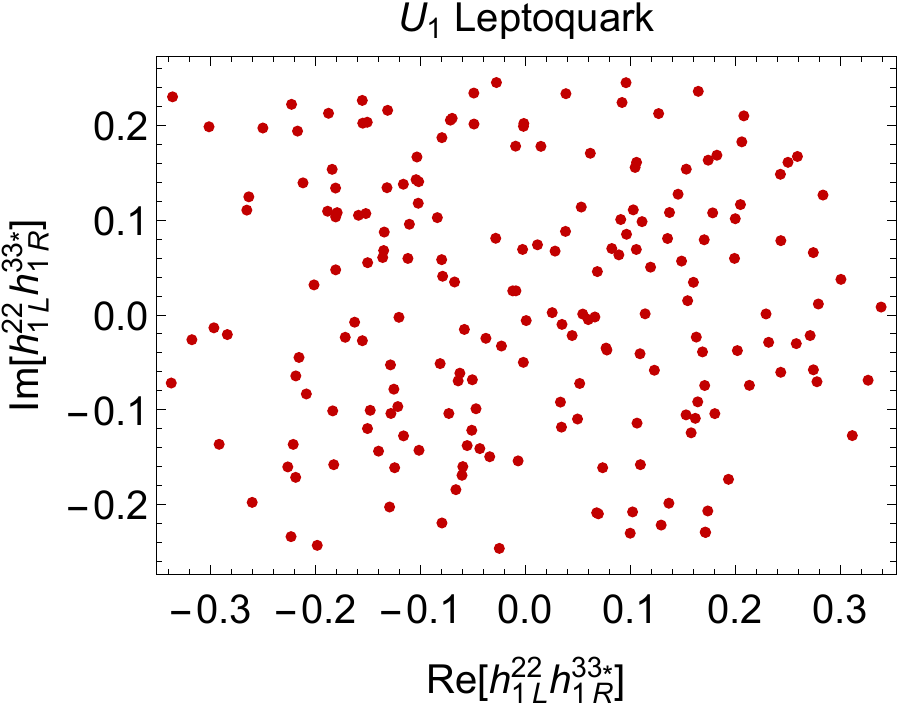}~~~
\includegraphics[width=5cm]{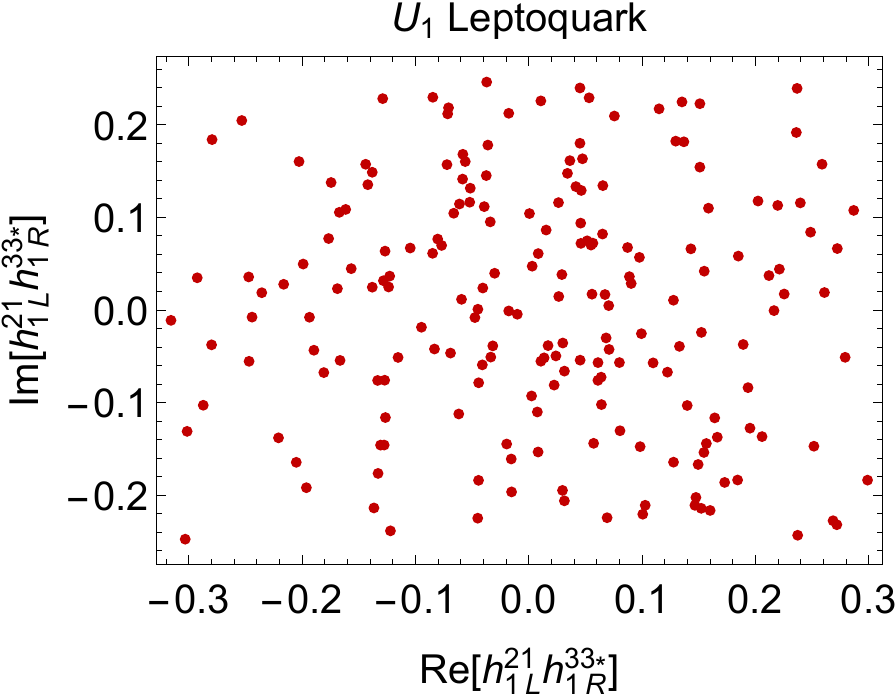}~~~\\

\vspace{2ex}

\includegraphics[width=5cm]{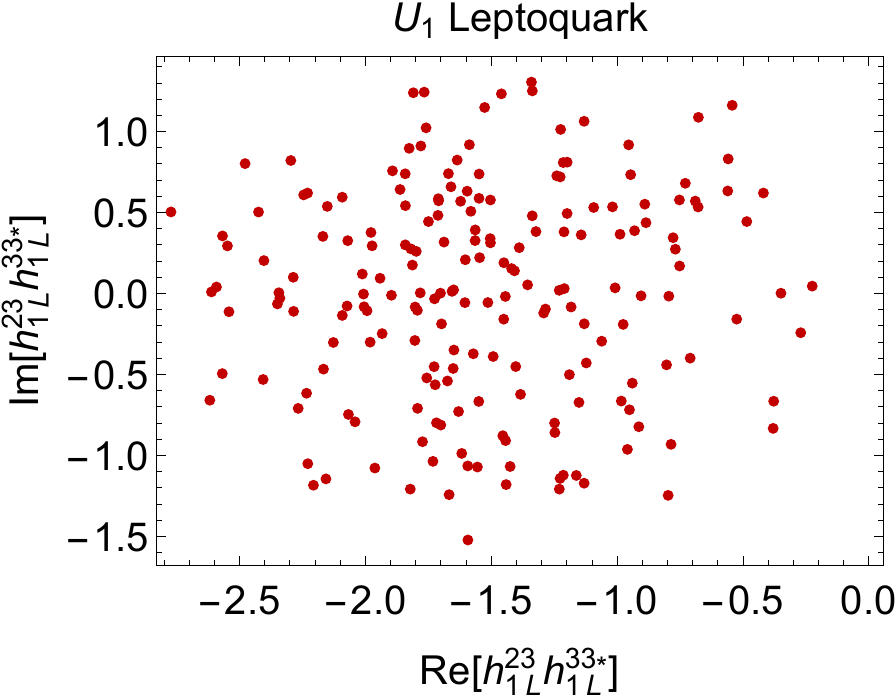}~~~
\includegraphics[width=5cm]{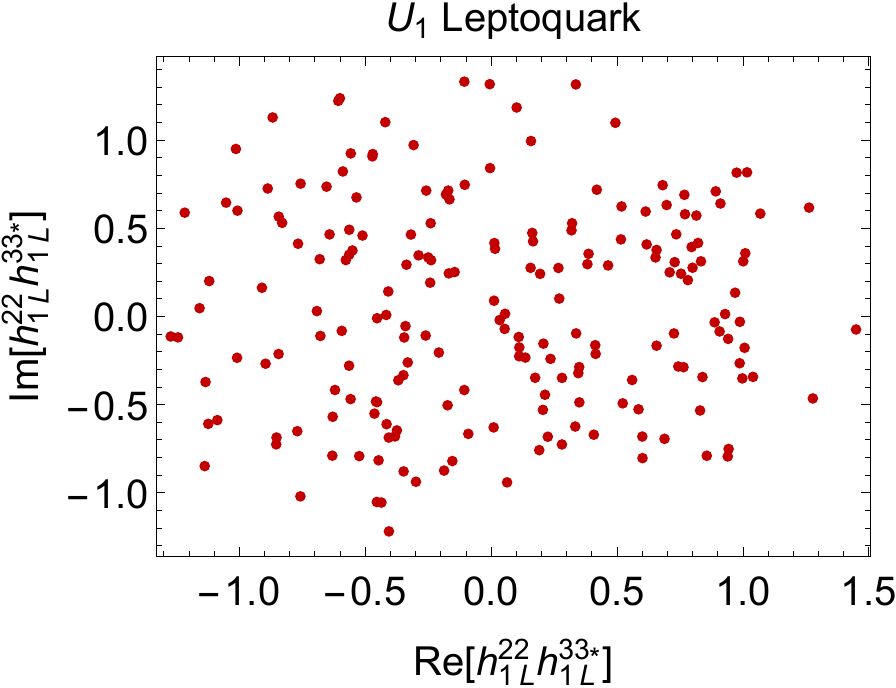}~~~
\includegraphics[width=5cm]{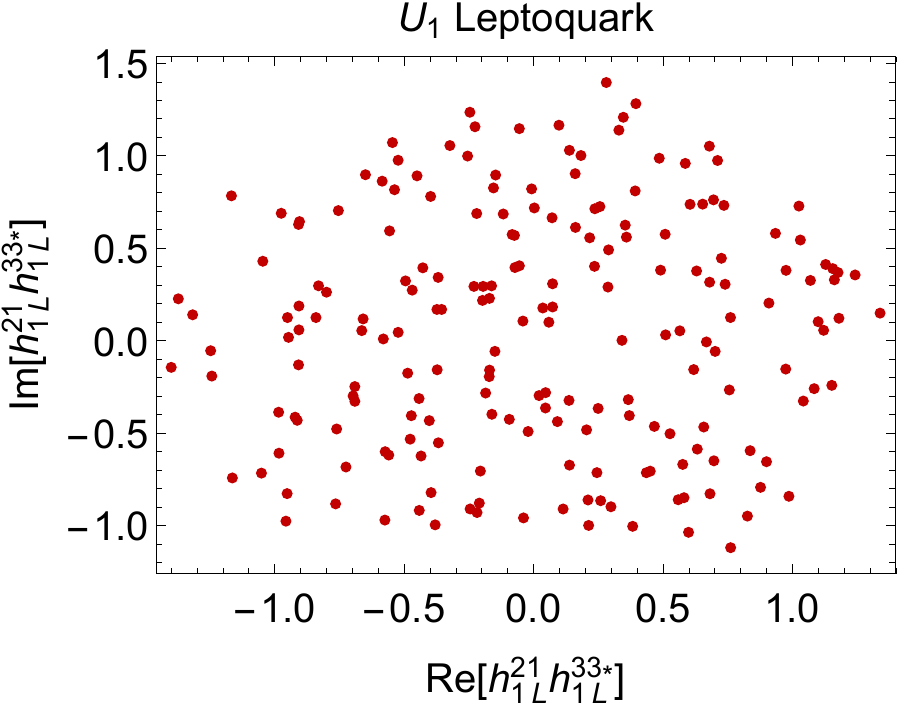}~~~
 \end{center}
 \end{adjustwidth}
\caption{Allowed regions for the couplings of the $S_1$ and $U_1$ leptoquark models in the case that all relevant couplings in each model are included simultaneously. We require that the couplings reproduce the measurements of $\RDr$ and $\RDrstar$ in Eqs.~(\ref{eq:RDrexp}) and (\ref{eq:RDstrexp}) within 3$\sigma$, satisfy $\mathcal{B}(B_c \to \tau^- \bar{\nu}_\tau) \le 30\%$, and are consistent with the upper bounds on $\mathcal{B}(B \to K^{(*)} \nu\bar\nu)$ at 90\% C.L (the latter is only relevant for the left-handed couplings in the $S_1$ model).}
\label{LQ-S1-U1}
\end{figure}

\begin{figure}
\begin{center}
\includegraphics[width=5cm]{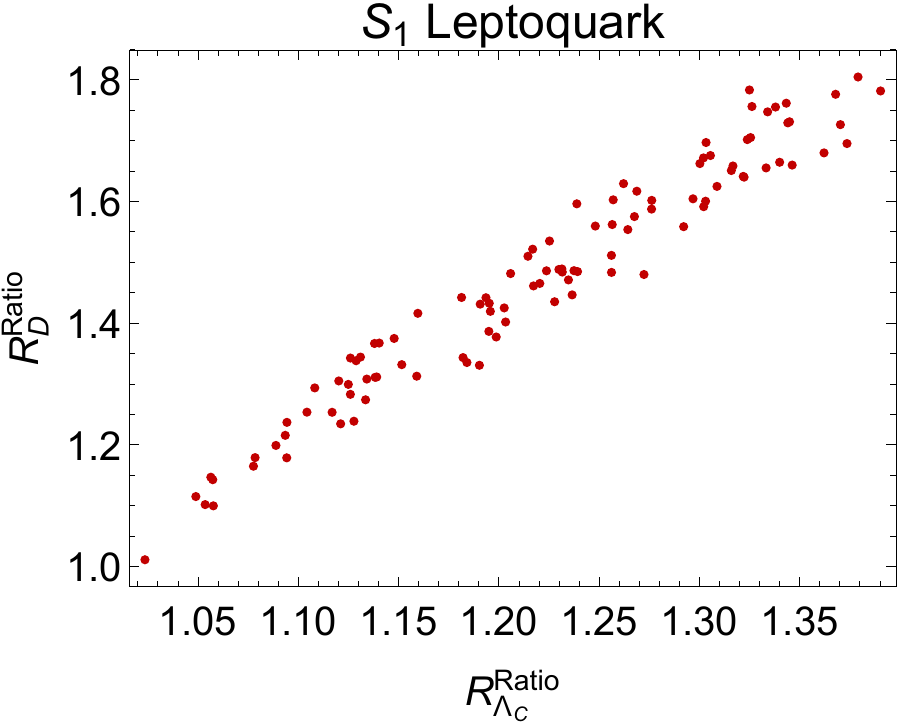}~~~
\includegraphics[width=5cm]{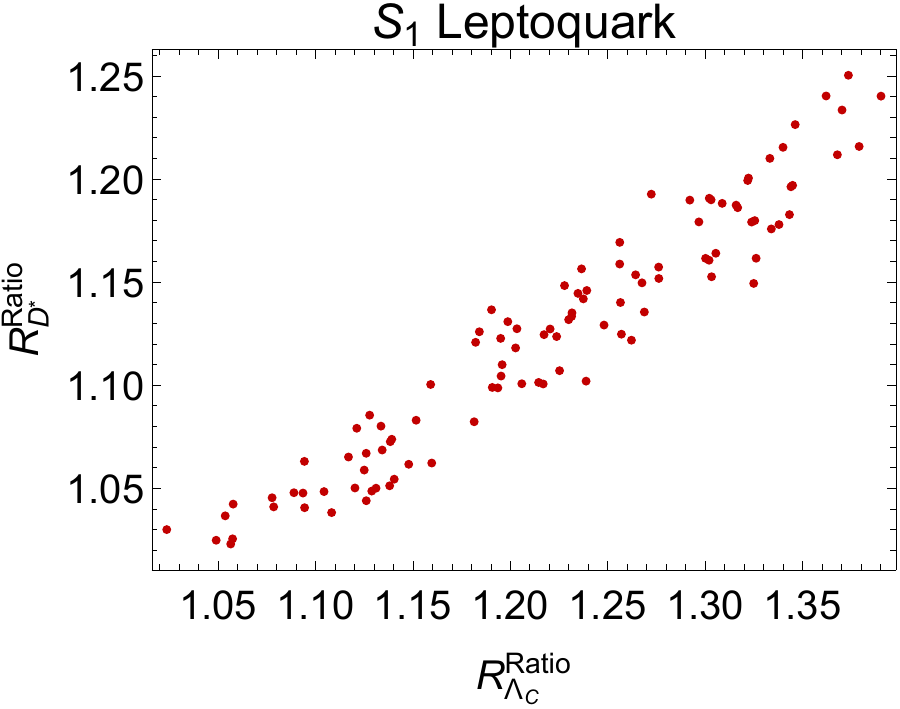}~~~\\

\vspace{2ex}

\includegraphics[width=5cm]{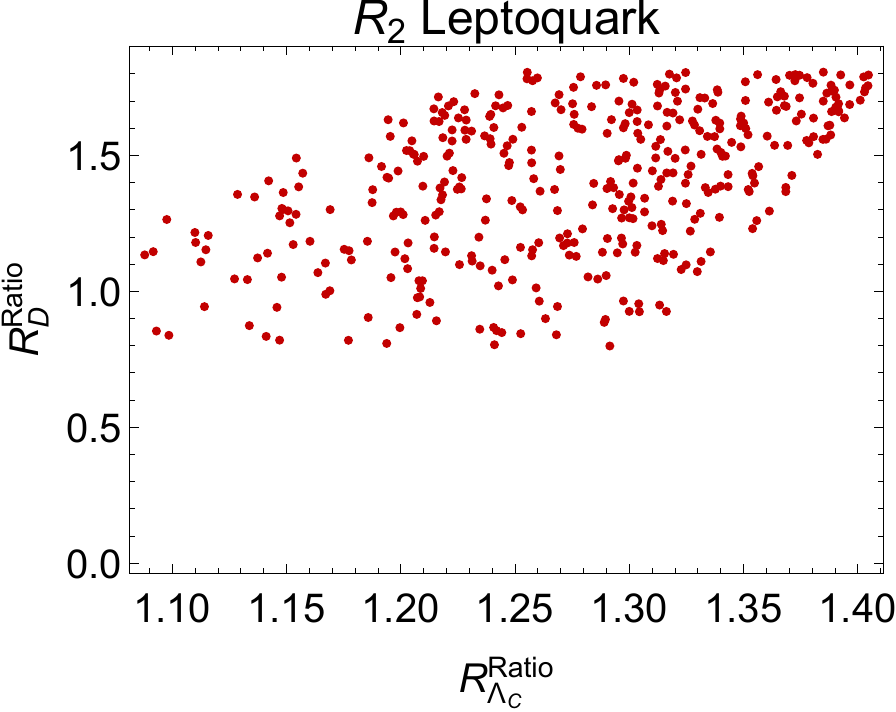}~~~
\includegraphics[width=5cm]{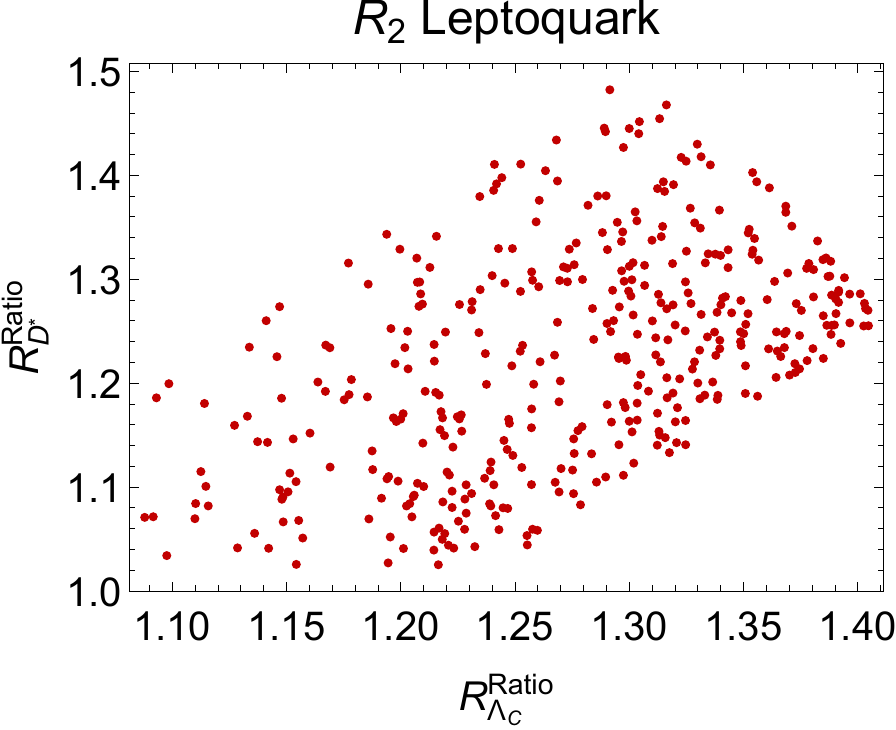}~~~\\

\vspace{2ex}

\includegraphics[width=5cm]{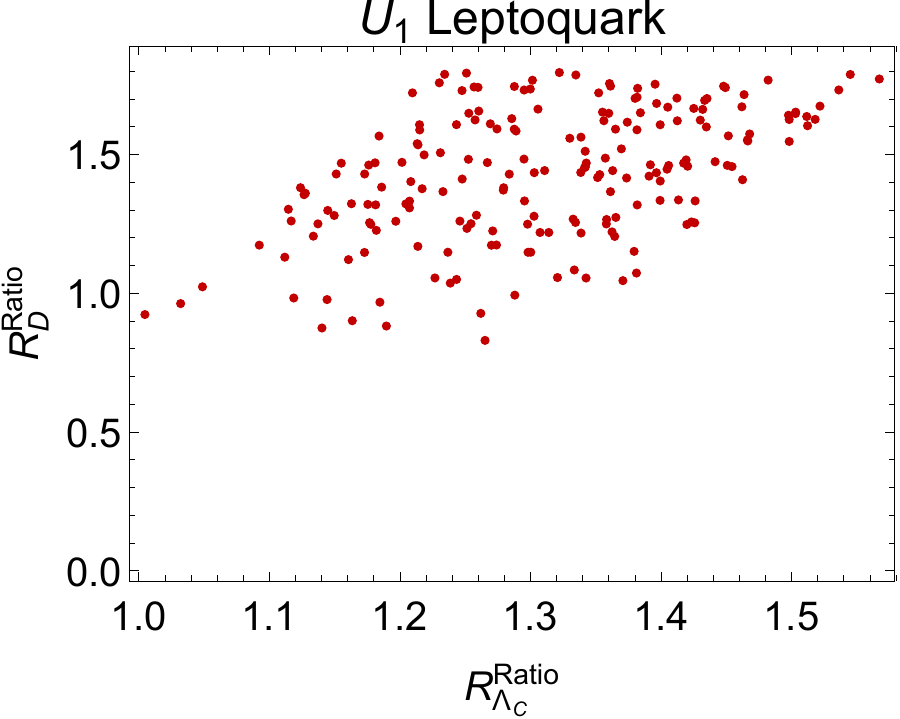}~~~
\includegraphics[width=5cm]{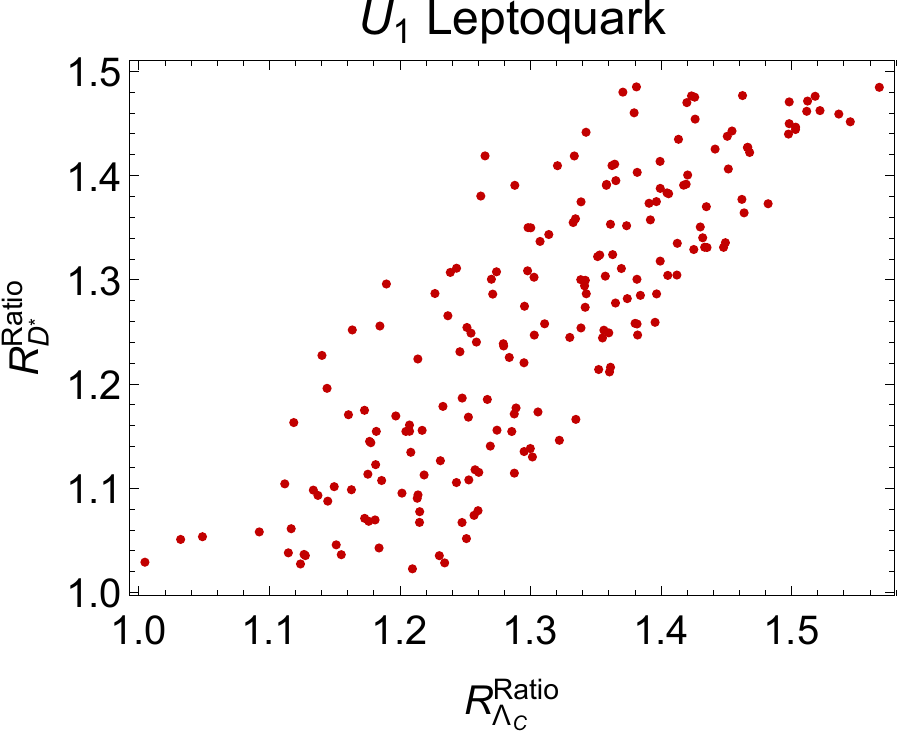}~~~\\

\vspace{2ex}

\includegraphics[width=5cm]{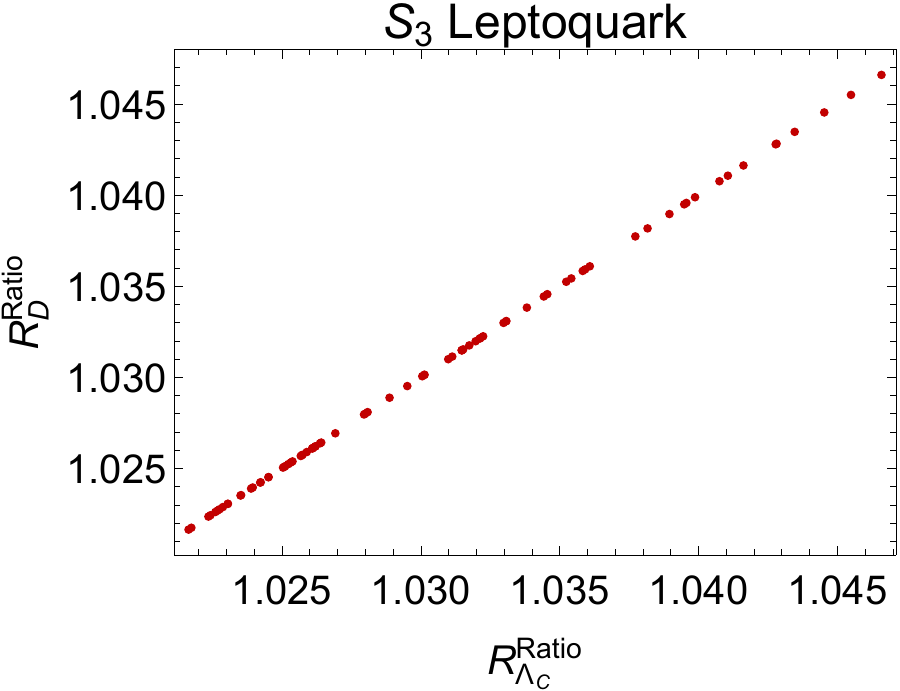}~~~
\includegraphics[width=5cm]{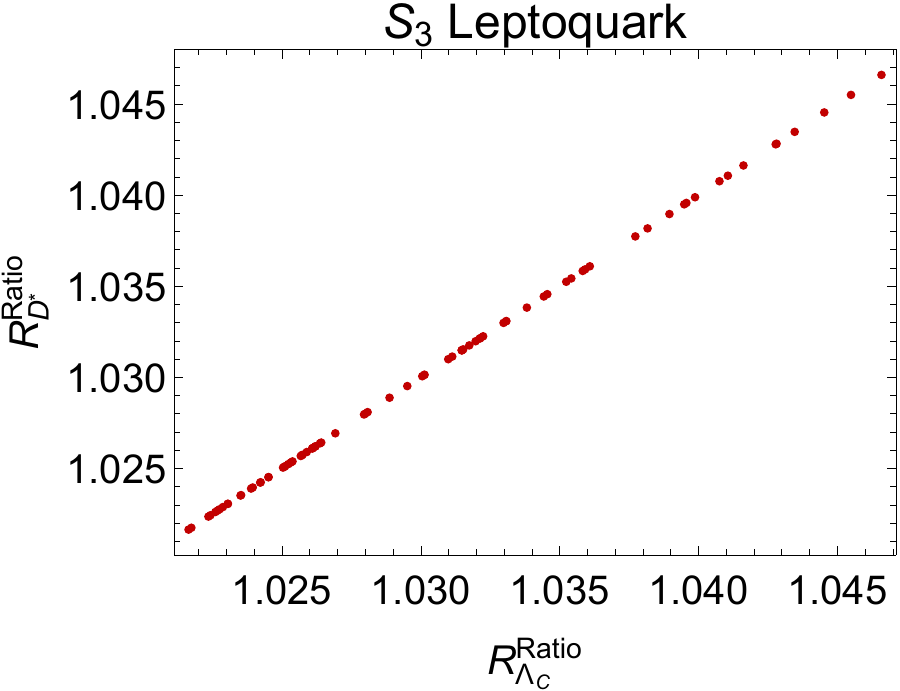}~~~\\

\vspace{2ex}

\includegraphics[width=5cm]{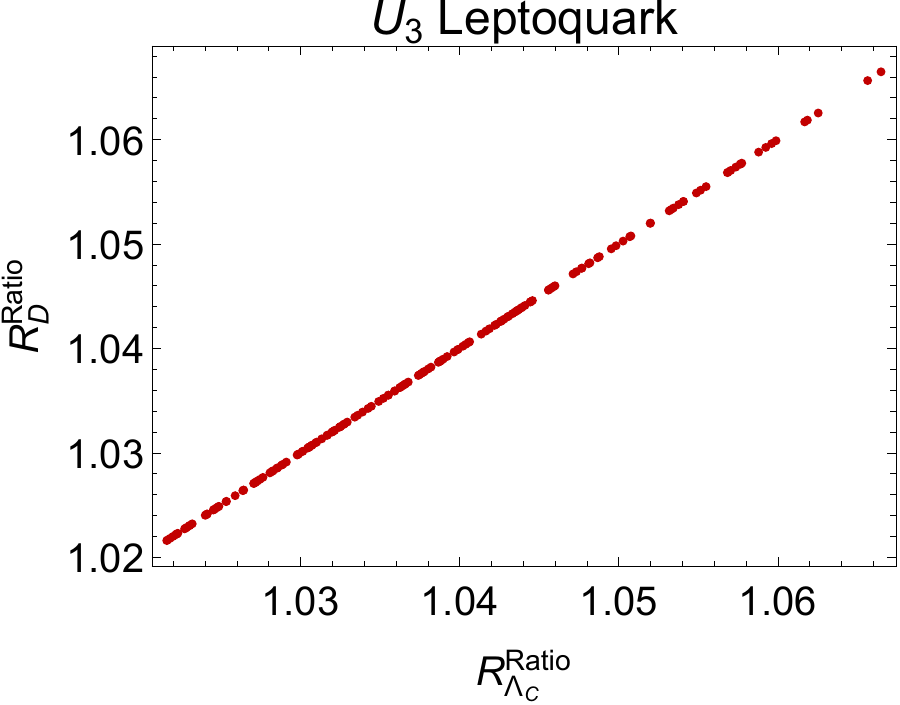}~~~
\includegraphics[width=5cm]{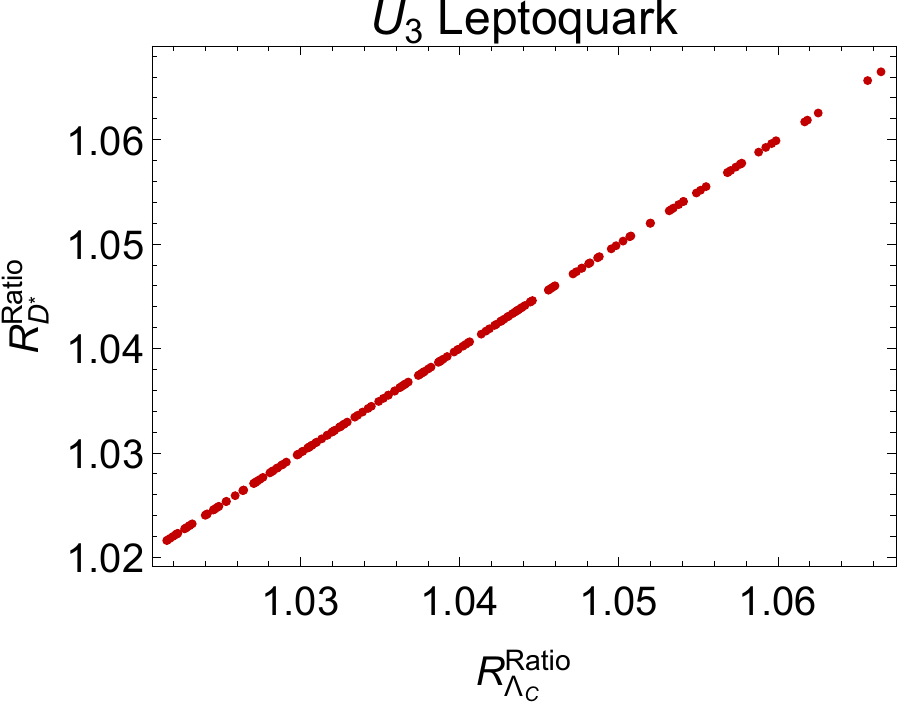}
\end{center}
\caption{The allowed regions in the $\Rlcr - R_D^{Ratio}$ and $\Rlcr - R_{D^*}^{Ratio}$ planes for each leptoquark model, given the allowed regions for
the couplings from Figs.~\protect\ref{LQ-R2-S3_U3} and \protect\ref{LQ-S1-U1}.}
\label{LQ-RatioRegions}
\end{figure}

Allowing all relevant couplings in each model to be nonzero simultaneously, we obtain the coupling regions sampled by the random points
in Figs.~\ref{LQ-R2-S3_U3} and \ref{LQ-S1-U1}. The corresponding allowed regions in the $\Rlcr - R_{D}^{Ratio}$ and $\Rlcr - R_{D^*}^{Ratio}$ planes
are shown in Fig.~\ref{LQ-RatioRegions}. Since the ${\bm S}_3$ and ${\bm U}_3$ leptoquarks produce only the vector coupling $g_L$, all ratios
get rescaled by the common factor of $|1+g_L|^2$. The ${\bm S}_3$ and ${\bm U}_3$ models are tighly constrained and only small effects are allowed. The other
leptoquark models can produce substantial effects in $\Rlcr$, with varying degrees of correlation between the mesonic and baryonic observables.

\begin{figure}
\begin{adjustwidth}{-0.5cm}{-0.5cm}
\begin{center}
\includegraphics[width=5cm]{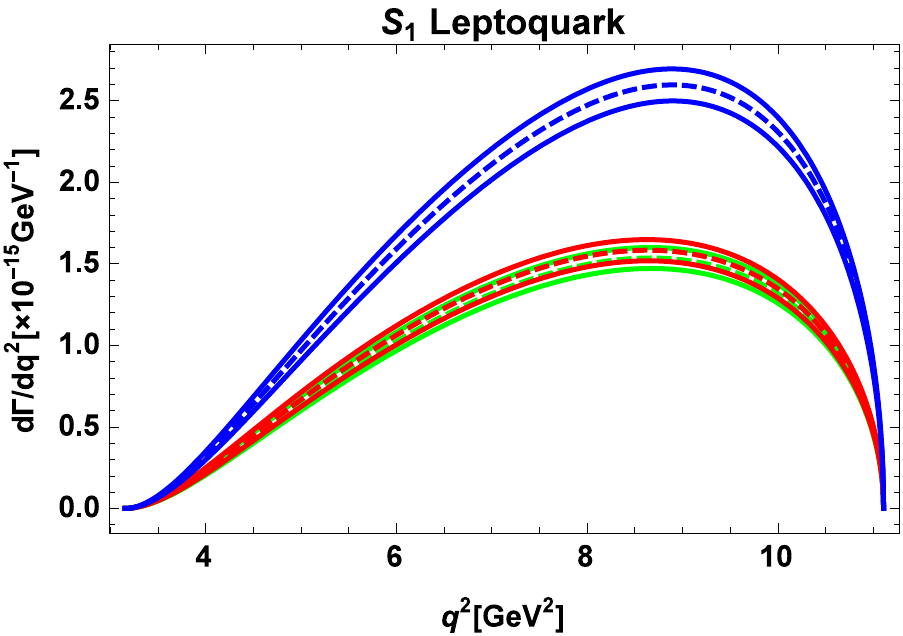}~~~
\includegraphics[width=5cm]{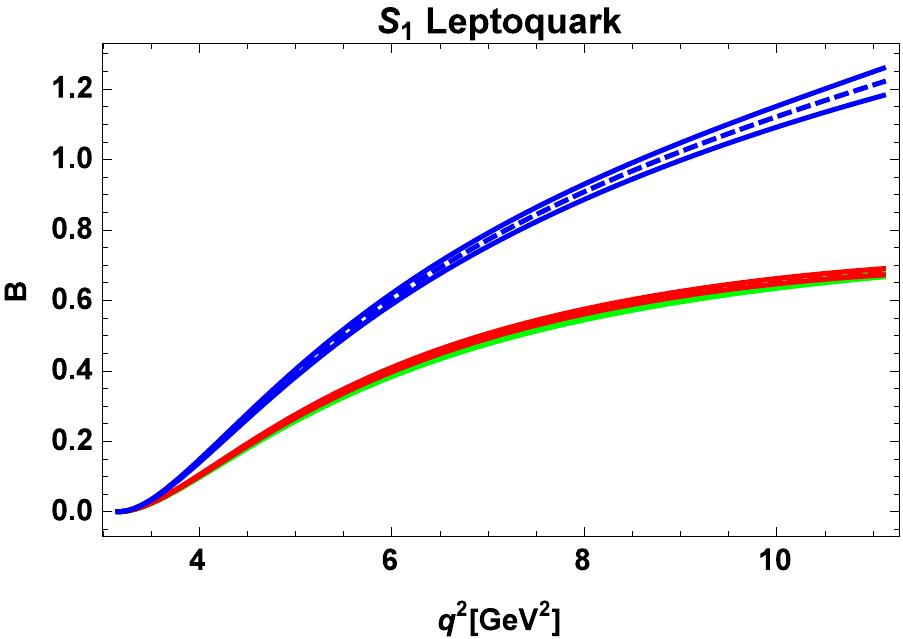}~~~
\includegraphics[width=5cm]{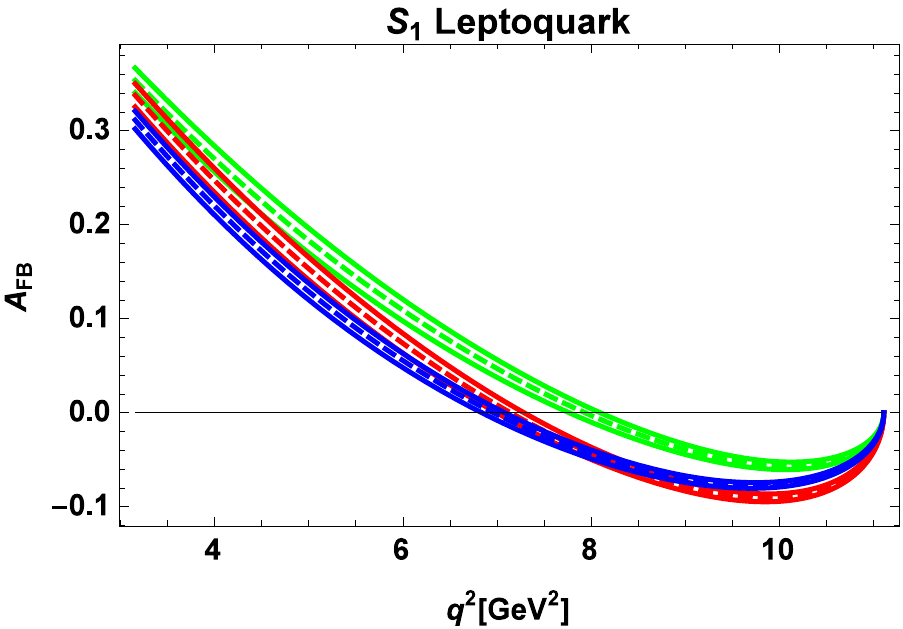}\\

\vspace{2ex}

\includegraphics[width=5cm]{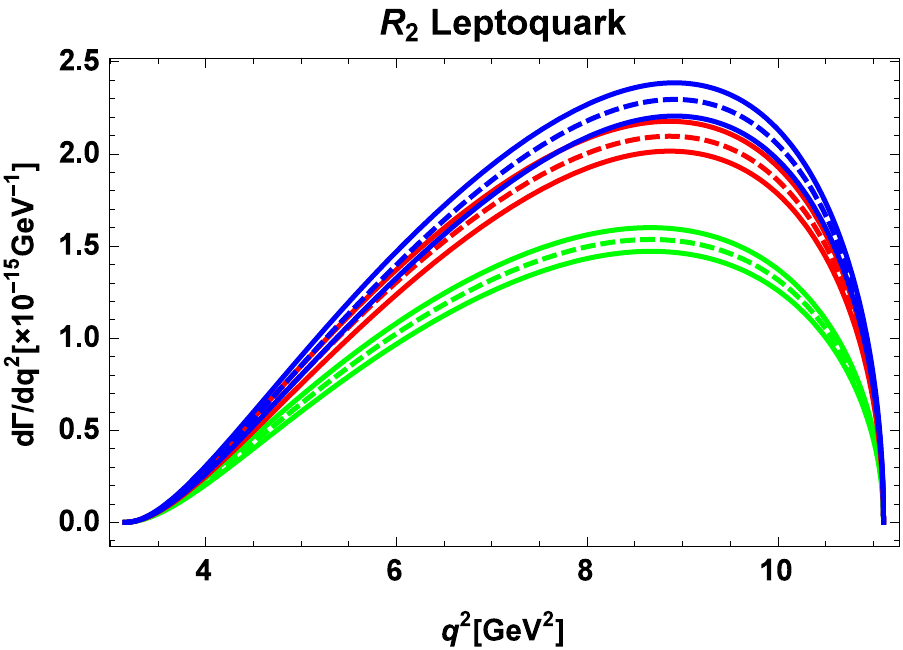}~~~
\includegraphics[width=5cm]{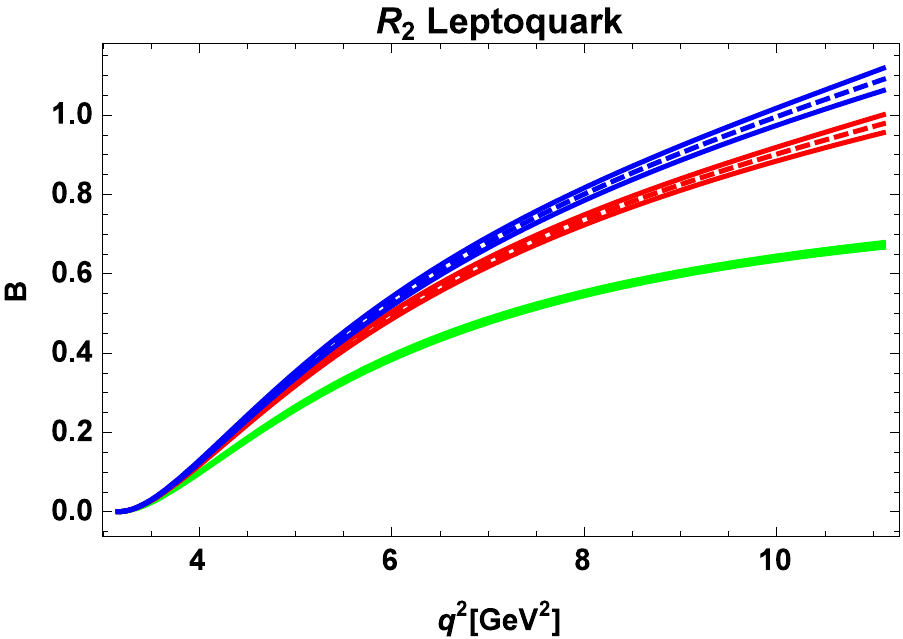}~~~
\includegraphics[width=5cm]{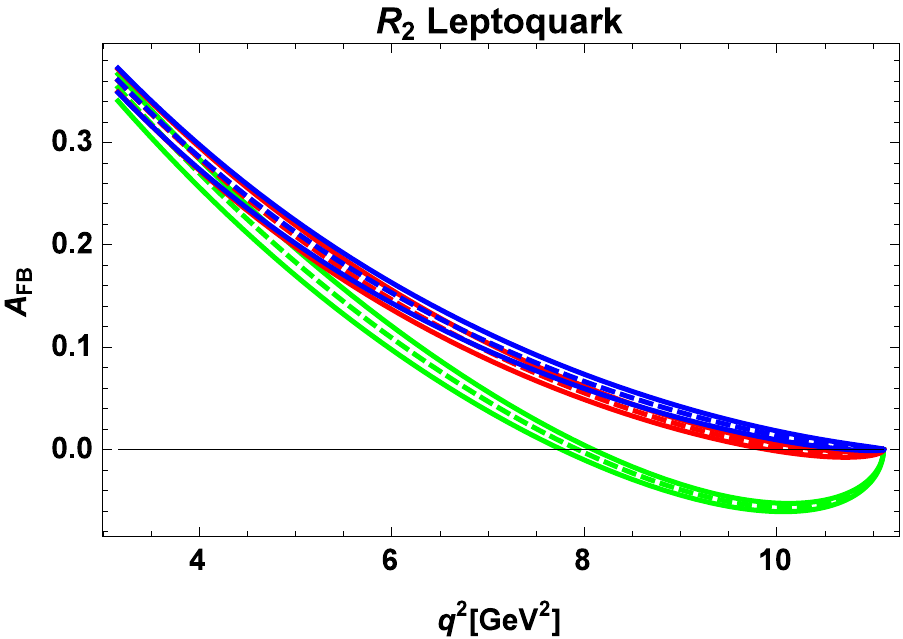}\\

\vspace{2ex}

\includegraphics[width=5cm]{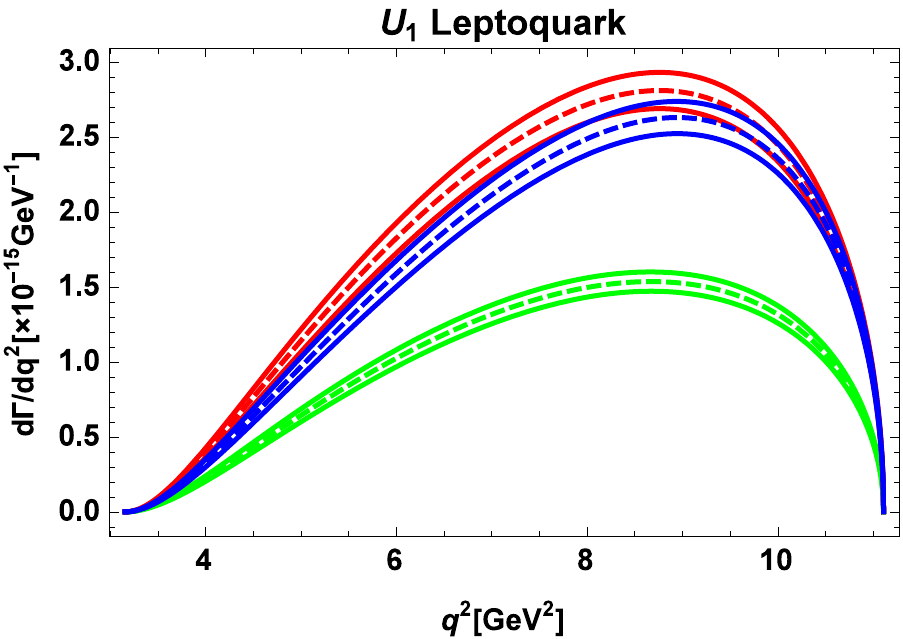}~~~
\includegraphics[width=5cm]{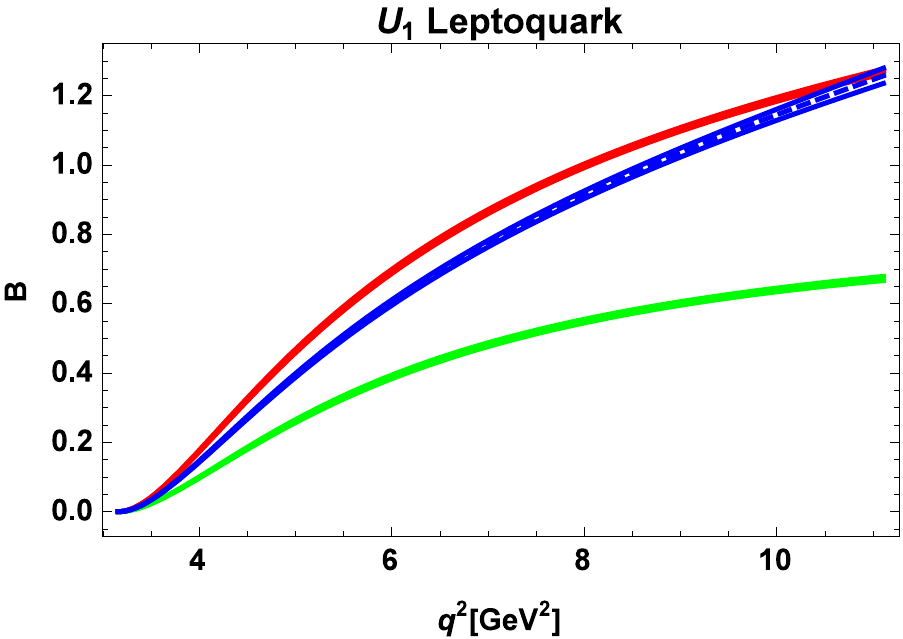}~~~
\includegraphics[width=5cm]{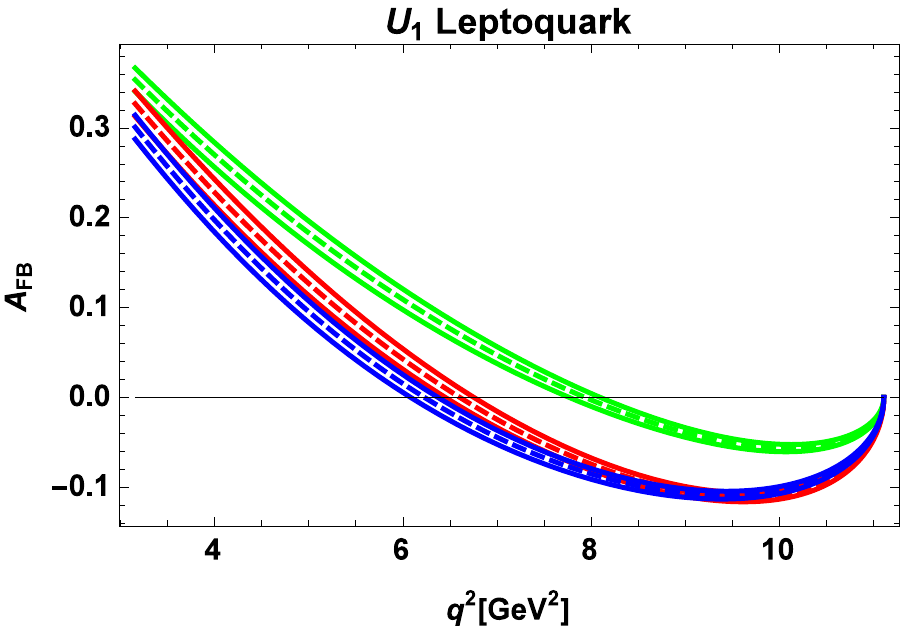}\\

\vspace{2ex}

\includegraphics[width=5cm]{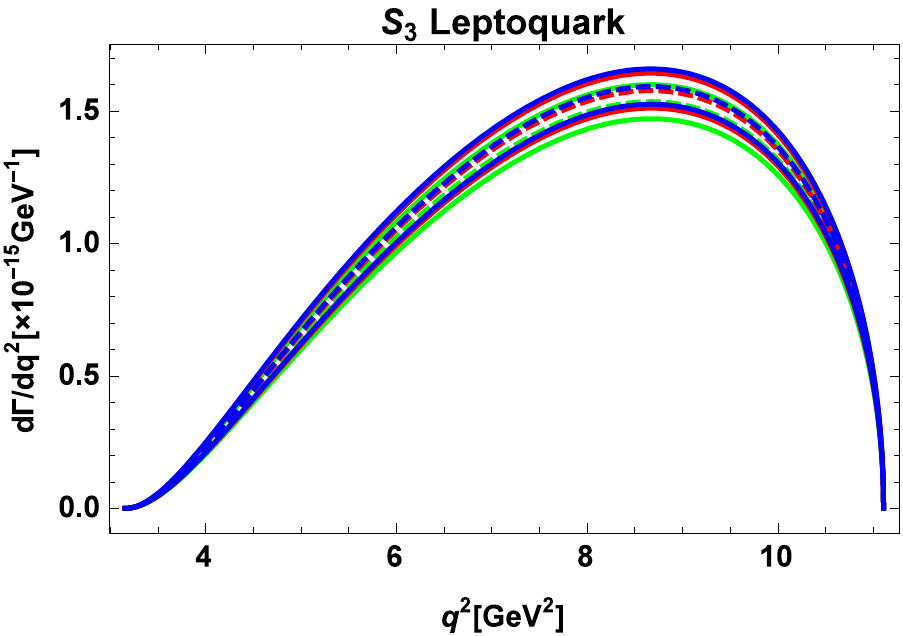}~~~
\includegraphics[width=5cm]{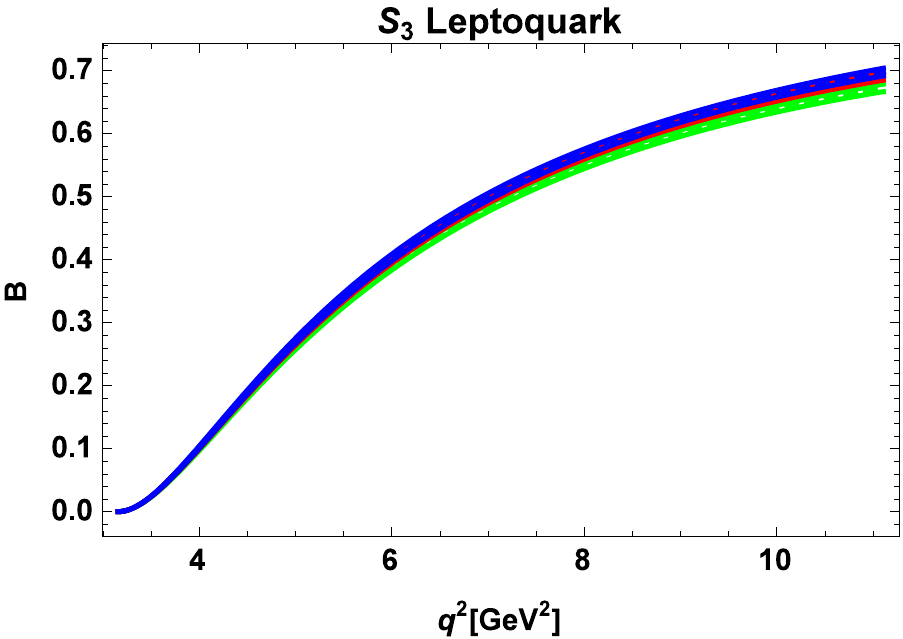}~~~
\includegraphics[width=5cm]{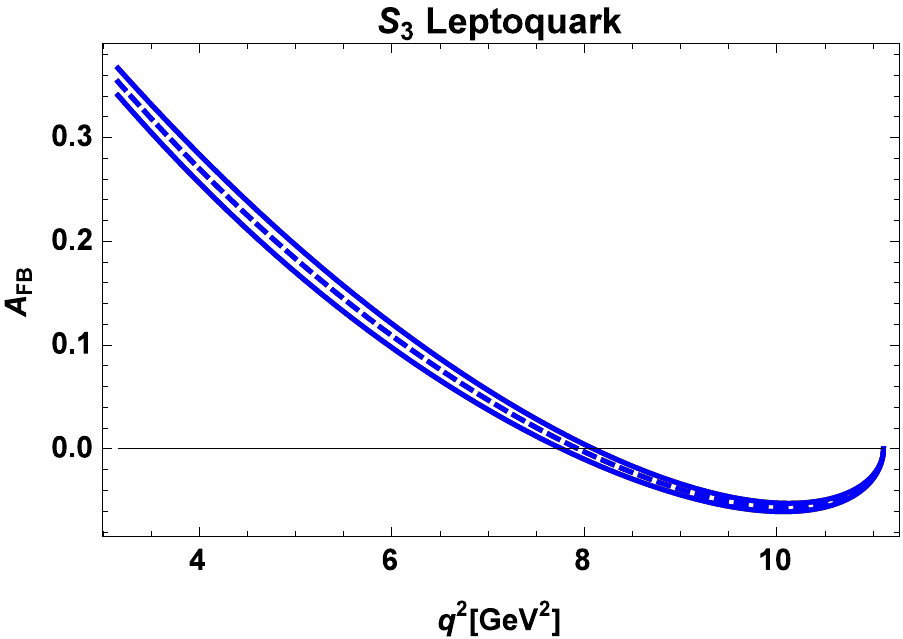}\\

\vspace{2ex}

\includegraphics[width=5cm]{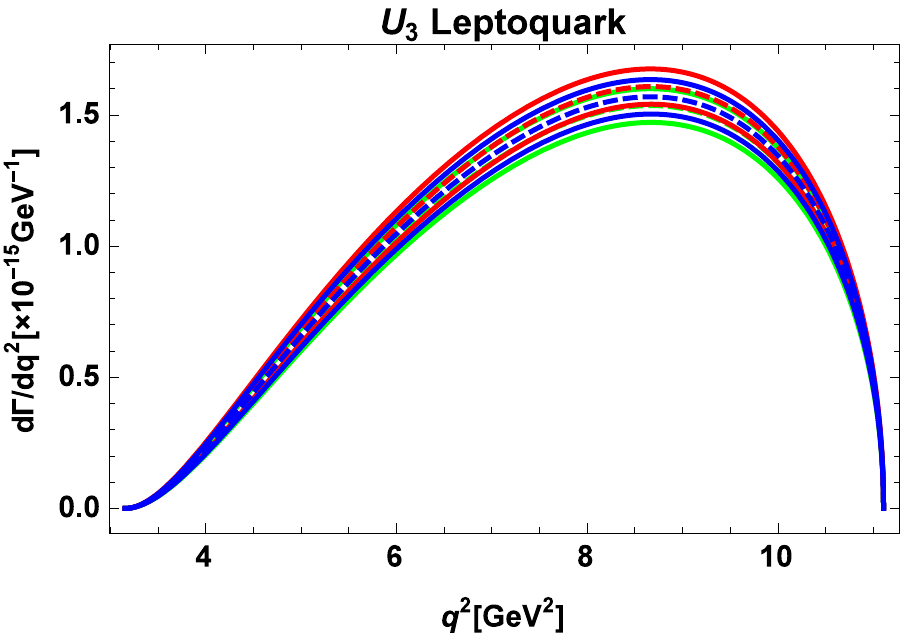}~~~
\includegraphics[width=5cm]{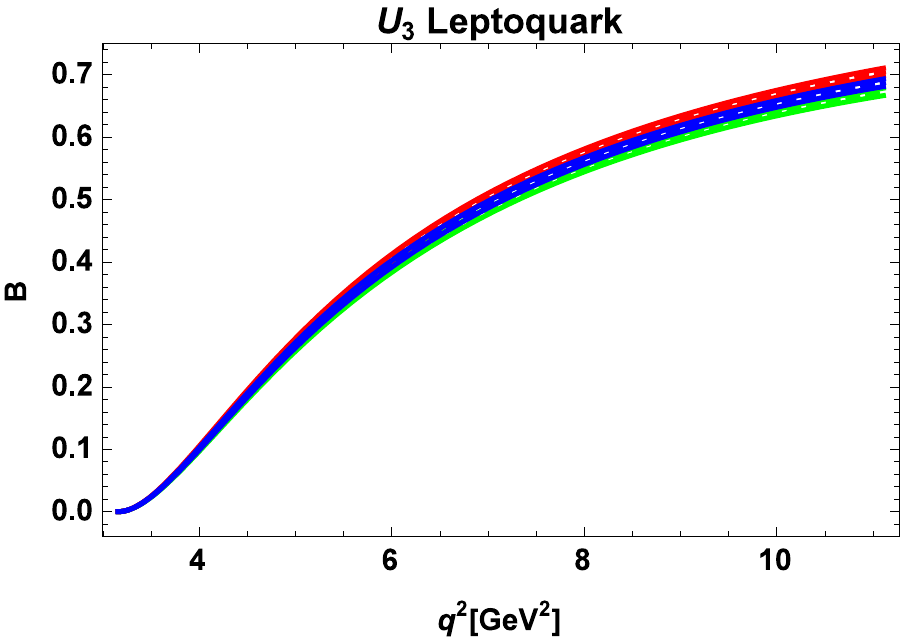}~~~
\includegraphics[width=5cm]{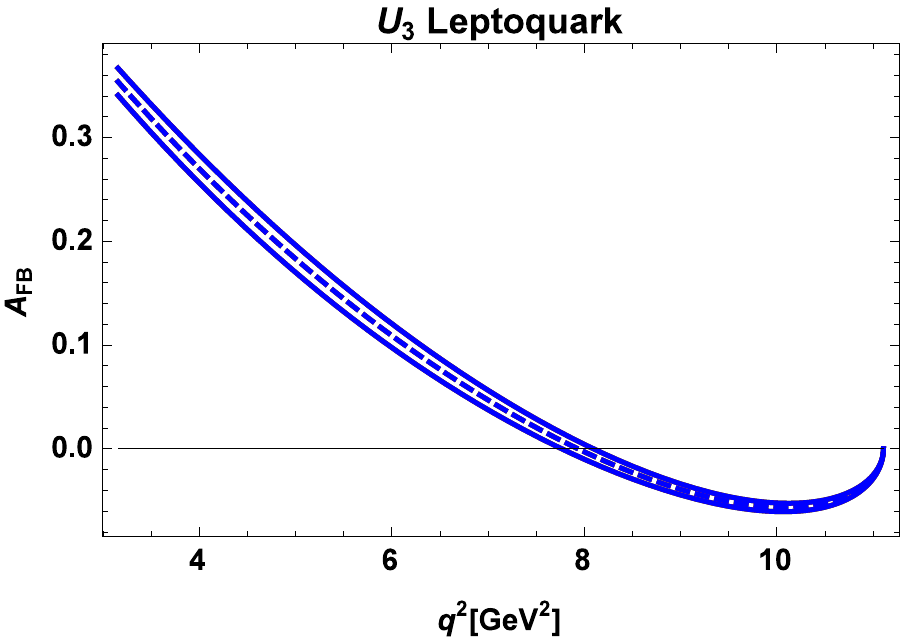}\\
\end{center}
\end{adjustwidth}
\caption{The effects of the different leptoquark models on the $\lbt$ differential decay rate (left), the ratio of the $\lbt$ and $\lbl$ differential decay rates (middle), and the $\lbt$ forward-backward asymmetry (right), for two representative choices of the couplings. The red and blue curves correspond
to the couplings from Cases 1 and 2 in Table \protect\ref{LQ_values}, respectively, while the green curves correspond to the Standard Model. Because
the ${\bm S}_3$ and ${\bm U}_3$ leptoquarks produce only the vector coupling $g_L$, the forward-backward asymmetry remains equal to the Standard Model in those cases. The bands indicate the $1\sigma$ uncertainties originating from the $\Lambda_b\to\Lambda_c$ form factors.}
\label{LQ-shapes}
\end{figure}

The values of $\Rlc$ and $\Rlcr$ for two typical allowed combinations of the couplings in each model are given in Table \ref{LQ_values}.
In Fig.~\ref{LQ-shapes}, we present plots of the observables $(d\Gamma/dq^2,\; B_{\Lambda_c},\; A_{FB})$ for the same values of the couplings.

\begin{table}
\begin{center}
 \begin{tabular}{|c|c|l|c|c|}
 \hline
 Model        & Case      & \parbox{0.3\linewidth}{\centerline{Couplings}} &  $\Rlc$ & $\Rlcr$ \\
 \hline       &           &           &               &               \\[-2ex]
 $S_1$        &    1      & \parbox{0.3\linewidth}{$g_{1L}^{33}g_{1R}^{23*}= 0.332 + 0.403 i$, $g_{1L}^{3i}g_{1R}^{23*}=0.417 - 0.311 i$, $g_{1L}^{33}g_{1L}^{23*}=0.015 - 0.037 i$, $g_{1L}^{3i}g_{1L}^{23*}= -0.079 - 0.002 i$}  & $0.343\pm 0.011$ & $1.032\pm 0.004$ \\
              &           &           &               &               \\[-2ex] 
 \hline       &           &           &               &               \\[-2ex] 
 $S_1$        &    2      & \parbox{0.3\linewidth}{$g_{1L}^{33}g_{1R}^{23*}=0.064 - 0.142 i$, $g_{1L}^{3i}g_{1R}^{23*}=-1.05 + 0.638 i$, $g_{1L}^{33}g_{1L}^{23*}=0.116 - 0.043 i$, $g_{1L}^{3i}g_{1L}^{23*}= 0.018 + 0.104 i$} & $0.549\pm 0.020$ & $1.648 \pm 0.025$ \\
              &           &           &               &               \\[-2ex] 
 \hline       &           &           &               &               \\[-2ex] 
 $R_2$        &    1      & \parbox{0.3\linewidth}{$h_{2L}^{23}h_{2R}^{33*}=0.373 - 0.118 i$, $h_{2L}^{2i}h_{2R}^{33*}=-0.846 - 0.191 i$} & $0.445 \pm 0.016$ & $1.337 \pm 0.016$ \\
              &           &           &               &               \\[-2ex] 
 \hline       &           &           &               &               \\[-2ex] 
 $R_2$        &    2      & \parbox{0.3\linewidth}{$h_{2L}^{23}h_{2R}^{33*}=0.753 - 0.199 i$, $h_{2L}^{2i}h_{2R}^{33*}=0.897 - 0.031 i$} & $0.485 \pm 0.018$ & $1.455 \pm 0.025$ \\
              &           &           &               &               \\[-2ex] 
 \hline       &           &           &               &               \\[-2ex] 
 $U_1$        &    1      & \parbox{0.3\linewidth}{$h_{1L}^{23}h_{1R}^{33*}=-0.115 - 0.021 i$, $h_{1L}^{2i}h_{1R}^{33*}=0.049 + 0.159 i$, $h_{1L}^{23}h_{1L}^{33*}=-1.468 + 0.271 i$, $h_{1L}^{2i}h_{1L}^{33*}=1.116 + 0.744 i$} & $0.605 \pm 0.019$ & $1.818 \pm 0.008$ \\
              &           &           &               &               \\[-2ex] 
 \hline       &           &           &               &               \\[-2ex] 
 $U_1$        &    2      & \parbox{0.3\linewidth}{$h_{1L}^{23}h_{1R}^{33*}=-0.059 + 0.236 i$, $h_{1L}^{2i}h_{1R}^{33*}=0.234 + 0.105 i$, $h_{1L}^{23}h_{1L}^{33*}=-2.002 + 0.854 i$, $h_{1L}^{2i}h_{1L}^{33*}=-0.135 + 0.940 i$} & $0.553 \pm 0.018$ & $1.663 \pm 0.005$ \\
              &           &           &               &               \\[-2ex] 
 \hline       &           &           &               &               \\[-2ex] 
 ${\bm S}_3$  &    1      & \parbox{0.3\linewidth}{$g_{3L}^{33}g_{3L}^{23*}=-0.035 + 0.032 i$, $g_{3L}^{3i}g_{3L}^{23*}=0.061 + 0.041 i$} & $0.342 \pm 0.010$ &  $1.027$ \\
              &           &           &               &               \\[-2ex] 
 \hline       &           &           &               &               \\[-2ex] 
 ${\bm S}_3$  &    2      & \parbox{0.3\linewidth}{$g_{3L}^{33}g_{3L}^{23*}=-0.049 - 0.038 i$, $g_{3L}^{3i}g_{3L}^{23*}=-0.01 - 0.019 i$} & $0.345 \pm 0.011$ & $1.037$ \\
              &           &           &               &               \\[-2ex] 
 \hline       &           &           &               &               \\[-2ex] 
 ${\bm U}_3$  &    1      & \parbox{0.3\linewidth}{$h_{3L}^{23}h_{3L}^{33*}=-0.032 - 0.014 i$, $h_{3L}^{2i}h_{3L}^{33*}=0.003 + 0.002 i$} & $0.349 \pm 0.011$ & $1.047$ \\
              &           &           &               &               \\[-2ex] 
 \hline       &           &           &               &               \\[-2ex] 
 ${\bm U}_3$  &    2      & \parbox{0.3\linewidth}{$h_{3L}^{23}h_{3L}^{33*}=-0.014 - 0.006 i$, $h_{3L}^{2i}h_{3L}^{33*}=0.017 - 0.007 i$} & $0.340 \pm 0.010$ & $1.022$ \\[2.5ex]
 \hline      
\end{tabular}
\caption{The values of the $\Rlc$ and $\Rlcr$ ratios for two representative cases of the couplings of the different leptoquark models. Above, the index $i=1,2$ denotes the electron and muon neutrinos.  The Standard-model value of the ratio is $\Rlc=0.333 \pm 0.010$ \cite{Detmold:2015aaa}. The uncertainties given are due to the $\Lambda_b \to \Lambda_c$ form factor uncertainties.}
\label{LQ_values}
\end{center}
\end{table}

\FloatBarrier
\section{Conclusions}
\label{sec:conclusion}

The baryonic decay $\lbt$ has the potential to shed new light on the $\RD$ puzzle. Here, we studied
the phenomenology of $\lbt$ in the presence of new-physics couplings with all relevant Dirac structures. In contrast
to the mesonic decays, the $\Lambda_b \to\Lambda_c$ form factors have not yet been determined from experimental data,
and it is even more important to use form factors from lattice QCD. Here, we presented new lattice QCD results for the
$\Lambda_b \to\Lambda_c$ tensor form factors, extending the analysis of Ref.~\cite{Detmold:2015aaa}. The
parameters and covariance matrices of the complete set of $\Lambda_b \to\Lambda_c$ form factors are provided as
supplemental material.

In the first part of our phenomenological analysis, we considered individual new-physics couplings
in the effective Hamiltonian in a model-independent way. After constraining these couplings using the $\RD$ measurements and the $B_c$ lifetime, we calculated
the effects of the NP couplings in $\lbt$ decays, focusing on the observables $R(\lc)$, $B_{\lc}(q^2)$, and $A_{FB}(q^2)$.
Measurements of these observables can help in distinguishing among the different NP operators. For instance, the forward-backward asymmetry
$A_{FB}(q^2)$ tends to be mostly above the SM value in the presence of right-handed ($g_R$) or tensor ($g_T$) couplings, but is lower than the SM value for most
allowed values of the scalar ($g_S$) coupling. To illustrate the impact of a future $R(\lc)$ measurement, we presented the constraints
on all couplings resulting from two possible ranges of $R(\lc)$. The baryonic decay can tightly constrain all of the couplings
$g_L$, $g_R$, $g_S$, $g_P$, and $g_T$. For example, we have shown that if $\Rlcr=R(\Lambda_c)/R(\Lambda_c)^{SM}$ is observed to have a value around 1.3,
the scenario with only $g_P$ becomes ruled out by the combined constraints from $R(\Lambda_c)$ and $\tau_{B_c}$.

In the second part of our phenomenological analysis, we considered explicit models in which multiple NP operators are present.
For the two-Higgs-doublet model we found significant contribution to $\lbt$. However, the full numerical analysis was not included in this work as we did not consider RGE evolution
which could  impact the phenomenology of the model. 
Models with $SU(2)$ gauge symmetry generally cannot produce large effects in $b \to c\tau\bar{\nu}_\tau$ transitions without violating bounds from other observables such
as $B_s$ mixing, and we therefore did not present their effects on $\lbt$. On the other hand, we have demonstrated that some of the leptoquark models can produce
large effects in the $\lbt$ observables, in particular through scalar and tensor couplings. We have presented
correlation plots of $R_D^{Ratio}$ and $R_{D^{*}}^{ Ratio}$ versus $\Rlcr$, which may be helpful in discriminating among the various models.

\bigskip
\noindent
{\bf Acknowledgments}:
We thank Shanmuka Shivashankara for early work on this project. This
work was financially supported by the National Science Foundation under Grant Nos.\ 
PHY-1414345 (AD and AR) and PHY-1520996 (SM). SM is also supported by the RHIC Physics Fellow Program of the RIKEN BNL Research Center.
AD acknowledges the hospitality of the Department of Physics and
Astronomy, University of Hawaii, where part of the work was done.
The lattice QCD calculations were carried out using
high-performance computing resources provided by XSEDE (supported by National Science Foundation Grant No.\ OCI-1053575)
and NERSC (supported by U.S.~Department of Energy Grant No.\ DE-AC02-05CH11231).

\pagebreak

\appendix

\section{Helicity spinors and polarization vectors}
\label{sec:spinorsandvectors}

In this appendix, we give explicit expressions for the spinors and polarization vectors used to calculate
the helicity amplitudes for the decay $\lbt$.

\subsection{$\Lambda_b$ rest frame}

To calculate the hadronic helicity amplitudes, we work in the $\Lambda_b$ rest frame and take the three-momentum of the $\Lambda_c$ along the $+z$ direction and
the three-momentum of the virtual vector boson along the $-z$ direction. The baryon spinors are then given by \cite{Auvil:1966eao}
\begin{eqnarray}
\label{spinor}
\bar{u}_2(\pm{\textstyle \frac{1}{2}},p_{\lc})&=& \sqrt{E_{\lc}+m_{\lc}}\left( \chi_\pm^\dagger,
\frac{\mp|\mathbf{p}_{\lc}|}{E_{\lc}+m_{\lc}}\chi_\pm^\dagger \right) \,,\nonumber \\
u_1(\pm{\textstyle \frac{1}{2}},p_{\lb}) &=& \sqrt{2m_{\lb}}
\left(\begin{array}{c}\chi_\pm \\ 0 \end{array}\right) \, ,
\end{eqnarray}
where $\chi_+=\left(\begin{array}{c}1 \\ 0 \end{array}\right)$ and 
$\chi_-=\left(\begin{array}{c}0 \\ 1 \end{array}\right)$ are the usual Pauli 
two-spinors. The polarization vectors of the virtual vector boson are \cite{Auvil:1966eao}
\begin{eqnarray}
\label{polvec}
\epsilon^{\mu *}(t) &=& \frac{1}{\sqrt{q^2}}\left(q_0;0,0,-|\mathbf{q}|\right)\,, \nonumber \\
\epsilon^{\mu *}(\pm 1) &=& \frac{1}{\sqrt{2}}\left(0;\pm 1,-i,0\right) \,,\nonumber \\
\epsilon^{\mu *}(0) &=& \frac{1}{\sqrt{q^2}}\left(|\mathbf{q}|;0,0,-q_0\right)\,, 
\end{eqnarray}
where $q^{\mu}=(q_{0};0,0,-|\mathbf{q}|)$ is the four-momentum of the virtual vector boson in the $\lb$ rest frame. We have
\begin{align}
q_0 &=\frac{1}{2m_{\lb}}(m_{\lb}^2-m_{\lc}^2+q^2)\,,\\
|\mathbf{q}| &=|\mathbf{p}_{\lc}|=\frac{1}{2m_{\lb}}\sqrt{Q_+Q_-}\,,
\end{align}
where
\begin{equation}
 Q_\pm = (m_{\lb} \pm m_{\lc})^2 - q^2.
\end{equation}

\subsection{Dilepton rest frame}

In the calculation of the lepton helicity amplitudes, we work in the rest frame of the virtual vector boson boson, which is
equal to the rest frame of the $\tau\bar{\nu}_\tau$ dilepton system. We define the angle $\theta_\tau$ as the angle between the three-momenta of the $\tau$ and the $\Lambda_c$ in this frame.

The lepton spinors for $\mathbf{p}_\tau$ pointing in the $+z$ direction and $\mathbf{p}_{\bar{\nu}_\tau}$ pointing in the $-z$ direction are
\begin{eqnarray}
\label{spinor2}
\bar{u}_\tau(\pm{\textstyle \frac{1}{2}},p_\tau)&=& \sqrt{E_\tau+m_\tau}\left( \chi_\pm^\dagger,
\frac{\mp|\mathbf{p}_\tau|}{E_\tau+m_\tau}\chi_\pm^\dagger \right) \,,\nonumber \\
v_{\bar{\nu}_\tau}({\textstyle \frac{1}{2}},p_{\bar{\nu}_\tau}) &=& \sqrt{E_\nu}
\left(\begin{array}{c}\chi_+ \\-\chi_+  \end{array}\right) \, .
\end{eqnarray}
We then rotate about the $y$ axis by the angle $\theta_\tau$ so that after the rotation, the three-momentum of the $\Lambda_c$ points in the $+z$ direction.
The two-spinors transform as
\bea
\chi^\prime_\pm &=& e^{-i\theta_\tau \sigma_2/2} \chi_\pm\nonumber\\
&=& \begin{pmatrix}
\cos (\theta_\tau/2) & -\sin (\theta_\tau/2) \cr
\sin (\theta_\tau/2) & \cos (\theta_\tau/2)
\end{pmatrix} \chi_\pm,
\eea
and
\bea
\chi^{\prime\dagger}_\pm &=& \chi^\dagger_\pm \begin{pmatrix}
\cos (\theta_\tau/2) & \sin (\theta_\tau/2) \cr
-\sin (\theta_\tau/2) & \cos (\theta_\tau/2)
\end{pmatrix} ,
\eea
and the full lepton spinors after the rotation are
\begin{eqnarray}
\label{spinor3}
\bar{u}_\tau(+{\textstyle \frac{1}{2}},p_\tau)&=& \sqrt{E_\tau+m_\tau}\left( \cos (\theta_\tau/2),\sin (\theta_\tau/2),
\frac{-|\mathbf{p}_\tau|}{E_\tau+m_\tau}\cos (\theta_\tau/2),\frac{-|\mathbf{p}_\tau|}{E_\tau+m_\tau}\sin (\theta_\tau/2) \right) \,,\nonumber \\
\bar{u}_\tau(-{\textstyle \frac{1}{2}},p_\tau)&=& \sqrt{E_\tau+m_\tau}\left( -\sin (\theta_\tau/2),\cos (\theta_\tau/2),
\frac{-|\mathbf{p}_\tau|}{E_\tau+m_\tau}\sin (\theta_\tau/2),\frac{|\mathbf{p}_\tau|}{E_\tau+m_\tau}\cos (\theta_\tau/2) \right) \,,\nonumber \\
v_{\bar{\nu}_\tau}({\textstyle \frac{1}{2}},p_{\bar{\nu}_\tau}) &=& \sqrt{E_\nu}
\left(\begin{array}{c}\cos (\theta_\tau/2) \\ \sin (\theta_\tau/2) \\ -\cos (\theta_\tau/2) \\ -\sin (\theta_\tau/2)  \end{array}\right) \, .
\end{eqnarray}
The polarization vectors of the virtual vector boson in this frame are
\begin{eqnarray}
\label{polvec2}
\epsilon^{\mu *}(t) &=& \left(1;0,0,0\right)\,, \nonumber \\
\epsilon^{\mu *}(\pm 1) &=& \frac{1}{\sqrt{2}}\left(0;\pm 1,-i,0\right) \,,\nonumber \\
\epsilon^{\mu *}(0) &=& \left(0;0,0,-1\right)\,.
\end{eqnarray}
The three-momentum and energy of the $\tau$ lepton in this frame can be written as
\bea
|\mathbf{p}_\tau| &=& \sqrt{q^2}\, v^2/2,\nonumber\\
E_\tau &=& |\mathbf{p}_\tau| + m_\tau^2/\sqrt{q^2},
\eea
where
\beq
v=\sqrt{1-\frac{m_\tau^2}{q^2}}.
\eeq

\pagebreak

\providecommand{\href}[2]{#2}\begingroup\raggedright\endgroup


\begin{thebibliography}{10}

\bibitem{Lees:2013uzd}
{\scshape BaBar} collaboration, J.~P. Lees et~al., \emph{{Measurement of an
  Excess of $\bar{B} \to D^{(*)}\tau^- \bar{\nu}_\tau$ Decays and Implications
  for Charged Higgs Bosons}},
  \href{http://dx.doi.org/10.1103/PhysRevD.88.072012}{\emph{Phys. Rev.} {\bf
  D88} (2013) 072012}, [\href{https://arxiv.org/abs/1303.0571}{{\tt
  1303.0571}}].

\bibitem{Huschle:2015rga}
{\scshape Belle} collaboration, M.~Huschle et~al., \emph{{Measurement of the
  branching ratio of $\bar{B} \to D^{(\ast)} \tau^- \bar{\nu}_\tau$ relative to
  $\bar{B} \to D^{(\ast)} \ell^- \bar{\nu}_\ell$ decays with hadronic tagging
  at Belle}}, \href{http://dx.doi.org/10.1103/PhysRevD.92.072014}{\emph{Phys.
  Rev.} {\bf D92} (2015) 072014}, [\href{https://arxiv.org/abs/1507.03233}{{\tt
  1507.03233}}].

\bibitem{Abdesselam:2016cgx}
{\scshape Belle} collaboration, A.~Abdesselam et~al., \emph{{Measurement of the
  branching ratio of $\bar{B}^0 \rightarrow D^{*+} \tau^- \bar{\nu}_{\tau}$
  relative to $\bar{B}^0 \rightarrow D^{*+} \ell^- \bar{\nu}_{\ell}$ decays
  with a semileptonic tagging method}},
  \href{https://arxiv.org/abs/1603.06711}{{\tt 1603.06711}}.

\bibitem{Aaij:2015yra}
{\scshape LHCb} collaboration, R.~Aaij et~al., \emph{{Measurement of the ratio
  of branching fractions $\mathcal{B}(\bar{B}^0 \to
  D^{*+}\tau^{-}\bar{\nu}_{\tau})/\mathcal{B}(\bar{B}^0 \to
  D^{*+}\mu^{-}\bar{\nu}_{\mu})$}},
  \href{http://dx.doi.org/10.1103/PhysRevLett.115.159901,
  10.1103/PhysRevLett.115.111803}{\emph{Phys. Rev. Lett.} {\bf 115} (2015)
  111803}, [\href{https://arxiv.org/abs/1506.08614}{{\tt 1506.08614}}].

\bibitem{Bailey:2012jg}
J.~A. Bailey et~al., \emph{{Refining new-physics searches in $B \to D \tau \nu$
  decay with lattice QCD}},
  \href{http://dx.doi.org/10.1103/PhysRevLett.109.071802}{\emph{Phys. Rev.
  Lett.} {\bf 109} (2012) 071802}, [\href{https://arxiv.org/abs/1206.4992}{{\tt
  1206.4992}}].

\bibitem{Bailey:2015rga}
{\scshape MILC} collaboration, J.~A. Bailey et~al., \emph{{$B \to D \ell \nu$
  form factors at nonzero recoil and $|V_{cb}|$ from 2+1-flavor lattice QCD}},
  \href{http://dx.doi.org/10.1103/PhysRevD.92.034506}{\emph{Phys. Rev.} {\bf
  D92} (2015) 034506}, [\href{https://arxiv.org/abs/1503.07237}{{\tt
  1503.07237}}].

\bibitem{Na:2015kha}
{\scshape HPQCD} collaboration, H.~Na, C.~M. Bouchard, G.~P. Lepage, C.~Monahan
  and J.~Shigemitsu, \emph{{$B \rightarrow D l \nu$ form factors at nonzero
  recoil and extraction of $|V_{cb}|$}},
  \href{http://dx.doi.org/10.1103/PhysRevD.93.119906,
  10.1103/PhysRevD.92.054510}{\emph{Phys. Rev.} {\bf D92} (2015) 054510},
  [\href{https://arxiv.org/abs/1505.03925}{{\tt 1505.03925}}].

\bibitem{Sakaki:2013bfa}
Y.~Sakaki, M.~Tanaka, A.~Tayduganov and R.~Watanabe, \emph{{Testing leptoquark
  models in $\bar B \to D^{(*)} \tau \bar\nu$}},
  \href{http://dx.doi.org/10.1103/PhysRevD.88.094012}{\emph{Phys. Rev.} {\bf
  D88} (2013) 094012}, [\href{https://arxiv.org/abs/1309.0301}{{\tt
  1309.0301}}].

\bibitem{Wang:2017jow}
Y.-M. Wang, Y.-B. Wei, Y.-L. Shen and C.-D. L$\text{\"u}$, \emph{{Perturbative
  corrections to $B \to D$ form factors in QCD}},
  \href{https://arxiv.org/abs/1701.06810}{{\tt 1701.06810}}.

\bibitem{Fajfer:2012vx}
S.~Fajfer, J.~F. Kamenik and I.~Nisandzic, \emph{{On the $B \to D^* \tau \bar
  \nu_{\tau}$ Sensitivity to New Physics}},
  \href{http://dx.doi.org/10.1103/PhysRevD.85.094025}{\emph{Phys. Rev.} {\bf
  D85} (2012) 094025}, [\href{https://arxiv.org/abs/1203.2654}{{\tt
  1203.2654}}].

\bibitem{HFAGWinter2016}
Y.~Amhis et~al., ``{Averages of $b$-hadron, $c$-hadron, and $\tau$-lepton
  properties as of winter 2016}.''
  \href{http://www.slac.stanford.edu/xorg/hfag/semi/winter16/winter16\_dtaunu.html}{http://www.slac.stanford.edu/xorg/hfag/semi/winter16/winter16\_dtaunu.html},
  2016.

\bibitem{Ricciardi:2016pmh}
G.~Ricciardi, \emph{{Semileptonic and leptonic $B$ decays, circa 2016}},
  \href{http://dx.doi.org/10.1142/S0217732317300051}{\emph{Mod. Phys. Lett.}
  {\bf A32} (2017) 1730005}, [\href{https://arxiv.org/abs/1610.04387}{{\tt
  1610.04387}}].

\bibitem{DeTar}
C.~DeTar, \emph{Private communication},  2016.

\bibitem{Aubert:2009ac}
{\scshape BaBar} collaboration, B.~Aubert et~al., \emph{{Measurement of
  $|V_{cb}|$ and the Form-Factor Slope in $\bar B \to D^{(*)} \tau \bar\nu$
  Decays in Events Tagged by a Fully Reconstructed $B$ Meson}},
  \href{http://dx.doi.org/10.1103/PhysRevLett.104.011802}{\emph{Phys. Rev.
  Lett.} {\bf 104} (2010) 011802}, [\href{https://arxiv.org/abs/0904.4063}{{\tt
  0904.4063}}].

\bibitem{Glattauer:2015teq}
{\scshape Belle} collaboration, R.~Glattauer et~al., \emph{{Measurement of the
  decay $B\to D\ell\nu_\ell$ in fully reconstructed events and determination of
  the Cabibbo-Kobayashi-Maskawa matrix element $|V_{cb}|$}},
  \href{http://dx.doi.org/10.1103/PhysRevD.93.032006}{\emph{Phys. Rev.} {\bf
  D93} (2016) 032006}, [\href{https://arxiv.org/abs/1510.03657}{{\tt
  1510.03657}}].

\bibitem{Fajfer:2012jt}
S.~Fajfer, J.~F. Kamenik, I.~Nisandzic and J.~Zupan, \emph{{Implications of
  Lepton Flavor Universality Violations in $B$ Decays}},
  \href{http://dx.doi.org/10.1103/PhysRevLett.109.161801}{\emph{Phys. Rev.
  Lett.} {\bf 109} (2012) 161801}, [\href{https://arxiv.org/abs/1206.1872}{{\tt
  1206.1872}}].

\bibitem{Crivellin:2012ye}
A.~Crivellin, C.~Greub and A.~Kokulu, \emph{{Explaining $B\to D\tau\nu$, $B\to
  D^*\tau\nu$ and $B\to \tau\nu$ in a 2HDM of type III}},
  \href{http://dx.doi.org/10.1103/PhysRevD.86.054014}{\emph{Phys. Rev.} {\bf
  D86} (2012) 054014}, [\href{https://arxiv.org/abs/1206.2634}{{\tt
  1206.2634}}].

\bibitem{Datta:2012qk}
A.~Datta, M.~Duraisamy and D.~Ghosh, \emph{{Diagnosing New Physics in $b \to c
  \, \tau \, \nu_\tau$ decays in the light of the recent BaBar result}},
  \href{http://dx.doi.org/10.1103/PhysRevD.86.034027}{\emph{Phys. Rev.} {\bf
  D86} (2012) 034027}, [\href{https://arxiv.org/abs/1206.3760}{{\tt
  1206.3760}}].

\bibitem{Becirevic:2012jf}
D.~Becirevic, N.~Kosnik and A.~Tayduganov, \emph{{$\bar B\to D\tau\bar
  \nu_\tau$ vs. $\bar B\to D\mu\bar \nu_\mu$}},
  \href{http://dx.doi.org/10.1016/j.physletb.2012.08.016}{\emph{Phys. Lett.}
  {\bf B716} (2012) 208--213}, [\href{https://arxiv.org/abs/1206.4977}{{\tt
  1206.4977}}].

\bibitem{Deshpande:2012rr}
N.~G. Deshpande and A.~Menon, \emph{{Hints of R-parity violation in B decays
  into $\tau \nu$}},
  \href{http://dx.doi.org/10.1007/JHEP01(2013)025}{\emph{JHEP} {\bf 01} (2013)
  025}, [\href{https://arxiv.org/abs/1208.4134}{{\tt 1208.4134}}].

\bibitem{Celis:2012dk}
A.~Celis, M.~Jung, X.-Q. Li and A.~Pich, \emph{{Sensitivity to charged scalars
  in ${B\to D^{(*)}\tau\nu_\tau}$ and ${B\to\tau\nu_\tau}$ decays}},
  \href{http://dx.doi.org/10.1007/JHEP01(2013)054}{\emph{JHEP} {\bf 01} (2013)
  054}, [\href{https://arxiv.org/abs/1210.8443}{{\tt 1210.8443}}].

\bibitem{Choudhury:2012hn}
D.~Choudhury, D.~K. Ghosh and A.~Kundu, \emph{{B decay anomalies in an
  effective theory}},
  \href{http://dx.doi.org/10.1103/PhysRevD.86.114037}{\emph{Phys. Rev.} {\bf
  D86} (2012) 114037}, [\href{https://arxiv.org/abs/1210.5076}{{\tt
  1210.5076}}].

\bibitem{Tanaka:2012nw}
M.~Tanaka and R.~Watanabe, \emph{{New physics in the weak interaction of $\bar
  B\to D^{(*)}\tau\bar\nu$}},
  \href{http://dx.doi.org/10.1103/PhysRevD.87.034028}{\emph{Phys. Rev.} {\bf
  D87} (2013) 034028}, [\href{https://arxiv.org/abs/1212.1878}{{\tt
  1212.1878}}].

\bibitem{Ko:2012sv}
P.~Ko, Y.~Omura and C.~Yu, \emph{{$B \to D^{(*)} \tau \nu$ and $B\to \tau \nu$
  in chiral $U(1)'$ models with flavored multi Higgs doublets}},
  \href{http://dx.doi.org/10.1007/JHEP03(2013)151}{\emph{JHEP} {\bf 03} (2013)
  151}, [\href{https://arxiv.org/abs/1212.4607}{{\tt 1212.4607}}].

\bibitem{Fan:2013qz}
Y.-Y. Fan, W.-F. Wang, S.~Cheng and Z.-J. Xiao, \emph{{Semileptonic decays $B
  \to D^{(*)} l\nu$ in the perturbative QCD factorization approach}},
  \href{http://dx.doi.org/10.1007/s11434-013-0049-9}{\emph{Chin. Sci. Bull.}
  {\bf 59} (2014) 125--132}, [\href{https://arxiv.org/abs/1301.6246}{{\tt
  1301.6246}}].

\bibitem{Biancofiore:2013ki}
P.~Biancofiore, P.~Colangelo and F.~De~Fazio, \emph{{On the anomalous
  enhancement observed in $B \to D^{(*)}\tau{\bar \nu}_\tau$ decays}},
  \href{http://dx.doi.org/10.1103/PhysRevD.87.074010}{\emph{Phys. Rev.} {\bf
  D87} (2013) 074010}, [\href{https://arxiv.org/abs/1302.1042}{{\tt
  1302.1042}}].

\bibitem{Celis:2013jha}
A.~Celis, M.~Jung, X.-Q. Li and A.~Pich, \emph{{ $B \to D^{(*)}\tau{\bar
  \nu}_\tau$ decays in two-Higgs-doublet models}},
  \href{http://dx.doi.org/10.1088/1742-6596/447/1/012058}{\emph{J. Phys. Conf.
  Ser.} {\bf 447} (2013) 012058}, [\href{https://arxiv.org/abs/1302.5992}{{\tt
  1302.5992}}].

\bibitem{Duraisamy:2013kcw}
M.~Duraisamy and A.~Datta, \emph{{The Full $B \to D^{*} \tau^{-}
  \bar{\nu_\tau}$ Angular Distribution and CP violating Triple Products}},
  \href{http://dx.doi.org/10.1007/JHEP09(2013)059}{\emph{JHEP} {\bf 09} (2013)
  059}, [\href{https://arxiv.org/abs/1302.7031}{{\tt 1302.7031}}].

\bibitem{Dorsner:2013tla}
I.~Dorsner, S.~Fajfer, N.~Kosnik and I.~Nisandzic, \emph{{Minimally flavored
  colored scalar in $\bar B \to D^{(*)} \tau \bar \nu$ and the mass matrices
  constraints}}, \href{http://dx.doi.org/10.1007/JHEP11(2013)084}{\emph{JHEP}
  {\bf 11} (2013) 084}, [\href{https://arxiv.org/abs/1306.6493}{{\tt
  1306.6493}}].

\bibitem{Sakaki:2014sea}
Y.~Sakaki, M.~Tanaka, A.~Tayduganov and R.~Watanabe, \emph{{Probing New Physics
  with $q^2$ distributions in $\bar{B} \to D^{(*)} \tau \bar\nu$}},
  \href{http://dx.doi.org/10.1103/PhysRevD.91.114028}{\emph{Phys. Rev.} {\bf
  D91} (2015) 114028}, [\href{https://arxiv.org/abs/1412.3761}{{\tt
  1412.3761}}].

\bibitem{Bhattacharya:2014wla}
B.~Bhattacharya, A.~Datta, D.~London and S.~Shivashankara, \emph{{Simultaneous
  Explanation of the $R_K$ and $R(D^{(*)})$ Puzzles}},
  \href{http://dx.doi.org/10.1016/j.physletb.2015.02.011}{\emph{Phys. Lett.}
  {\bf B742} (2015) 370--374}, [\href{https://arxiv.org/abs/1412.7164}{{\tt
  1412.7164}}].

\bibitem{Aaij:2014ora}
{\scshape LHCb} collaboration, R.~Aaij et~al., \emph{{Test of lepton
  universality using $B^{+}\rightarrow K^{+}\ell^{+}\ell^{-}$ decays}},
  \href{http://dx.doi.org/10.1103/PhysRevLett.113.151601}{\emph{Phys. Rev.
  Lett.} {\bf 113} (2014) 151601}, [\href{https://arxiv.org/abs/1406.6482}{{\tt
  1406.6482}}].

\bibitem{Aaij:2013qta}
{\scshape LHCb} collaboration, R.~Aaij et~al., \emph{{Measurement of
  Form-Factor-Independent Observables in the Decay $B^{0} \to K^{*0} \mu^+
  \mu^-$}}, \href{http://dx.doi.org/10.1103/PhysRevLett.111.191801}{\emph{Phys.
  Rev. Lett.} {\bf 111} (2013) 191801},
  [\href{https://arxiv.org/abs/1308.1707}{{\tt 1308.1707}}].

\bibitem{Aaij:2015oid}
{\scshape LHCb} collaboration, R.~Aaij et~al., \emph{{Angular analysis of the
  $B^{0} \to K^{*0} \mu^{+} \mu^{-}$ decay using 3 fb$^{-1}$ of integrated
  luminosity}}, \href{http://dx.doi.org/10.1007/JHEP02(2016)104}{\emph{JHEP}
  {\bf 02} (2016) 104}, [\href{https://arxiv.org/abs/1512.04442}{{\tt
  1512.04442}}].

\bibitem{Blake:2016olu}
T.~Blake, G.~Lanfranchi and D.~M. Straub, \emph{{Rare $B$ Decays as Tests of
  the Standard Model}},
  \href{http://dx.doi.org/10.1016/j.ppnp.2016.10.001}{\emph{Prog. Part. Nucl.
  Phys.} {\bf 92} (2017) 50--91}, [\href{https://arxiv.org/abs/1606.00916}{{\tt
  1606.00916}}].

\bibitem{Calibbi:2015kma}
L.~Calibbi, A.~Crivellin and T.~Ota, \emph{{Effective Field Theory Approach to
  $b\to s\ell\ell^{(\prime)}$, $B\to K^{(*)}\nu\overline{\nu}$ and $B\to
  D^{(*)}\tau\nu$ with Third Generation Couplings}},
  \href{http://dx.doi.org/10.1103/PhysRevLett.115.181801}{\emph{Phys. Rev.
  Lett.} {\bf 115} (2015) 181801},
  [\href{https://arxiv.org/abs/1506.02661}{{\tt 1506.02661}}].

\bibitem{Greljo:2015mma}
A.~Greljo, G.~Isidori and D.~Marzocca, \emph{{On the breaking of Lepton Flavor
  Universality in $B$ decays}},
  \href{http://dx.doi.org/10.1007/JHEP07(2015)142}{\emph{JHEP} {\bf 07} (2015)
  142}, [\href{https://arxiv.org/abs/1506.01705}{{\tt 1506.01705}}].

\bibitem{Boucenna:2016wpr}
S.~M. Boucenna, A.~Celis, J.~Fuentes-Martin, A.~Vicente and J.~Virto,
  \emph{{Non-abelian gauge extensions for $B$-decay anomalies}},
  \href{http://dx.doi.org/10.1016/j.physletb.2016.06.067}{\emph{Phys. Lett.}
  {\bf B760} (2016) 214--219}, [\href{https://arxiv.org/abs/1604.03088}{{\tt
  1604.03088}}].

\bibitem{Bhattacharya:2016mcc}
B.~Bhattacharya, A.~Datta, J.-P. Guevin, D.~London and R.~Watanabe,
  \emph{{Simultaneous Explanation of the $R_K$ and $R_{D^{(*)}}$ Puzzles: a
  Model Analysis}},
  \href{http://dx.doi.org/10.1007/JHEP01(2017)015}{\emph{JHEP} {\bf 01} (2017)
  015}, [\href{https://arxiv.org/abs/1609.09078}{{\tt 1609.09078}}].

\bibitem{Barbieri:2016las}
R.~Barbieri, C.~W. Murphy and F.~Senia, \emph{{B-decay Anomalies in a Composite
  Leptoquark Model}},
  \href{http://dx.doi.org/10.1140/epjc/s10052-016-4578-7}{\emph{Eur. Phys. J.}
  {\bf C77} (2017) 8}, [\href{https://arxiv.org/abs/1611.04930}{{\tt
  1611.04930}}].

\bibitem{Gutsche:2015mxa}
T.~Gutsche, M.~A. Ivanov, J.~G. Korner, V.~E. Lyubovitskij, P.~Santorelli and
  N.~Habyl, \emph{{Semileptonic decay $\Lambda_b \to \Lambda_c + \tau^- +
  \bar{\nu_\tau}$ in the covariant confined quark model}},
  \href{http://dx.doi.org/10.1103/PhysRevD.91.074001,
  10.1103/PhysRevD.91.119907}{\emph{Phys. Rev.} {\bf D91} (2015) 074001},
  [\href{https://arxiv.org/abs/1502.04864}{{\tt 1502.04864}}].

\bibitem{Woloshyn:2014hka}
R.~M. Woloshyn, \emph{{Semileptonic decay of the $\Lambda_b$ baryon}},
  {\emph{PoS} {\bf Hadron2013} (2013) 203}.

\bibitem{Shivashankara:2015cta}
S.~Shivashankara, W.~Wu and A.~Datta, \emph{{$\Lambda_b \to \Lambda_c \tau
  \bar{\nu}_{\tau}$ Decay in the Standard Model and with New Physics}},
  \href{http://dx.doi.org/10.1103/PhysRevD.91.115003}{\emph{Phys. Rev.} {\bf
  D91} (2015) 115003}, [\href{https://arxiv.org/abs/1502.07230}{{\tt
  1502.07230}}].

\bibitem{Dutta:2015ueb}
R.~Dutta, \emph{{$\Lambda_b \to (\Lambda_c,\,p)\,\tau\,\nu$ decays within
  standard model and beyond}},
  \href{http://dx.doi.org/10.1103/PhysRevD.93.054003}{\emph{Phys. Rev.} {\bf
  D93} (2016) 054003}, [\href{https://arxiv.org/abs/1512.04034}{{\tt
  1512.04034}}].

\bibitem{Faustov:2016pal}
R.~N. Faustov and V.~O. Galkin, \emph{{Semileptonic decays of $\Lambda_b$
  baryons in the relativistic quark model}},
  \href{http://dx.doi.org/10.1103/PhysRevD.94.073008}{\emph{Phys. Rev.} {\bf
  D94} (2016) 073008}, [\href{https://arxiv.org/abs/1609.00199}{{\tt
  1609.00199}}].

\bibitem{Li:2016pdv}
X.-Q. Li, Y.-D. Yang and X.~Zhang,
  \emph{{$\Lambda_b\to\Lambda_c\tau\bar\nu_\tau$ decay in scalar and vector
  leptoquark scenarios}},  \href{https://arxiv.org/abs/1611.01635}{{\tt
  1611.01635}}.

\bibitem{Celis:2016azn}
A.~Celis, M.~Jung, X.-Q. Li and A.~Pich, \emph{{Scalar contributions to $b\to c
  (u) \tau \nu$ transitions}},  \href{https://arxiv.org/abs/1612.07757}{{\tt
  1612.07757}}.

\bibitem{Detmold:2015aaa}
W.~Detmold, C.~Lehner and S.~Meinel, \emph{{$\Lambda_b \to p \ell^-
  \bar{\nu}_\ell$ and $\Lambda_b \to \Lambda_c \ell^- \bar{\nu}_\ell$ form
  factors from lattice QCD with relativistic heavy quarks}},
  \href{http://dx.doi.org/10.1103/PhysRevD.92.034503}{\emph{Phys. Rev.} {\bf
  D92} (2015) 034503}, [\href{https://arxiv.org/abs/1503.01421}{{\tt
  1503.01421}}].

\bibitem{Li:2016vvp}
X.-Q. Li, Y.-D. Yang and X.~Zhang, \emph{{Revisiting the one leptoquark
  solution to the R(D$^{(∗)}$) anomalies and its phenomenological
  implications}}, \href{http://dx.doi.org/10.1007/JHEP08(2016)054}{\emph{JHEP}
  {\bf 08} (2016) 054}, [\href{https://arxiv.org/abs/1605.09308}{{\tt
  1605.09308}}].

\bibitem{Alonso:2016oyd}
R.~Alonso, B.~Grinstein and J.~Martin~Camalich, \emph{{The lifetime of the
  $B_c^-$ meson and the anomalies in $B\to D^{(*)}\tau\nu$}},
  \href{https://arxiv.org/abs/1611.06676}{{\tt 1611.06676}}.

\bibitem{Chen:2005gr}
C.-H. Chen and C.-Q. Geng, \emph{{Lepton angular asymmetries in semileptonic
  charmful $B$ decays}},
  \href{http://dx.doi.org/10.1103/PhysRevD.71.077501}{\emph{Phys. Rev.} {\bf
  D71} (2005) 077501}, [\href{https://arxiv.org/abs/hep-ph/0503123}{{\tt
  hep-ph/0503123}}].

\bibitem{Bhattacharya:2011qm}
T.~Bhattacharya, V.~Cirigliano, S.~D. Cohen, A.~Filipuzzi, M.~Gonzalez-Alonso,
  M.~L. Graesser et~al., \emph{{Probing Novel Scalar and Tensor Interactions
  from (Ultra)Cold Neutrons to the LHC}},
  \href{http://dx.doi.org/10.1103/PhysRevD.85.054512}{\emph{Phys. Rev.} {\bf
  D85} (2012) 054512}, [\href{https://arxiv.org/abs/1110.6448}{{\tt
  1110.6448}}].

\bibitem{Feruglio:2016gvd}
F.~Feruglio, P.~Paradisi and A.~Pattori, \emph{{Revisiting Lepton Flavor
  Universality in B Decays}},
  \href{http://dx.doi.org/10.1103/PhysRevLett.118.011801}{\emph{Phys. Rev.
  Lett.} {\bf 118} (2017) 011801},
  [\href{https://arxiv.org/abs/1606.00524}{{\tt 1606.00524}}].

\bibitem{Feruglio:2017rjo}
F.~Feruglio, P.~Paradisi and A.~Pattori, \emph{{On the Importance of
  Electroweak Corrections for B Anomalies}},
  \href{https://arxiv.org/abs/1705.00929}{{\tt 1705.00929}}.

\bibitem{Feldmann:2011xf}
T.~Feldmann and M.~W.~Y. Yip, \emph{{Form Factors for $\Lambda_b \to \Lambda$
  Transitions in {SCET}}}, \href{http://dx.doi.org/10.1103/PhysRevD.85.014035,
  10.1103/PhysRevD.86.079901}{\emph{Phys. Rev.} {\bf D85} (2012) 014035},
  [\href{https://arxiv.org/abs/1111.1844}{{\tt 1111.1844}}].

\bibitem{Olive:2016xmw}
{\scshape Particle Data Group} collaboration, C.~Patrignani et~al.,
  \emph{{Review of Particle Physics}},
  \href{http://dx.doi.org/10.1088/1674-1137/40/10/100001}{\emph{Chin. Phys.}
  {\bf C40} (2016) 100001}.

\bibitem{Detmold:2016pkz}
W.~Detmold and S.~Meinel, \emph{{$\Lambda_b \to \Lambda \ell^+ \ell^-$ form
  factors, differential branching fraction, and angular observables from
  lattice QCD with relativistic $b$ quarks}},
  \href{http://dx.doi.org/10.1103/PhysRevD.93.074501}{\emph{Phys. Rev.} {\bf
  D93} (2016) 074501}, [\href{https://arxiv.org/abs/1602.01399}{{\tt
  1602.01399}}].

\bibitem{Bourrely:2008za}
C.~Bourrely, I.~Caprini and L.~Lellouch, \emph{{Model-independent description
  of $B \to \pi \ell \nu$ decays and a determination of $|V_{ub}|$}},
  \href{http://dx.doi.org/10.1103/PhysRevD.82.099902,
  10.1103/PhysRevD.79.013008}{\emph{Phys. Rev.} {\bf D79} (2009) 013008},
  [\href{https://arxiv.org/abs/0807.2722}{{\tt 0807.2722}}].

\bibitem{Abdesselam:2016xqt}
A.~Abdesselam et~al., \emph{{Measurement of the $\tau$ lepton polarization in
  the decay ${\bar B} \rightarrow D^* \tau^- {\bar \nu_{\tau}}$}},
  \href{https://arxiv.org/abs/1608.06391}{{\tt 1608.06391}}.

\bibitem{Beneke:1996xe}
M.~Beneke and G.~Buchalla, \emph{{The $B_c$ Meson Lifetime}},
  \href{http://dx.doi.org/10.1103/PhysRevD.53.4991}{\emph{Phys. Rev.} {\bf D53}
  (1996) 4991--5000}, [\href{https://arxiv.org/abs/hep-ph/9601249}{{\tt
  hep-ph/9601249}}].

\bibitem{Colquhoun:2015oha}
{\scshape HPQCD} collaboration, B.~Colquhoun, C.~T.~H. Davies, R.~J. Dowdall,
  J.~Kettle, J.~Koponen, G.~P. Lepage et~al., \emph{{$B$-meson decay constants:
  a more complete picture from full lattice QCD}},
  \href{http://dx.doi.org/10.1103/PhysRevD.91.114509}{\emph{Phys. Rev.} {\bf
  D91} (2015) 114509}, [\href{https://arxiv.org/abs/1503.05762}{{\tt
  1503.05762}}].

\bibitem{Alonso:2015sja}
R.~Alonso, B.~Grinstein and J.~Martin~Camalich, \emph{{Lepton universality
  violation and lepton flavor conservation in $B$-meson decays}},
  \href{http://dx.doi.org/10.1007/JHEP10(2015)184}{\emph{JHEP} {\bf 10} (2015)
  184}, [\href{https://arxiv.org/abs/1505.05164}{{\tt 1505.05164}}].

\bibitem{Ferreira:2010xe}
P.~M. Ferreira, L.~Lavoura and J.~P. Silva, \emph{{Renormalization-group
  constraints on Yukawa alignment in multi-Higgs-doublet models}},
  \href{http://dx.doi.org/10.1016/j.physletb.2010.04.033}{\emph{Phys. Lett.}
  {\bf B688} (2010) 341--344}, [\href{https://arxiv.org/abs/1001.2561}{{\tt
  1001.2561}}].

\bibitem{Pich:2009sp}
A.~Pich and P.~Tuzon, \emph{{Yukawa Alignment in the Two-Higgs-Doublet Model}},
  \href{http://dx.doi.org/10.1103/PhysRevD.80.091702}{\emph{Phys. Rev.} {\bf
  D80} (2009) 091702}, [\href{https://arxiv.org/abs/0908.1554}{{\tt
  0908.1554}}].

\bibitem{Dumont:2016xpj}
B.~Dumont, K.~Nishiwaki and R.~Watanabe, \emph{{LHC constraints and prospects
  for $S_1$ scalar leptoquark explaining the $\bar B \to D^{(*)} \tau \bar\nu$
  anomaly}}, \href{http://dx.doi.org/10.1103/PhysRevD.94.034001}{\emph{Phys.
  Rev.} {\bf D94} (2016) 034001}, [\href{https://arxiv.org/abs/1603.05248}{{\tt
  1603.05248}}].

\bibitem{Lees:2013kla}
{\scshape BaBar} collaboration, J.~P. Lees et~al., \emph{{Search for $B \to
  K^{(*)} \nu \overline \nu$ and invisible quarkonium decays}},
  \href{http://dx.doi.org/10.1103/PhysRevD.87.112005}{\emph{Phys. Rev.} {\bf
  D87} (2013) 112005}, [\href{https://arxiv.org/abs/1303.7465}{{\tt
  1303.7465}}].

\bibitem{Lutz:2013ftz}
{\scshape Belle} collaboration, O.~Lutz et~al., \emph{{Search for $B \to
  h^{(*)} \nu \bar{\nu}$ with the full Belle $\Upsilon(4S)$ data sample}},
  \href{http://dx.doi.org/10.1103/PhysRevD.87.111103}{\emph{Phys. Rev.} {\bf
  D87} (2013) 111103}, [\href{https://arxiv.org/abs/1303.3719}{{\tt
  1303.3719}}].

\bibitem{Buras:2014fpa}
A.~J. Buras, J.~Girrbach-Noe, C.~Niehoff and D.~M. Straub, \emph{{$ B\to
  {K}^{\left(\ast \right)}\nu \overline{\nu} $ decays in the Standard Model and
  beyond}}, \href{http://dx.doi.org/10.1007/JHEP02(2015)184}{\emph{JHEP} {\bf
  02} (2015) 184}, [\href{https://arxiv.org/abs/1409.4557}{{\tt 1409.4557}}].

\bibitem{Buras:1998raa}
A.~J. Buras, \emph{{Weak Hamiltonian, CP violation and rare decays}},  in
  \emph{{Probing the standard model of particle interactions. Proceedings,
  Summer School in Theoretical Physics, NATO Advanced Study Institute, 68th
  session, Les Houches, France, July 28-September 5, 1997. Pt. 1, 2}},
  pp.~281--539, 1998.
\newblock \href{https://arxiv.org/abs/hep-ph/9806471}{{\tt hep-ph/9806471}}.

\bibitem{Auvil:1966eao}
P.~R. Auvil and J.~J. Brehm, \emph{{Wave Functions for Particles of Higher
  Spin}}, \href{http://dx.doi.org/10.1103/PhysRev.145.1152}{\emph{Phys. Rev.}
  {\bf 145} (1966) 1152}.

\end{thebibliography}
\end{document}